\documentclass[amsmath,amssymb,aps,prd,11pt,tightenlines,superscriptaddress,
nofootinbib,preprintnumbers,notitlepage]{revtex4-1}
\usepackage{multirow}
\usepackage{graphicx}
\usepackage{tabularx}
\usepackage{slashed}
\usepackage{url}
\usepackage{hyperref}
\usepackage{feynmp-auto}
\usepackage[maxfloats=100]{morefloats}
\usepackage{color,xcolor}
\usepackage{booktabs}
\usepackage[normalem]{ulem}

\newcommand{\yu}{\ensuremath{y_{\tilde{u}}}}
\newcommand{\msu}{\ensuremath{m_{\tilde{u}}}}
\newcommand{\mst}{\ensuremath{m_{\tilde{t}}}}
\newcommand{\mmed}{\ensuremath{m_\text{med}}}

\newcommand{\lag}{\ensuremath{\mathcal{L}}}

\newcommand{\met}{\ensuremath{\slashed{E}_T}}

\newcommand{\br}{\text{BR}}

\newcommand{\qqqquad}{\qquad \qquad \qquad}

\newcommand{\gev}{{\ensuremath\rm GeV}}
\newcommand{\tev}{{\ensuremath\rm TeV}}

\newcommand{\ifb}{{\ensuremath\rm fb^{-1}}}

\def\slashchar#1{\setbox0=\hbox{$#1$}           
   \dimen0=\wd0                                 
   \setbox1=\hbox{/} \dimen1=\wd1               
   \ifdim\dimen0>\dimen1                        
      \rlap{\hbox to \dimen0{\hfil/\hfil}}      
      #1                                        
   \else                                        
      \rlap{\hbox to \dimen1{\hfil$#1$\hfil}}   
      /                                         
   \fi}

\def\eg{\textsl{e.g. }}
\def\ie{\textsl{i.e. }}

\begin{document}


\title{On the Validity of Dark Matter Effective Theory} 

\author{Martin Bauer}
\affiliation{Institut f\"ur Theoretische Physik, Universit\"at Heidelberg, Germany}

\author{Anja Butter}
\affiliation{Institut f\"ur Theoretische Physik, Universit\"at Heidelberg, Germany}

\author{Nishita Desai}
\affiliation{Laboratoire Charles Coulomb (L2C) \& Laboratoire Univers et Particules de Montpellier (LUPM), CNRS-Universit\'e de Montpellier, France}
\author{J. Gonzalez-Fraile}
\affiliation{Institut f\"ur Theoretische Physik, Universit\"at Heidelberg, Germany}

\author{Tilman Plehn}
\affiliation{Institut f\"ur Theoretische Physik, Universit\"at Heidelberg, Germany}

\begin{abstract} 
  An effective theory of dark matter offers an attractive framework
  for global analyses of dark matter. In the light of global fits we
  test the validity of the link between the non-relativistic dark
  matter annihilation, or the predicted relic density, and LHC
  signatures. Specifically, we study how well the effective theory
  describes the main features of simple models with s-channel and
  t-channel mediators coupling to the Standard Model at tree level or
  through one-loop diagrams. Our results indicate that global dark
  matter analyses in terms of effective Lagrangians are highly
  non-trivial to interpret in term of actual models.
\end{abstract}

\maketitle
\tableofcontents


\section{Introduction}
\label{sec:intro}

An effective field theory~\cite{eftfoundations,eftorig} of dark matter
can systematically describe the effects of heavy, non-propagating
mediators between the dark sector and the Standard Model. In
particular for light dark matter with $m_\chi \lesssim v$, an
effective Lagrangian can describe the effects of heavier
mediators~\cite{tim,eft_dm_2,eft_dm_fit}, always assuming
\begin{align}
 \mmed > m_\chi \; .
\label{eq:hierarchy}
\end{align}
The corresponding effective Lagrangian then includes a new physics
scale $\Lambda \sim \mmed$ and the corresponding coupling $g$. This
description is generally accepted for processes with a low external
energy scale and provided the effective Lagrangian expansion does not
interfere with the velocity suppression. This includes dark matter
annihilation in the early universe~\cite{eft_dm_2,eft_relic}, dark
matter annihilation with decay products observed
today~\cite{eft_indir}, or direct dark matter searches based on
scattering with nuclei~\cite{tim,eft_dir}. Issues with this picture
occur for dark matter production at the LHC~\cite{eft_coll}. In this
paper we attempt a comprehensive, quantitative study of the
limitations of an effective theory of dark matter.\medskip

The main question is \textsl{if an effective theory can be used to
  link very different observations as part of a global dark matter
  analysis, and if in this link it correctly represents the patterns
  of a full model behind it.}  From detailed studies of effective
theories in the Higgs sector at the LHC~\cite{yr4} we know that simple
arguments based on dimensional analysis and scale
estimates~\cite{validity_123,last_gasp} are not always well-suited for
LHC physics.  Instead, we should establish the (non-)applicability of
effective theory approximations model-by-model and
observable-by-observable~\cite{too_long,mvh,square_dim_6_others,comparison}. Obviously,
we will focus on standard LHC processes as the limiting factors. We
will study mono-jet or related signatures, where the relevant
observables are the transverse momentum distribution of the missing
particles recoiling against one or more jets and, to some degree, the
total rate.  The only additional ingredient we will use is the relic
density as a rough guideline if our model for the dark sector could be
responsible for the observed dark matter
density~\cite{eft_dm_fit,eft_dm_2}.\medskip

All models in our study will include simple dark matter sectors,
allowing for a straightforward test if an observed missing energy
signal is likely to be related to the dark matter relic density in our
universe. The reasoning behind this link is that from a practical
perspective the LHC either adds some relevant piece of information to
a global dark matter analysis --- or the question if the LHC observable
is well described by the effective Lagrangian is
irrelevant\footnote{To search for missing transverse momentum at the
  LHC neither requires a well-defined model framework nor an effective
  theory justification.}.  To get an idea what kind of thermal dark
matter signal we would be looking for at the LHC, we can estimate the
observed relic density, for example assuming the usual $2 \to 2$
annihilation process mediated by a dimension-6 operator
\begin{align}
 \langle \sigma_\text{ann} v \rangle 
\propto \frac{g^4 E^2}{4 \pi \, \mmed^4} 
\sim \frac{g^4 m_\chi^2}{4 \pi \, \mmed^4} \; .
\label{eq:eft_sigann}
\end{align}
We will at least roughly let the observed relic density guide us
through the dark sector's parameter space.  Typically, the above
scaling gives us a lower limit on the ratio $m_\chi/\mmed^2$, an upper
limit on the mediator mass, or a lower limit on LHC production cross
sections.  The rough relation between the mediator and dark matter
masses is
\begin{alignat}{5}
\frac{\mmed^2}{g^2 m_\chi} = 8~\tev
&\qquad \stackrel{m_\chi = 10~\gev}{\Rightarrow} \qquad 
&\frac{\mmed}{g} &=  150~\gev \notag \\
&\qquad \stackrel{m_\chi = \mmed/2}{\Rightarrow} \qquad 
&\frac{\mmed}{g} &=  2~\tev \; .
\label{eq:relic_rough}
\end{alignat}
In this simple model the dark matter agent can be very light, but the
mediator will typically be heavy. For weakly interacting models the
appropriate dark matter and mediator mass scales will
decrease. Similarly, if we only require our dark matter agent to be
responsible for part of the observed relic density, the annihilation
cross section can be larger, or the mass scales can be (slightly)
smaller.\medskip

Just as a side remark, for an $s$-channel mediator with the pole
condition $\mmed = 2 m_\chi$ the annihilation rate scales like $g^4
m_\chi^2/(\mmed^2 \Gamma_\text{med}^2)$, introducing the width as an
additional scale and strongly reducing the required couplings.  We
will not follow this well-known path in Higgs portal scenarios and
instead study generic $2 \to 2$ annihilation processes.

\subsubsection*{Effective Lagrangian for LHC}

Additional external energy scales are introduced by the hadron
collider environment of a given dark matter process, now leading to
three relevant energy scales
\begin{align}
\{ \, m_\chi, \mmed, \sqrt{s} \, \} \; .
\end{align}
For dark matter annihilation, indirect detection, and direct detection
these additional energy scales are no challenge to the validity of the
effective theory. At the LHC the situation is less obvious, because
kinematic cuts or a potential kinematic link between the mediator mass
$\mmed$ and the energy transfer can push the momentum transfer of the
process to energy scales $\sqrt{s} > m_\chi$.  One promising, general dark
matter search targets jet production with a pair of dark matter
particles. The key observables here are the $\met$ and $p_{T,j}$
distributions. For simple hard processes the two transverse momentum
distributions are rapidly dropping and strongly correlated. They define
the relevant energy scales
\begin{align}
\{ \, m_\chi, \mmed, \met^\text{min} \, \} \; .
\end{align}
The experimentally relevant $\met$ or $p_{T,j}$ regime is given by a
combination of the signal mass scale, the kinematics of the
dominant $Z_{\nu \nu}$+jets background, and triggering. Our effective
theory has to reproduce two key observables,
\begin{align}
\sigma_\text{tot}(m_\chi,\mmed) \Bigg|_\text{acceptance}
\qquad \text{and} \qquad
\frac{d \, \sigma(m_\chi,\mmed)}{d \, \met} 
\sim \frac{d \, \sigma(m_\chi,\mmed)}{d \, p_{T,j}} \; ,
\end{align}
the latter over the relevant phase space regime. These two LHC
observables will guide us through the different models in our study.

Finally, the hadronic LHC energy of 13~TeV, combined with reasonable
parton momentum fractions defines an absolute upper limit, above which
for example a particle in the $s$-channel cannot be produced as a
propagating state. This adds another energy scale to our
setup,
\begin{align}
\{ \, m_\chi, \mmed, \met^\text{min}, \sqrt{s}_\text{max} \, \} \; .
\end{align}
These four scales define the framework of our effective theory
considerations at the LHC.

\subsubsection*{Simplified models}

As described above, any effective theory description of dark matter
relies on some fundamental assumptions on the new physics
spectrum. For example, if the hierarchy in Eq.\eqref{eq:hierarchy} is
inverted,
\begin{align}
 \mmed < m_\chi \; ,
\label{eq:hierarchy2}
\end{align}
the effective theory is often not applicable. We can instead describe
dark matter annihilation and production through a simplified model
with a dynamic mediator field~\cite{simplified,comparison}. The relic
density becomes largely independent of the light mediator mass, and
the usual WIMP condition gives the dark matter mass through
\begin{align}
 \langle \sigma_\text{ann} v \rangle 
\sim \frac{g^4}{4 \pi \, m_\chi^2} 
\qquad \Rightarrow \qquad 
\frac{m_\chi}{g^2} \sim 4~\tev \; .
\end{align}
The phenomenology of such a model can be illustrated with the Higgs
portal story: the light mediator will be produced directly, and it will
decay to two light Standard Model states through its production
coupling. A decay to two dark matter agents will only occur for
off-shell mediator production and will be strongly suppressed.  In
other words, the LHC we will look for light new resonances, and this
signature will not allow us to identify the mediator nature of this
state or say anything about dark matter.\medskip

In terms of LHC energy scales this situation reads more specifically
\begin{align}
 \mmed < \sqrt{s}_\text{max}, m_\chi  \; .
\label{eq:hierarchy3}
\end{align}
The mediator appears as the propagating degree of freedom at the
LHC.  The spin and quantum numbers of the mediator states play a
significant role; vector mediators can couple to light
quarks~\cite{zprime}, while scalar mediators typically couple to top
quarks~\cite{top_scalar,matt_dorival,Mattelaer:2015haa}. Sizeable light-quark
couplings of scalar mediators in the $s$-channel are strongly
constrained by flavor physics~\cite{Dolan:2014ska}. Realistic scalar
mediators arise from mixing with the SM Higgs~\cite{higgs_mixing}.
This scenario is strongly constrained by Higgs physics, but allows for
an electroweak mono-jet signal comparable in size to the QCD-induced mono-jet
signal~\cite{Bauer:2016gys}.  Colored scalar mediators with
$t$-channel interactions can have large, flavor-specific couplings,
because the mediator is a triplet in flavor space~\cite{flavored}.\medskip

While at the LHC the search for jets plus missing energy --- or
mono-jet signatures --- is fully established, the nature of mediators
can also be studied in a variety of mono-X final states ($X=\gamma, Z,
W$). These rates are typically at least an order of magnitude smaller
than mono-jet cross-sections.  However, for gluon or photon
radiation the up-type and down-type quark initial states do not
interfere, and the sign of the relative couplings are irrelevant. In
contrast, the mono-$W$ signature with opposite-sign couplings to
up-type and down-type quarks can exceed the mono-jet signal for
large transverse momenta~\cite{bell}. This enhancement can be linked
to an eventually unitarity-violating part of the
amplitude~\cite{Haisch:2016usn}.  A gauge-invariant implementation of
such a model will be tamed by $Z-Z'$ mixing, or more general new
physics through higher-dimensional operators. Similar, power-enhanced
unitarity violation also exist in mono-$Z$ and mono-Higgs signatures
for vector mediators with axial-vector
couplings~\cite{simplified_complete}.  A logarithmic divergence exists
in mono-$Z$ production mediated through scalar mediators at one
loop~\cite{Mattelaer:2015haa}.  However, aside from mono-$W$
production, these divergences do not lead to actual problems at the
LHC~\cite{Englert:2016joy}, and mono-jet searches are still the key
search strategy for mediators at the LHC.

\clearpage
\section{Tree-level scalar in t-channel}
\label{sec:t_channel}

 One dark matter scenario, inspired for example by supersymmetry, is a
fermionic dark matter candidate combined with a scalar mediator.  The
mediator $\tilde{u}$ couples the dark matter fermion $\chi$ to an
up-type quark, implying that it carries a color charge in the
fundamental representation. Any model for such a $t$-channel mediator
has to assume at least two interactions,
\begin{align}
u-\tilde{u}-\chi \qquad \text{dark matter annihilation}
\qqqquad 
\tilde{u}-\tilde{u}-g(-g) \qquad \text{QCD} \; .
\end{align}
The first interaction has no structural counterpart in the Standard
Model, so we are free to choose its size. In supersymmetric models it
is fixed by the corresponding electroweak gauge or Yukawa
couplings~\cite{martin_vaughn}.  Because of this interaction the
$t$-channel mediator always has to be heavier than then dark matter
fermion,
\begin{align}
\msu > m_\chi + m_q \approx m_\chi \; , 
\end{align}
otherwise the dark matter agent decays. The second interaction occurs
through the QCD part of the covariant derivative. The mediator will
also have $Z$ and $\gamma$ interactions from the covariant derivative,
but we assume them to be sub-leading at the LHC.  In our very simple
toy model the dark matter interaction is described by 
\begin{align}
\lag \supset \yu\ \left( \bar{u}_R \chi \right) \ \tilde{u} + \text{h.c.}
\label{eq:t_model}
\end{align}
We assume $\yu =1$, unless the value is specified. Naively,
the mono-jet production rate scales like $\sigma_{\met+j}
\propto g_s^2 \yu^4$.  The mediator width
becomes~\cite{bai_berger}
\begin{align}
\frac{\Gamma_{\tilde{u}}}{\msu}
= \frac{\yu^2}{16\pi} \; 
  \frac{\left(\msu^2-m_\chi^2\right)^2}{\msu^4}
< \frac{\yu^2}{16\pi} 
\lesssim 2\% \; .
\label{eq:t_width}
\end{align}
As long as its coupling to SM fermions matches the flavor structure of
the Standard Model, the $t$-channel mediator is a narrow resonance
over our entire relevant parameter space.\medskip

\begin{figure}[t]
\includegraphics[width=0.35\textwidth]{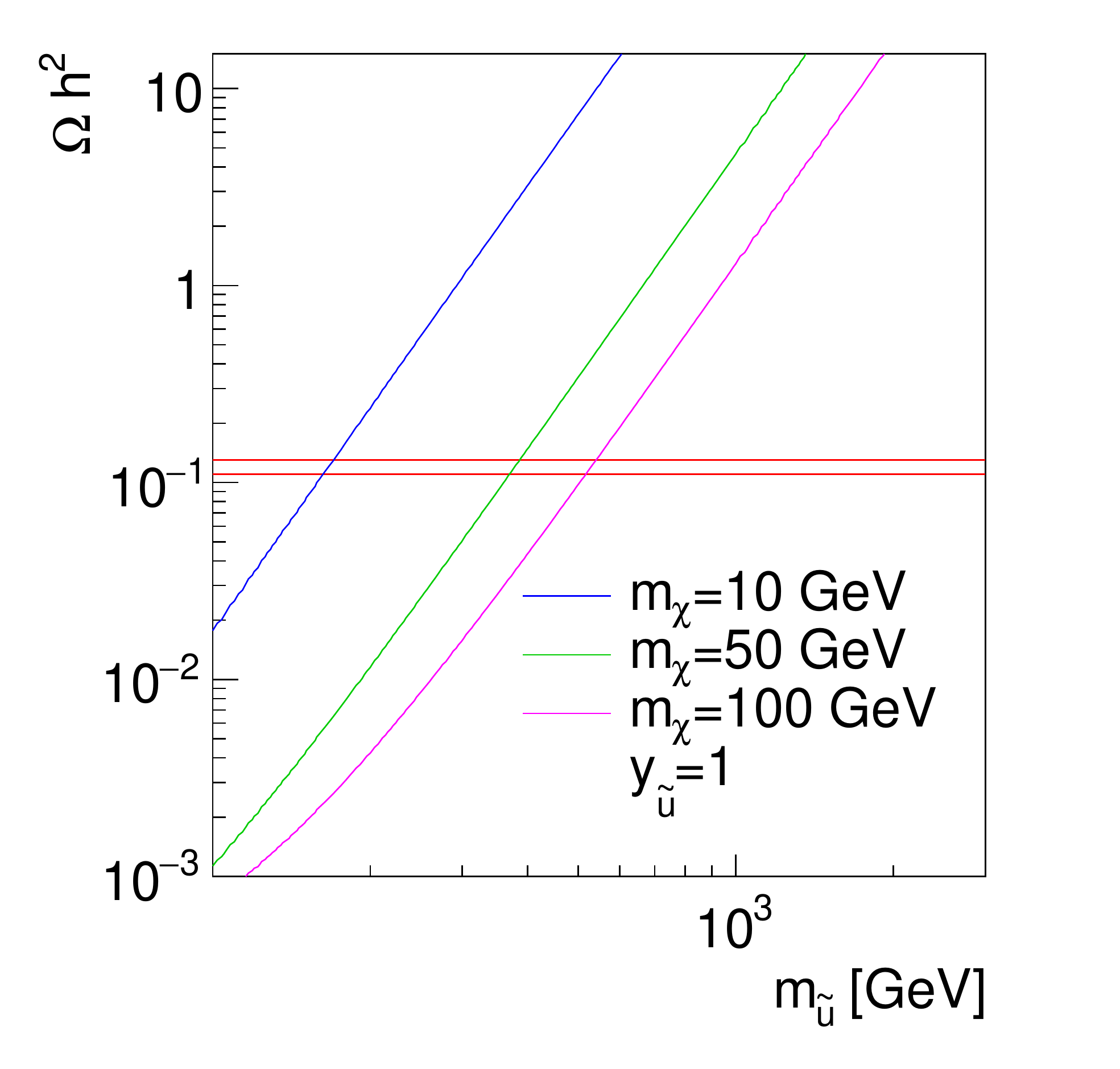}\hspace*{-0.4cm}
\includegraphics[width=0.35\textwidth]{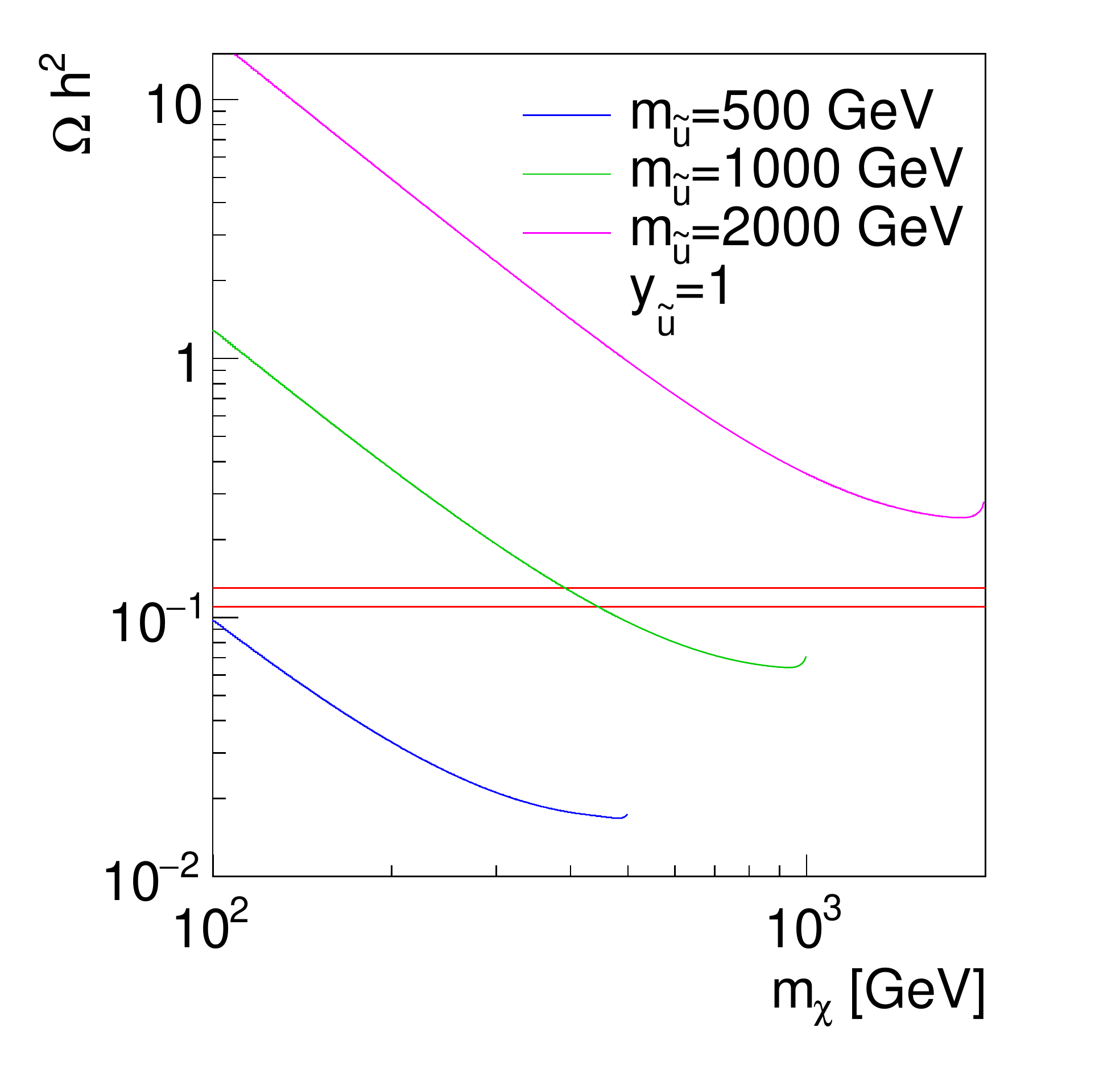}\hspace*{-0.4cm}
\includegraphics[width=0.35\textwidth]{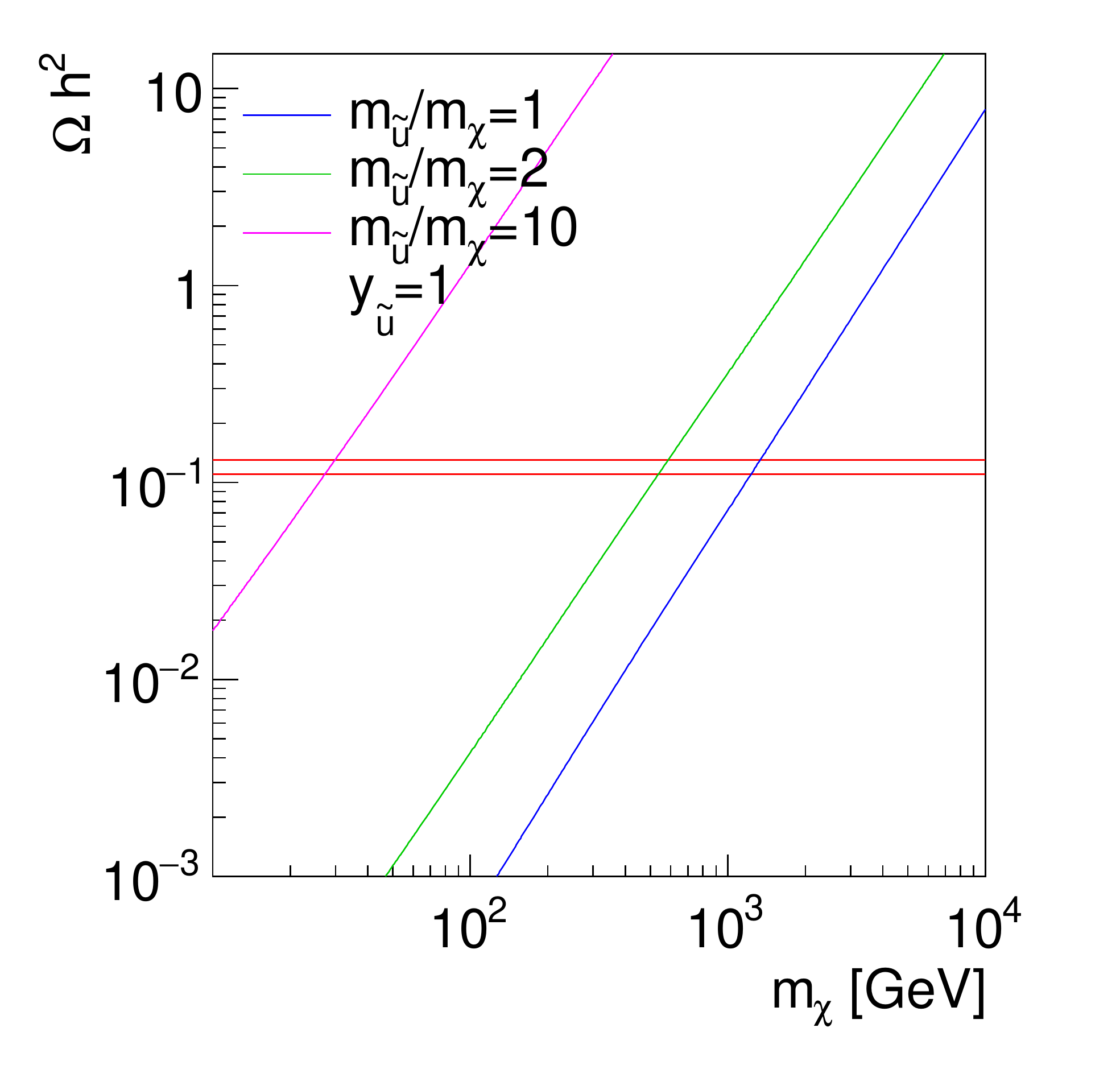}
\caption{Relic density for the $t$-channel mediator model as a
  function of the mediator mass for constant dark matter mass (left),
  as a function of the dark matter mass for constant mediator mass
  (center) and as a function of the dark matter mass for a constant
  ratio of mediator to dark matter mass. We assume $\yu=1$.}
\label{fig:relic_tchannel}
\end{figure}

From this Lagrangian we can compute the relic density using
\textsc{FeynRules}~\cite{feynrules} and
\textsc{Micromegas}~\cite{micromegas}. In
Fig.~\ref{fig:relic_tchannel} we show the predicted relic density for
different cuts in the two-dimensional dark matter vs mediator
parameter space. In the left panel we see that for light dark matter
the measured relic density requires a comparably light mediator within
the reach of the LHC. The numbers are slightly different than in
Eq.\eqref{eq:relic_rough} because of the color factor $N_c$ in the
annihilation rate.  In the second panel we see that for too heavy
mediators there is no dark matter mass which predicts the correct
relic density. This occurs because the required dark matter mass
becomes too large, eventually exceeding the mediator mass and leading
to dark matter decay channels. Finally, in the right panel we confirm
the inverse quadratic scaling with a linked dark matter and mediator
mass scale and the clear preference for ratios $\msu > m_\chi$. In
essence, the observed relic density is in accordance with an effective
theory description of the $t$-channel mediator model. Note that if we
choose a smaller coupling, for example $\yu \lesssim 1/2$, the model
will not give the correct relic density for parameter choices allowed
by LHC. If the $t$-channel mediator only couples along the lines of
the Standard Model flavor structure, additional mediators for example
carrying lepton number can even dominate the dark matter
annihilation rate.\medskip

\begin{figure}[b!]
\begin{center}
\begin{fmffile}{feyn3}
\begin{fmfgraph*}(100,60)
\fmfset{arrow_len}{2mm}
\fmfleft{i1,i2}
\fmfright{o1,o2,o3}
\fmf{fermion,tension=0.4,width=0.6}{i2,v2}
\fmf{fermion,tension=2.5,width=0.6}{v3,v1}
\fmf{fermion,tension=3,width=0.6}{v1,i1}
\fmf{scalar,tension=0.1,label=$\tilde{u}$,l.side=left,width=0.6}{v2,v3}
\fmf{plain,tension=1.5,width=0.6}{v3,o2} 
\fmf{plain,tension=0.6,width=0.6}{v2,o3}
\fmf{gluon,tension= 0.1,width=0.6}{v1,o1}
\fmflabel{$u$}{i1}
\fmflabel{$u$}{i2}
\fmflabel{$g$}{o1}
\fmflabel{$\chi$}{o2}
\fmflabel{$\chi$}{o3}
\fmfv{decor.shape=circle,decor.filled=full,decor.size=2thick}{v2}
\fmfv{decor.shape=circle,decor.filled=full,decor.size=2thick}{v3}
\end{fmfgraph*}
\hspace{1.5cm}
\begin{fmfgraph*}(100,60)
\fmfset{arrow_len}{2mm}
\fmfleft{i1,i2}
\fmfright{o1,o2,o3}
\fmf{plain,tension=0.35,width=0.6}{o1,v2}
\fmf{fermion,width=0.6}{v1,v2}
\fmf{fermion,width=0.6}{i2,v1}
\fmf{gluon,width=0.6}{i1,v1}
\fmf{scalar,tension=1.0,label=$\tilde{u}$,l.side=left,width=0.6}{v2,v3}
\fmf{plain,tension=1.2,width=0.6}{v3,o3}
\fmf{fermion,tension=1.2,width=0.6}{v3,o2}
\fmflabel{$g$}{i1}
\fmflabel{$u$}{i2}
\fmflabel{$\chi$}{o1}
\fmflabel{$u$}{o2}
\fmflabel{$\chi$}{o3}
\fmfv{decor.shape=circle,decor.filled=full,decor.size=2thick}{v2}
\fmfv{decor.shape=circle,decor.filled=full,decor.size=2thick}{v3}
\end{fmfgraph*}
\hspace{1.5cm}
\begin{fmfgraph*}(100,60)
\fmfset{arrow_len}{2mm}
\fmfleft{i1,i2}
\fmfright{o1,o2,o3,o4}
\fmf{fermion,width=0.6}{i1,v1}
\fmf{fermion,width=0.6}{v1,i2}
\fmf{gluon,width=0.6}{v1,v2}
\fmf{scalar,tension=1,label=$\tilde{u}$,l.side=left,width=0.6}{v2,v3}
\fmf{fermion,tension=1,width=0.6}{v3,o4}
\fmf{plain,tension=1,width=0.6}{v3,o3}
\fmf{scalar,tension=1,label=$\tilde{u}$,l.side=left,width=0.6}{v4,v2}
\fmf{plain,tension=1,width=0.6}{v4,o2}
\fmf{fermion,tension=1,width=0.6}{o1,v4}
\fmflabel{$u$}{i1}
\fmflabel{$u$}{i2}
\fmflabel{$\chi$}{o3}
\fmflabel{$\chi$}{o2}
\fmflabel{$u$}{o4}
\fmflabel{$u$}{o1}
\fmfv{decor.shape=circle,decor.filled=full,decor.size=2thick}{v4}
\fmfv{decor.shape=circle,decor.filled=full,decor.size=2thick}{v3}
\end{fmfgraph*}
\end{fmffile}
\end{center}
\caption{Feynman diagrams describing dark matter production in the
  $t$-channel mediator model defined in Eq.\eqref{eq:t_model}.}
\label{fig:feyn_tchannel}
\end{figure}
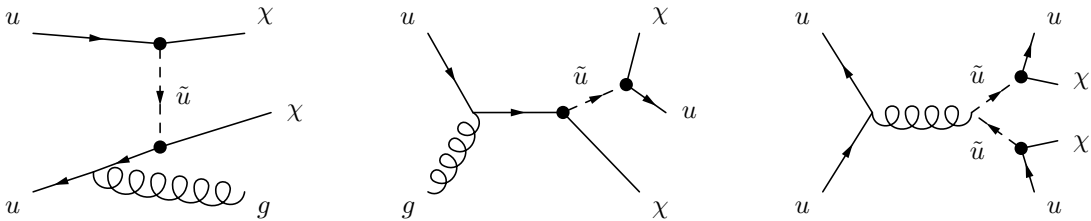

At the LHC the $t$-channel mediator model has very distinct features,
illustrated in Fig.~\ref{fig:feyn_tchannel}. The 
two partonic processes:
\begin{align}
u \bar{u} \to \chi \bar \chi g 
\qquad \text{and} \qquad 
u g \to \chi \bar \chi u 
\label{eq:t_monojet}
\end{align}
are of the same order in perturbation theory and experimentally
indistinguishable. The second of these two processes can be dominated
by an on-shell production of the mediator,
\begin{align}
u g \to \chi \tilde{u} \to \chi \; (\bar \chi u) \; .
\label{eq:t_single_pole}
\end{align}
As for the usual $2 \to 2$ annihilation process, we can cross this
amplitude into an annihilation process, now describing the
co-annihilation $\chi \tilde{u} \to u g$. The main difference between
the (co-) annihilation and LHC interpretations of this amplitude is
that in the prediction of the relic density it only contributes for
$\msu \lesssim m_\chi + 10\%$, while at the LHC it dominates the
mono-jet rates even for $\msu \gg m_\chi$. This challenges the
theoretical link between LHC production and dark matter annihilation.

Finally, we can pair-produce the strongly interacting mediators
through the third Feynman diagram in Fig.~\ref{fig:feyn_tchannel},
\begin{align} 
q\bar{q}/gg 
\quad \stackrel{\text{QCD}}{\longrightarrow} \quad \tilde{u} \tilde{u}^* 
\stackrel{\text{dark matter}}{\longrightarrow} (\bar \chi u) \, (\chi \bar u) \; . 
\label{eq:t_double_pole}
\end{align}
From many studies in the supersymmetry framework we know that for a
wide range of mediator masses this pair production process completely
dominates the $\chi \chi$+jets process. If we can identify the $\chi
\chi jj$ final state we can use the $m_{T2}$ distribution to extract
the masses of both the mediator and the dark matter candidate.  On the
other hand, the process is entirely QCD-mediated and the 100\%
branching fraction gives us no information about the $\bar{u} - \chi -
\tilde{u}$ interaction.  In other words, in the presence of this
on-shell production process there is no link between LHC observables
and dark matter properties of our $t$-channel (simplified)
model.\medskip

For the pair-production process we can apply the search results for
a single supersymmetric squark. With $20~\ifb$ data at 8~TeV ATLAS
excludes $\msu < 470$ GeV for $m_\chi < 100$ GeV~\cite{Aad:2014wea},
comparable to the 13~TeV limits using
$3.2~\ifb$~\cite{Aaboud:2016zdn}. For small mass difference between
the mediator and the DM particle, there exists an ATLAS mono-jet
search which considers pair production in the simplified model and
sets a limit $\msu \lesssim 260~...~300$~GeV for $\msu - m_\chi
\sim 20$ GeV~\cite{Aaboud:2016tnv}. Single-resonant or direct $\chi
\chi$ production should lead to at least comparable results.  Similar
limits from CMS rule out $\msu < 450$~GeV for $m_\chi <
100$~GeV with $19.2~\ifb$ at
13~TeV~\cite{CMS-PAS-SUS-16-014}. Mediator masses below 300~GeV
are ruled out altogether.

\subsubsection*{Total rate}

\begin{figure}[t]
  \includegraphics[width=0.33\textwidth]{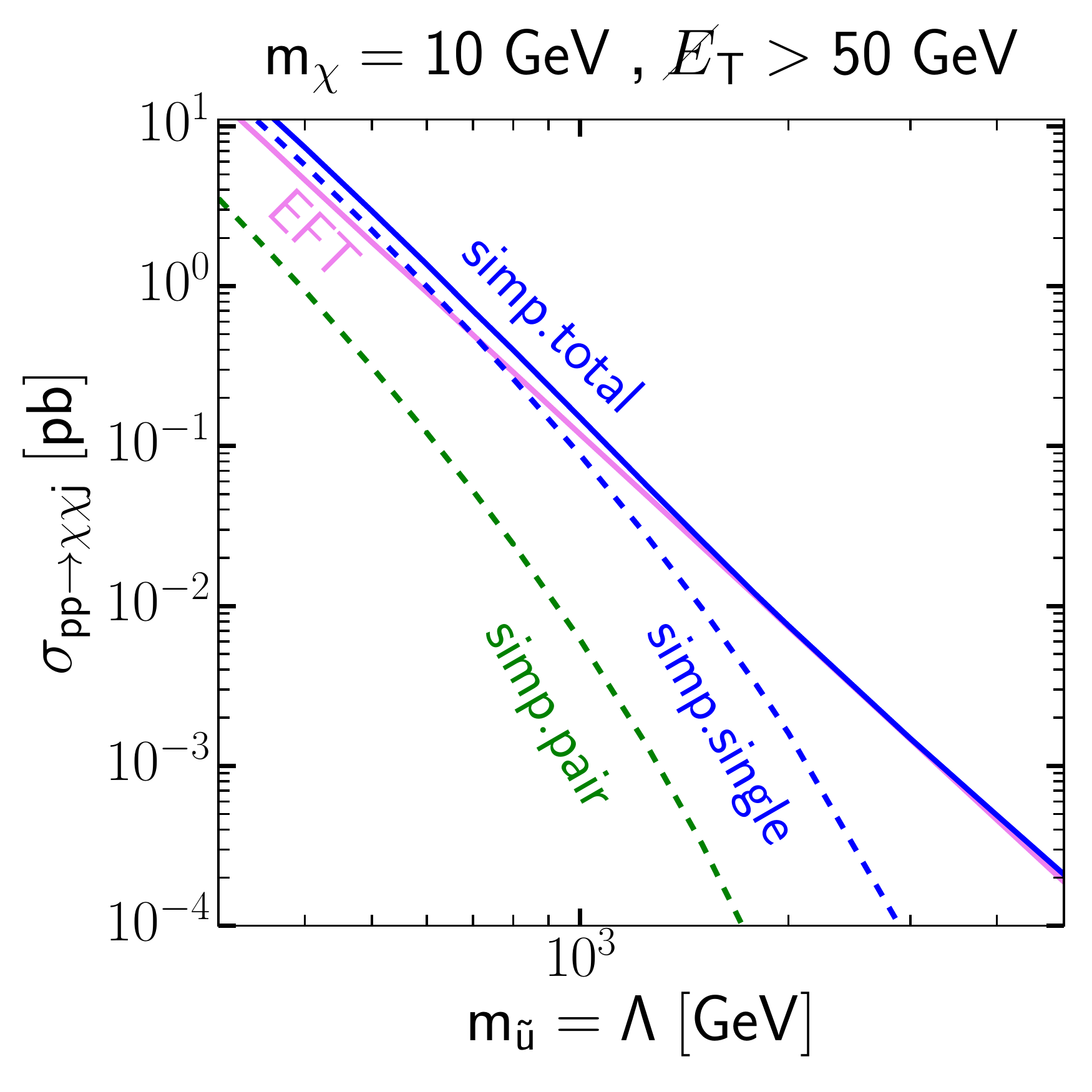}\hspace*{-0.2cm}
  \includegraphics[width=0.33\textwidth]{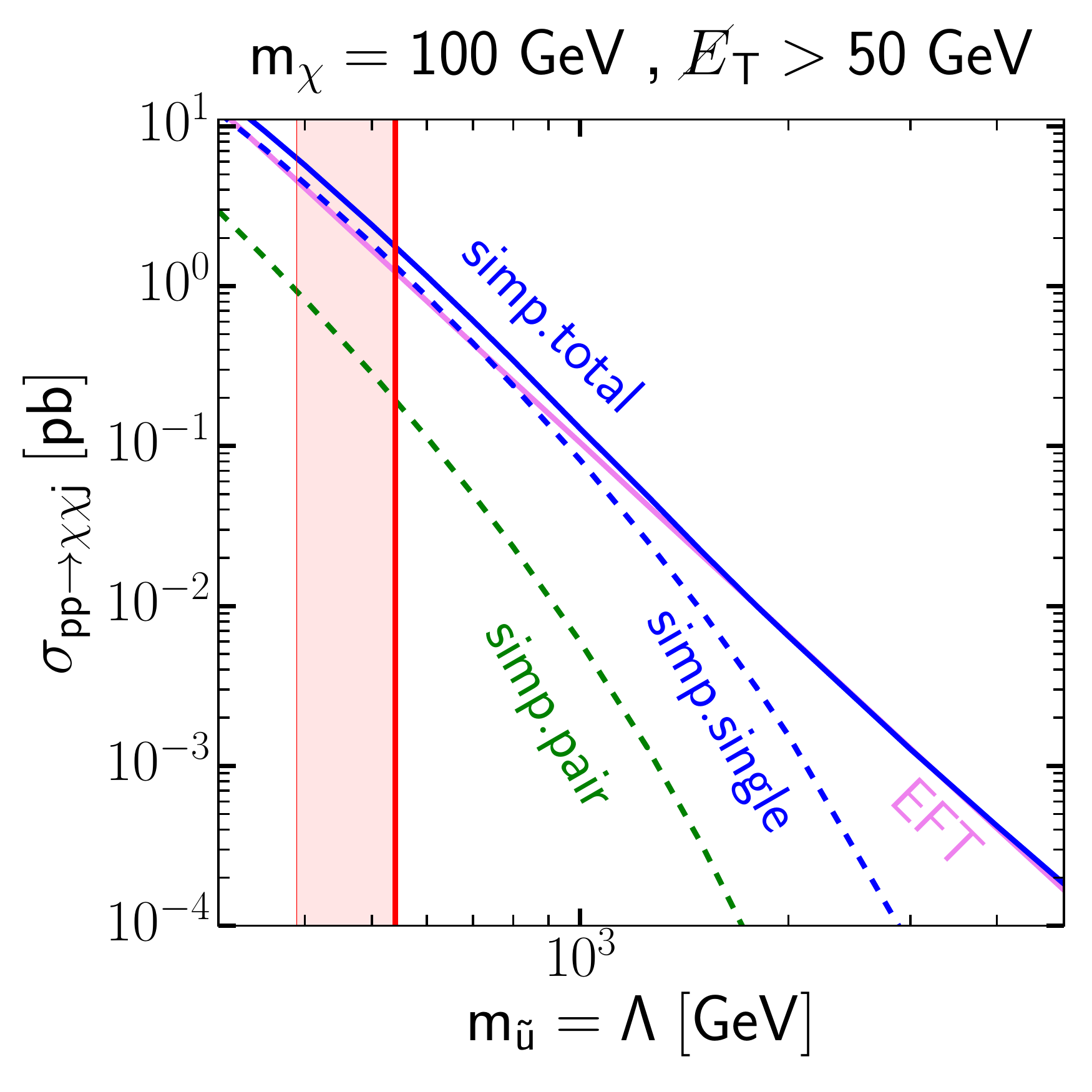}\hspace*{-0.2cm}
  \includegraphics[width=0.33\textwidth]{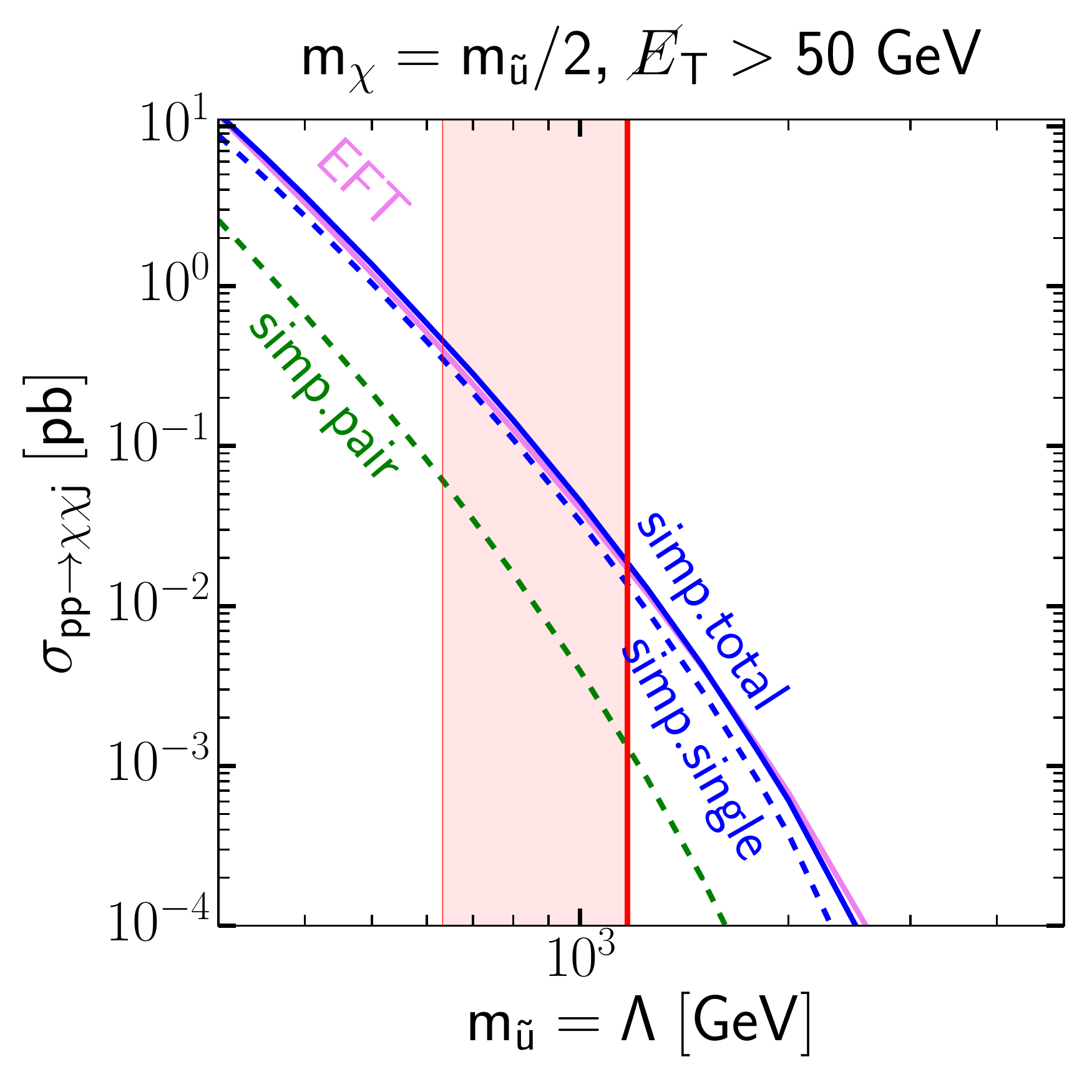} \\
  \includegraphics[width=0.33\textwidth]{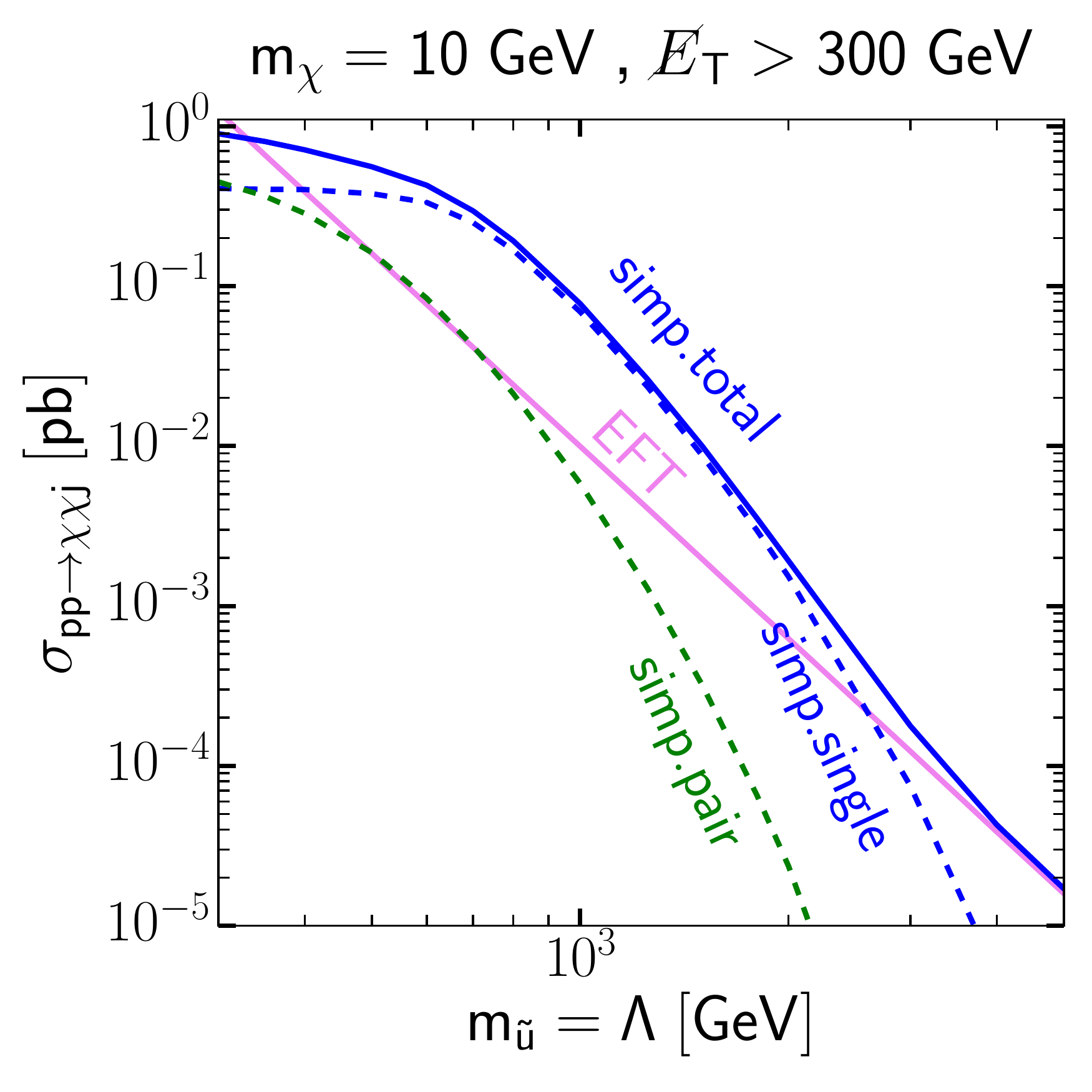}\hspace*{-0.2cm}
  \includegraphics[width=0.33\textwidth]{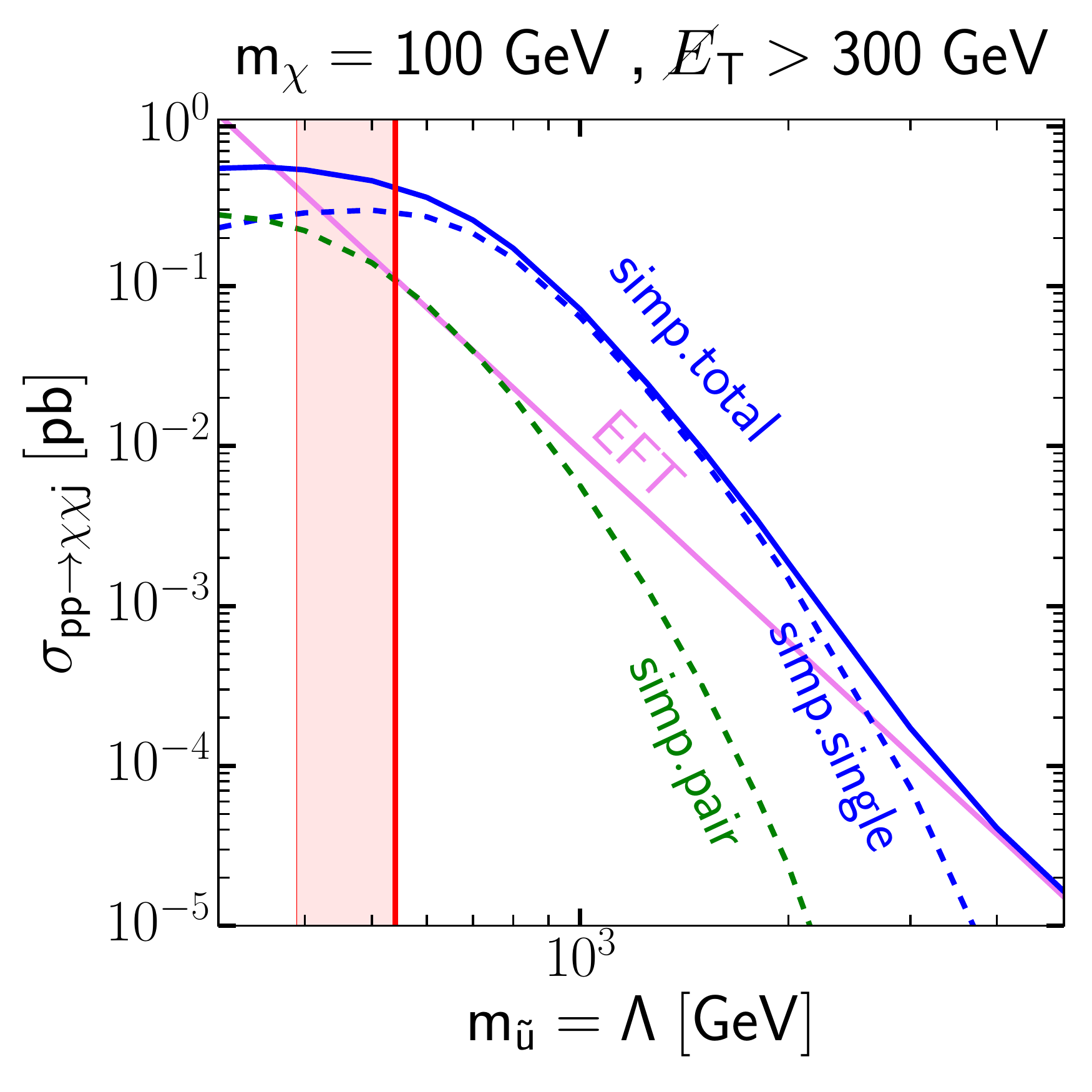}\hspace*{-0.2cm}
  \includegraphics[width=0.33\textwidth]{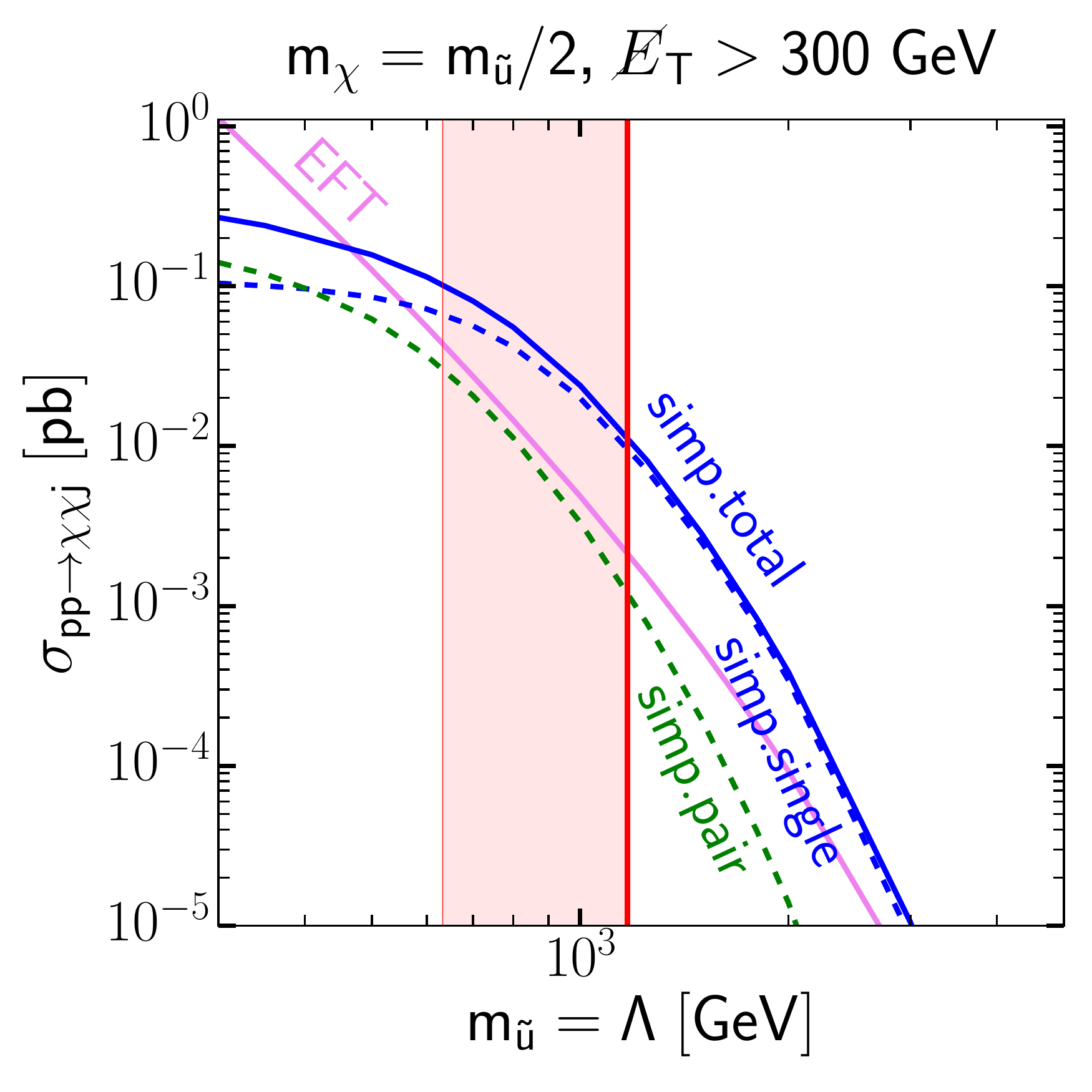} 
\caption{Total production rate in the $t$-channel model as a function
  of the mediator mass. The cut on $\met$ corresponds to a cut on the
  leading jet at parton level for all processes, except for mediator
  pair production. The vertical bands show the mediator masses
  predicting the observed relic density: upper edge for
  $\Omega_\chi^\text{obs}+10\%$ and lower edge for
  $\Omega_\chi^\text{obs}/3$.}
\label{fig:t_cross}
\end{figure}

One of the key questions in our $t$-channel model is how the different
production mechanisms shown in Fig.~\ref{fig:feyn_tchannel} decouple
towards large mediator masses.  First, we study the total production
cross section
\begin{align}
\sigma_{\met+j}(m_\chi, \msu, \yu) \; ,
\end{align}
for fixed dark matter mass and coupling, and as a function of the
mediator mass. All our models are implemented in
\textsc{FeynRules}~\cite{feynrules}, and we use
\textsc{MadGraph5}~\cite{madgraph} to compute mono-jet production at
the parton level with $|\eta_{j_1}|<2.5$.  Additional QCD jets will
not survive the hard analysis cut on $p_{T,j_1} \sim \met$, except for
the specific case of double-resonant mediator pair production. There,
we require the second jet to fulfill $p_{T,j_2}>20$~GeV,
$|\eta_{j_2}|<4.5$, and $\Delta R_{j_1 j_2}>0.4$. For illustration
purposes our results are based on parton-level simulations. In the
appendix we show that using \textsc{Pythia}~\cite{pythia} for parton
shower and hadronization and \textsc{Delphes}~\cite{delphes} for the
detector simulation, the decoupling patterns remain
unchanged.\medskip

The panels shown in Fig.~\ref{fig:t_cross} include three choices of
$m_\chi$ combined with two minimum values for the missing transverse energy
--- at the parton level equivalent to the transverse momentum
of the hard jet. The blue solid lines
include all Feynman diagrams leading to at least one hard jet, as
shown in Fig.~\ref{fig:feyn_tchannel}. For illustration purposes we
separate mediator pair production and single-resonant contribution
in the $ug$-initiated sub-process from the continuum process. The
regions predicting the correct relic density are also indicated;
smaller mediator masses lead to larger annihilation cross section and
require another dark matter component, while larger mediator masses
require an additional annihilation channel after thermal production.

The different production topologies show a distinctive dependence on
the mediator mass. The dependence on the dark matter mass is mild, as
long as $m_\chi$ does not become too large. For a light mediator both
the single-resonant production, and the slightly suppressed pair
production, dominate the combined rate. Towards larger mediator masses
they decouple more rapidly than the generic $t$-channel contribution.
As a consequence, with only a mild cut $\met>50$ GeV the $t$-channel
contribution starts to dominate the total cross section for $\msu >
1$~TeV.  The crossing point between the two regimes depends on the
dark matter mass. Finally, the cross sections show a dependence on the
minimum transverse momentum cut. For the simplified model we find that
a harder cut on $\met$ increases the region of mediator masses for
which single-resonant production dominates, extending to the multi-TeV
mediator regime. We will study the transverse momentum in more detail
below.\medskip

When the mediator becomes heavy, mono-jet production can be described
by an effective Lagrangian including the dimension-6 four-fermion
operator,
\begin{align}
\lag_\text{eff} \supset \frac{c_{u \chi}}{\Lambda^2}  \; 
                       \left(\bar{u}_R \chi \right) \; 
                       \left( \bar{\chi} u_R\right) \; .
\label{eq:t_eft}
\end{align}
The matching scale is chosen as $\Lambda=\msu$, and corresponding to
the choice $\yu=1$ we find for the Wilson coefficient
$c_{\tilde{u}\chi}=1$. This operator mediates the $t$-channel as well
as the single-resonant production topologies shown in
Fig.~\ref{fig:feyn_tchannel}. In contrast, pair production requires a
higher-dimensional operator involving the gluon field strength, like
for example
\begin{align}
\lag_\text{eff} \supset 
 \frac{c}{\Lambda^3}(\bar\chi\chi) \, G_{\mu\nu}G^{\mu\nu} \; .
\end{align}
This is consistent with its faster decoupling pattern as observed in
Fig.~\ref{fig:t_cross}.

The predictions of the effective Lagrangian in Eq.~\eqref{eq:t_eft}
for the mono-jet rate are also included in Fig.~\ref{fig:t_cross}.  We
find good agreement between the effective Lagrangian approximation and
the full model even for mediator masses below 1~TeV, as long as the
acceptance cut on the transverse momenta is low. This changes when we
require globally $\met > 300$~GeV, \ie within a factor ten of the
mediator mass, motivating a study of the transverse momentum
distribution in the mono-jet production process.

\subsubsection*{Kinematic distributions}

\begin{figure}[t]
  \includegraphics[width=0.33\textwidth]{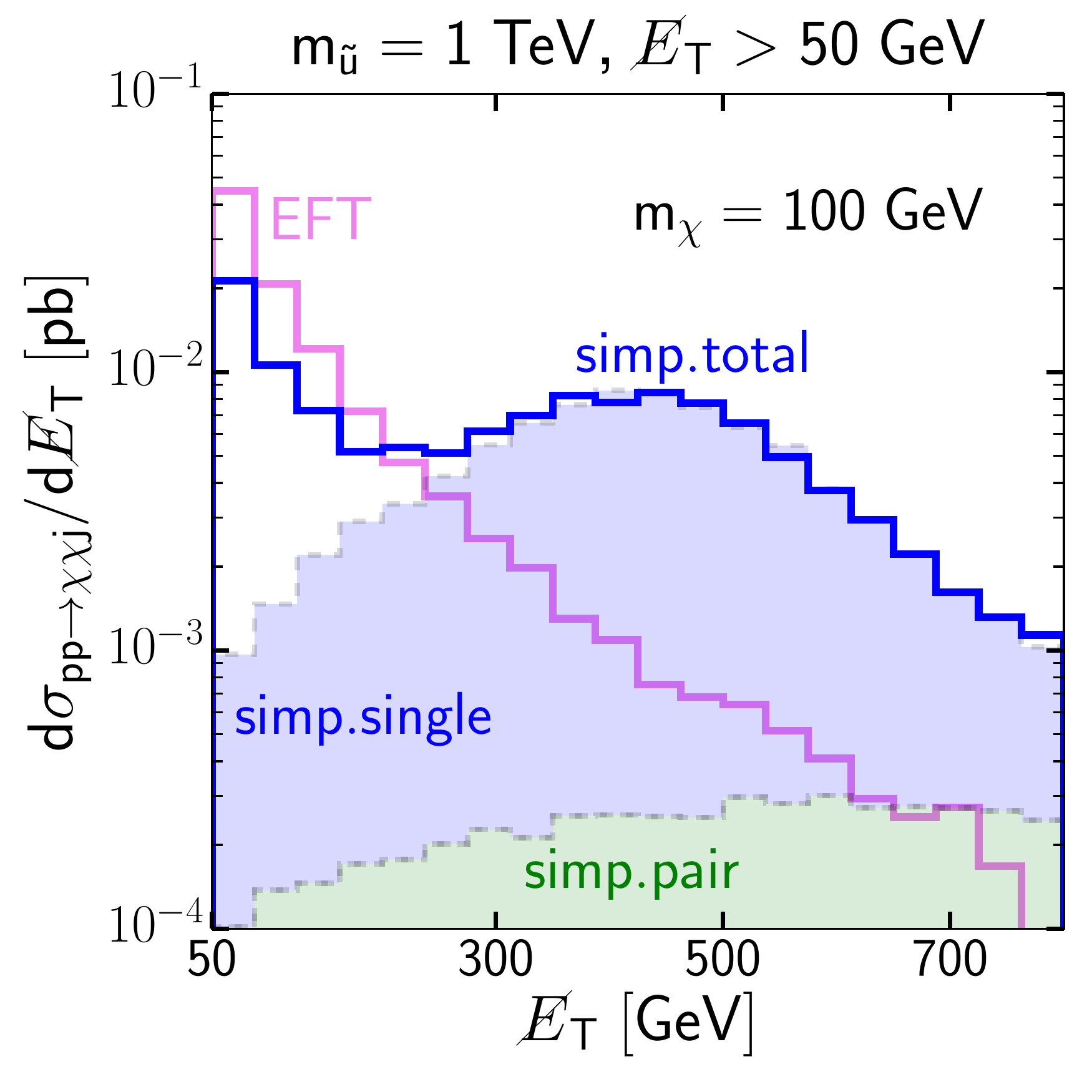}\hspace*{-0.2cm}
  \includegraphics[width=0.33\textwidth]{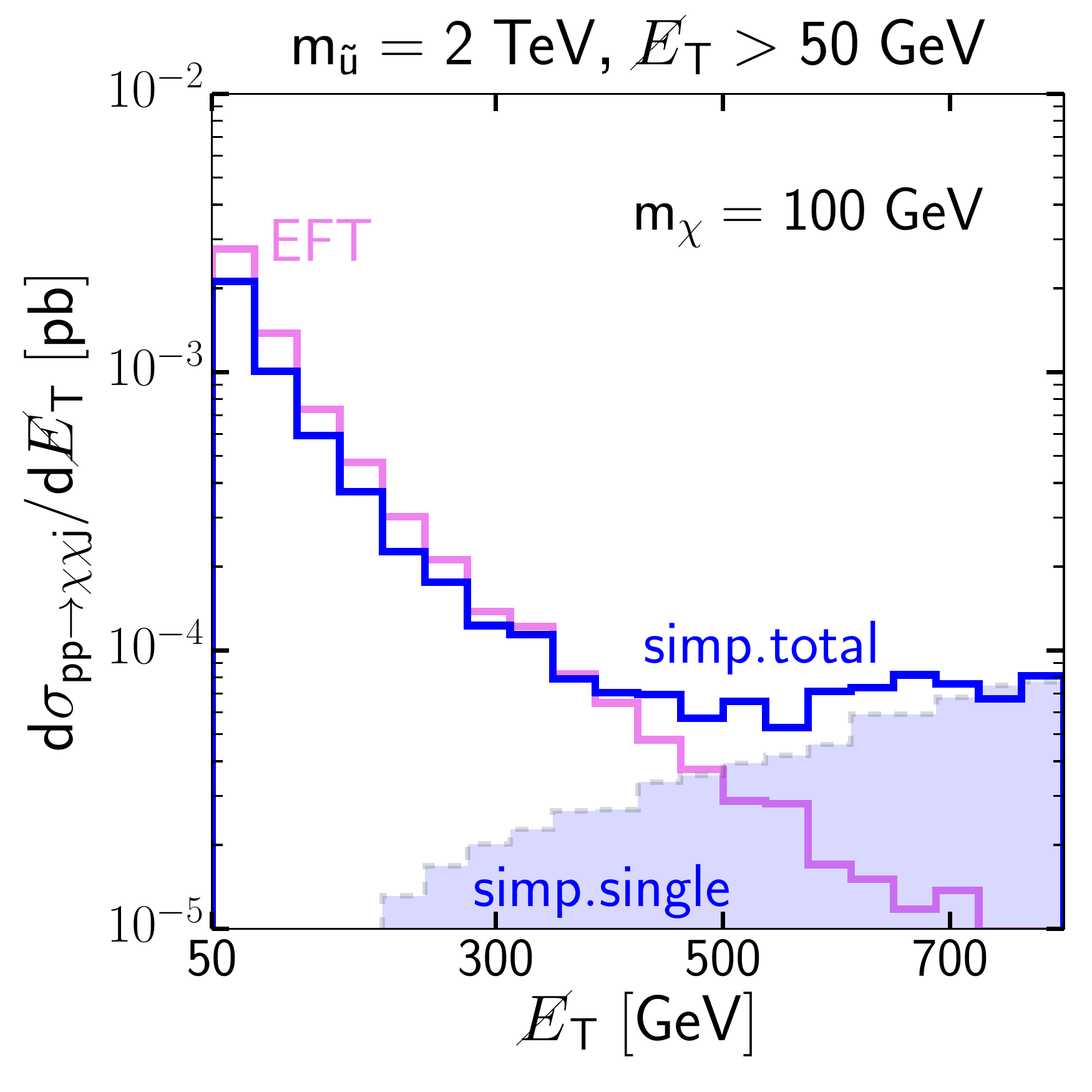}\hspace*{-0.12cm}
  \includegraphics[width=0.33\textwidth]{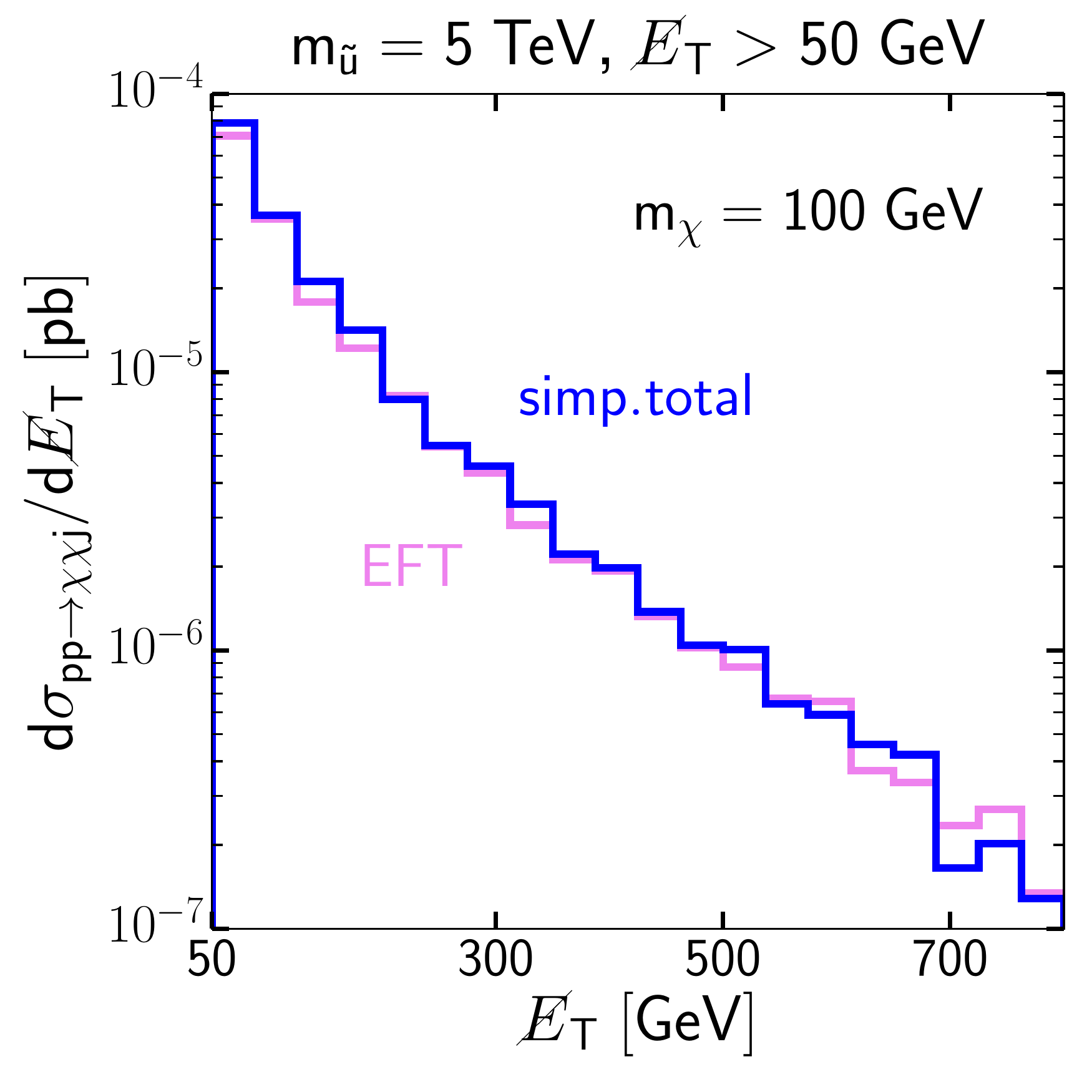} \\
  \includegraphics[width=0.33\textwidth]{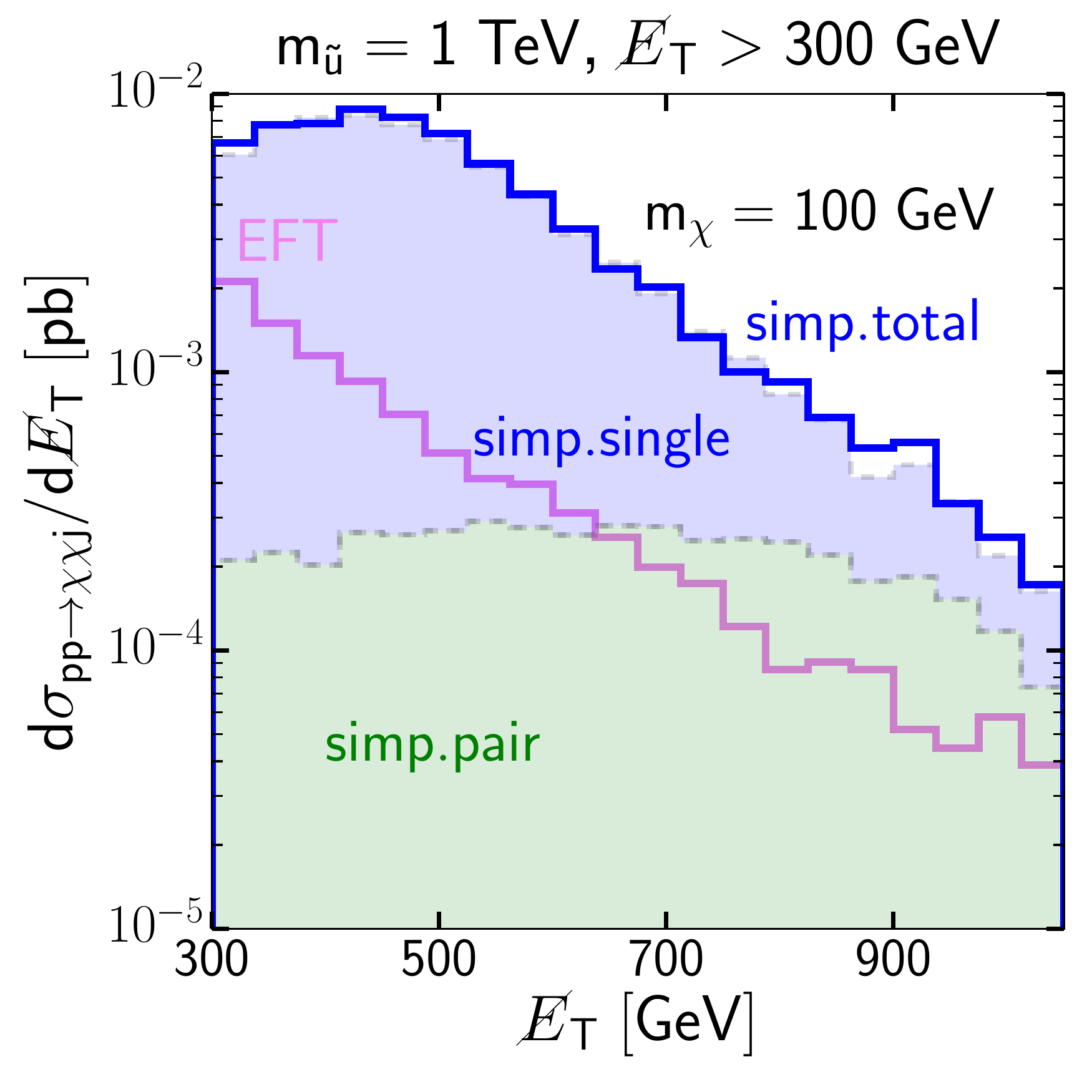}\hspace*{-0.2cm}
  \includegraphics[width=0.33\textwidth]{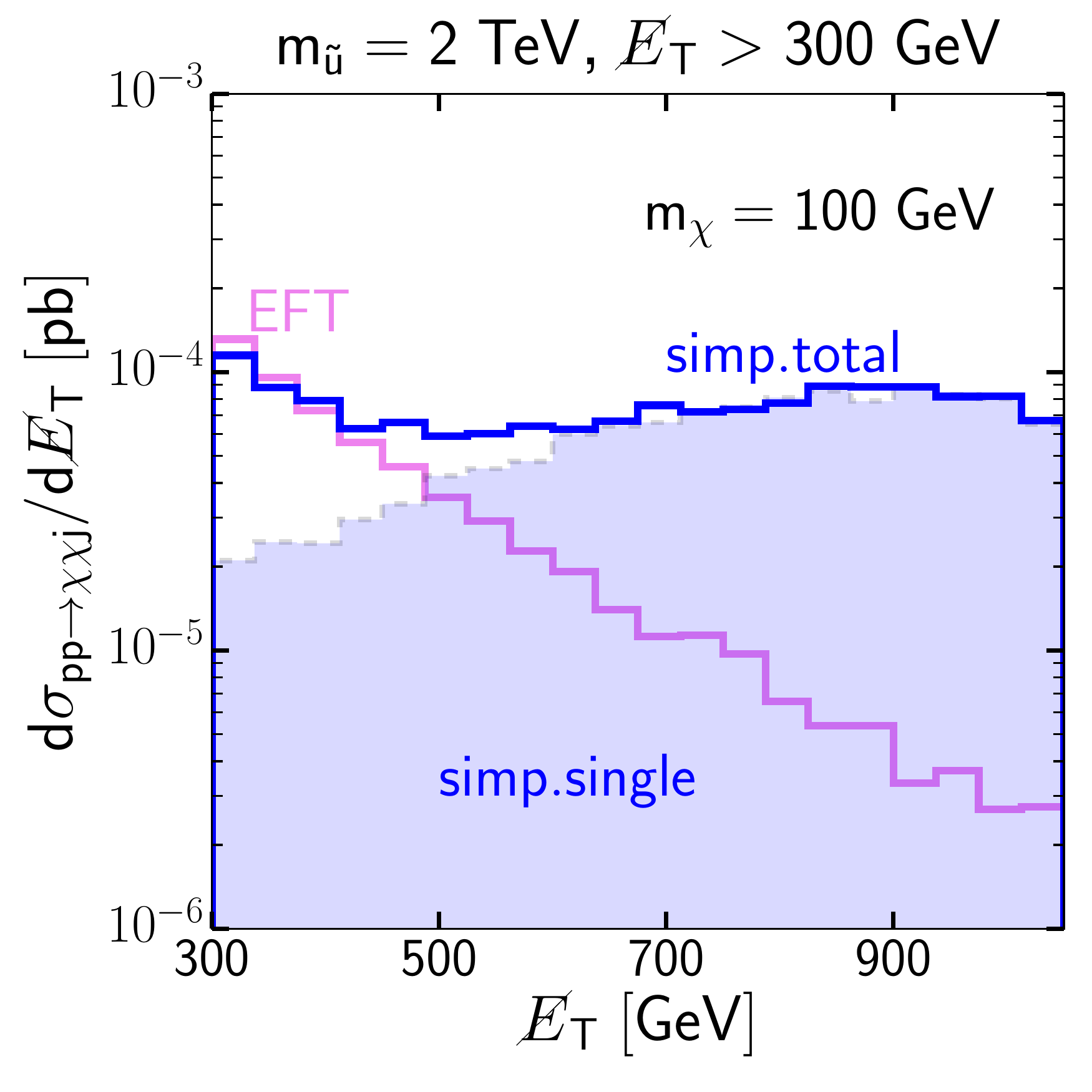}\hspace*{-0.12cm}
  \includegraphics[width=0.33\textwidth]{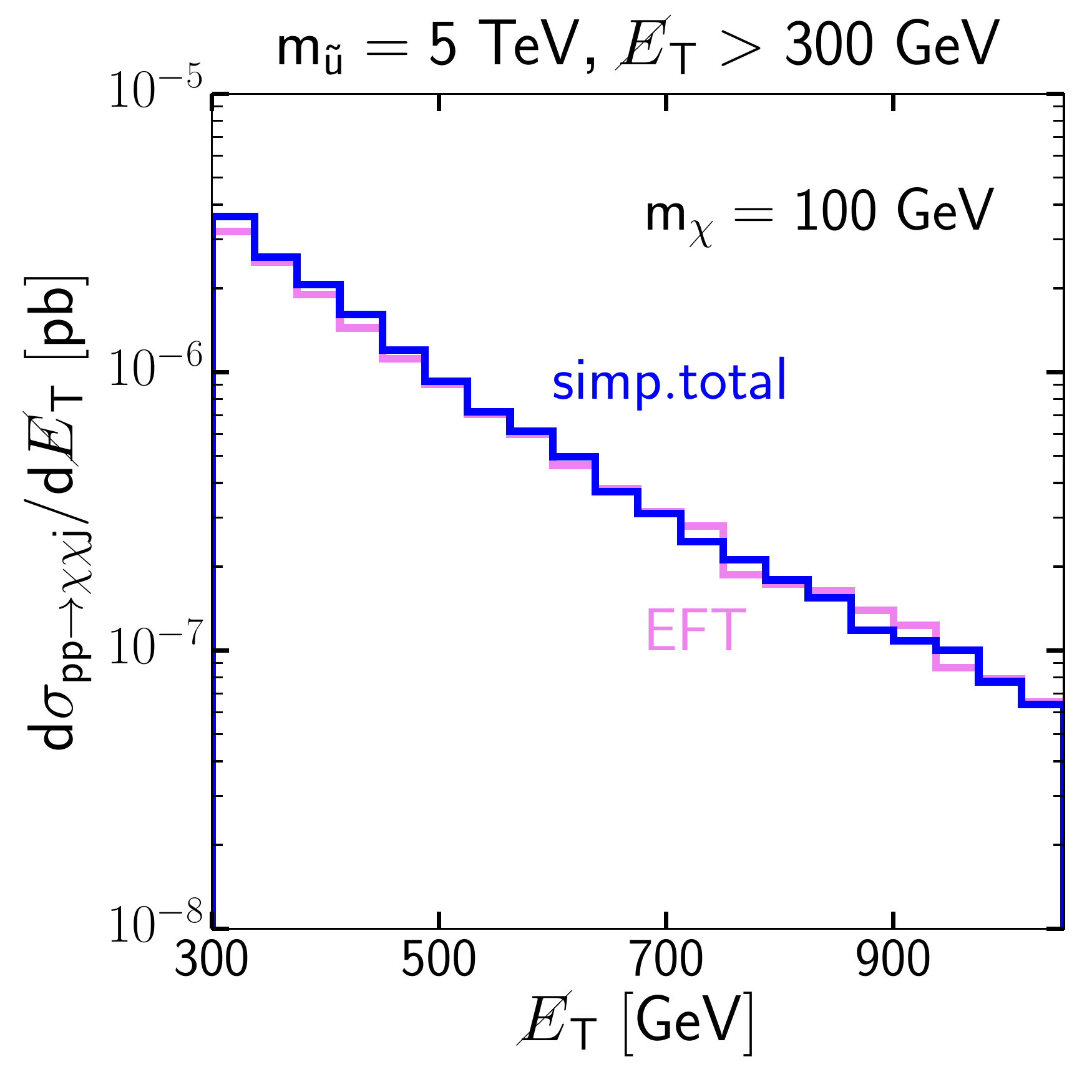}
\caption{$\met$ distributions in the $t$-channel
    model. The different topologies are stacked to combine to the
    total rate given in Fig.~\ref{fig:t_cross}.}
\label{fig:t_ptj}
\end{figure}

For the $t$-channel mediator model both, the total rate and the
kinematic patterns include information about the mediator mass and
couplings. To understand the decoupling patterns better we also study
the one kinematic observable which we can use to analyze the hard
process in mono-jet production, $\met \approx p_{T,j_1}$. In
Fig.~\ref{fig:t_ptj} we show stacked histograms for $\met$ split by 
tree-level topologies: mediator pair
production (dashed green), single-resonant
production contribution (dashed blue), and the full mono-jet channel
(solid blue), for a selection of $\msu$ and $m_\chi$ values.

The distributions for a moderate mediator mass of $\msu=1$ TeV are
shown in the left panels. Here, the mediator will be copiously
produced on-shell, leading to a distinctive $\met$ dependence: while
the resonant contributions peak for large values $\met \lesssim
\msu/2$, the generic $t$-channel contributions prefer low $\met$
values.  Requiring a sizeable $\met$ then causes a larger reduction of
the $t$-channel cross section, extending the region of dominant
resonant production shown in Fig.~\ref{fig:t_cross}.  For larger
values of the mediator masses, also shown in Fig.~\ref{fig:t_ptj},
this distinctive $t$-channel versus resonant cross section behavior is
further enhanced. However, given the much faster decoupling pattern of
the resonant processes, both the cross section rate and the $\met$
distribution are eventually dominated by the generic $t$-channel
contributions.

Finally, in Fig.~\ref{fig:t_ptj} we also show the EFT results, as
defined in Eq.\eqref{eq:t_eft}. Its cross section dependence on $\met$
is similar to the generic $t$-channel contributions in the simplified
model. The single-resonant contributions are formally included, but
implicitly suppressed.  Therefore, regardless of the agreement in the
inclusive cross section prediction, the EFT will only approximate the
$\met$ distribution of the full model for regions with generic
$t$-channel dominance, \ie in the right panels in
Fig.~\ref{fig:t_ptj}. Realistic mono-jet LHC searches require large
$\met$, pushing the region where the EFT gives a good LHC description
to very heavy mediator masses.

\subsubsection*{Effective Lagrangian vs model}

For the $t$-channel mediator model it only makes sense to study the
half plane with $m_\chi < \msu$; otherwise the dark matter agent would
decay. Because the $t$-channel mediator carries color charge, LHC
constraints typically force us into the regime $\msu \gtrsim
1$~TeV. In this situation the dark matter candidate must have a mass
at least around 400 GeV, to avoid over-closing the universe; for the
same reason the $\tilde{u}-u-\chi$ coupling should be larger than for
example in supersymmetric realizations. Both of these requirements
point to a valid effective Lagrangian.

One issue with the $t$-channel model and its effective theory
description at moderate mediator masses is the pair production process
at the LHC. Because the corresponding production rate does not include
any information on the dark sector, an additional branching ratio will
only become interesting when we expand the model to allow for two
competing decays. However, pair production decouples
rapidly with a heavier mediator mass, leading into a regime where
single-resonant production plays a more important role.
The single-resonant production process is in
principle described by the same effective theory as the generic
$t$-channel diagrams. Towards large mediator masses also the
single-resonant topology decouples rapidly. Again, this behavior does
not signal problems for the effective theory, except for some issues in
describing the transverse momentum distributions for still
moderate mediator masses.

The key problem of the $t$-channel model as well as its effective
theory is that in the kinematically defined decoupling regime the dark
matter annihilation rate and the LHC rates are both small. The model
exceeds the measured relic density unless we postulate
non-perturbative couplings.  Note, however, that this is not a problem
caused by the effective Lagrangian.  It is a well-known problem with
the $t$-channel mediator model, which for example in supersymmetric
models is usually resurrected through co-annihilation. In essence, the
effective theory description of the $t$-channel mediator suffers from
many issues of the poorly working full model, leading to a good
effective theory approximation only for not very interesting regions
of the model parameter space.

\clearpage
\section{Tree-level vector in s-channel}
\label{sec:s_channel}

An alternative dark matter scenario is a typically fermionic dark
matter candidate combined with a $s$-channel vector mediator $V$. In
supersymmetric models a similar mediator role is played by the
Standard Model $Z$-boson.  In general, we need to postulate two
interactions,
\begin{align}
\chi-\chi-V \qquad \text{dark matter}
\qqqquad 
u-u-V \qquad \text{Standard Model fermions} \; .
\end{align}
These two couplings induce two competing mediator decays, into
Standard Model particles and into dark matter.  We assume the mediator
to at least couple to quarks in the Standard Model, so we can test the
model at the LHC; a tree-level coupling to two gluons becomes a
serious issue in setting up the model. There are three
different mass regimes in the $m_\chi - m_V$ mass plane,
\begin{alignat}{5}
m_V & > 2 m_\chi  \qqqquad && \text{EFT description possible} \notag \\
m_V &\approx 2 m_\chi && \text{on-shell} \notag \\
m_V & < 2 m_\chi && \text{light mediator}\; .
\label{eq:s_regimes}
\end{alignat}
This mass relation determines if the generic $2 \to 2$ process $u
\bar{u} \to \chi \chi$ factorizes into a $2 \to 1$ kinematics or
not. From the LHC perspective the upper two regimes lead to a phase
space enhancement. For non-relativistic processes like dark matter
annihilation, a phase-space enhancement is limited to the on-shell
case. This translates into small couplings predicting the correct
relic density, essentially turning off all LHC signatures.  For our
effective theory considerations at the LHC we will therefore limit
ourselves to the first case. The light mediator case is, in general,
very interesting because its suppressed $2 \to 2$ rate will include
direct information on both mediator couplings.\medskip

We define an overly simple toy model including an $s$-channel vector to
fit our purpose; the mediator $V^\mu$ couples only to $u$-quark
pairs and to a dark matter fermion $\chi$,
\begin{align}
\lag \supset g_u\ \bar{u}\ \gamma^\mu V_\mu \ u
            + g_\chi\ \bar{\chi}\ \gamma^\mu V_\mu \ \chi \; .
\label{eq:s_model}
\end{align}
At this stage the mediator does not have any link to the Standard
Model gauge groups, but in principle it could. Adding a coupling to
$d$-quarks is trivial. Different observables scale differently with
these two couplings,
\begin{align}
\sigma_\text{mono-jet} \propto \frac{g_u^2 g_\chi^2}{\Gamma_V}
\qqqquad 
\sigma_\text{resonance} = \sigma_V \times \frac{\Gamma_{uu}}{\Gamma_V}
                       \propto \frac{g_u^4}{\Gamma_V}
\qqqquad 
\Gamma_V \sim \Gamma_{\chi \chi} + \Gamma_{uu} \; .
\end{align}
The mono-jet rate can factorize into $\sigma_{V+j} \times \br_{\chi\chi}$.
The di-jet decay signature is essentially limited
to the factorized kinematics in the presence of large QCD
backgrounds~\cite{qcd_eft}.  Throughout our analysis we assume $g_u =
g_\chi$, reflecting some kind of comparable charges in the visible and dark
sectors. As long as $m_V \gg m_\chi$ this maximizes a potential dark
matter signal at the LHC, minimizes the mediator width, and removes
the focus from Standard Model resonance searches\footnote{If the LHC
  should discover a mediator-like resonance without a missing energy
  signature, an interpretation in terms of dark matter will hardly go
  beyond the stage of pure speculation.}. On the other hand, such a
resonance signal will allow us to extract information
from a measurement of $\sigma_{V+j} \times \br_{uu}$. 

The typical total decay width of the mediator becomes
\begin{align}
\frac{\Gamma_V}{m_V} \lesssim 0.4~...~10\% 
\qquad \text{for} \quad 
g_u = g_\chi=0.2~...~1 \; ,
\end{align}
and $m_\chi \ll m_V$.  Unlike for the $t$-channel, the $s$-channel
mediator does not have to be very narrow, in particular if we
include additional, flavor-universal couplings. Still, at least for
the vector case we do not expect theoretical issues related to an
increasing width-to-mass ratio.\medskip

\begin{figure}[t]
\includegraphics[width=0.35\textwidth]{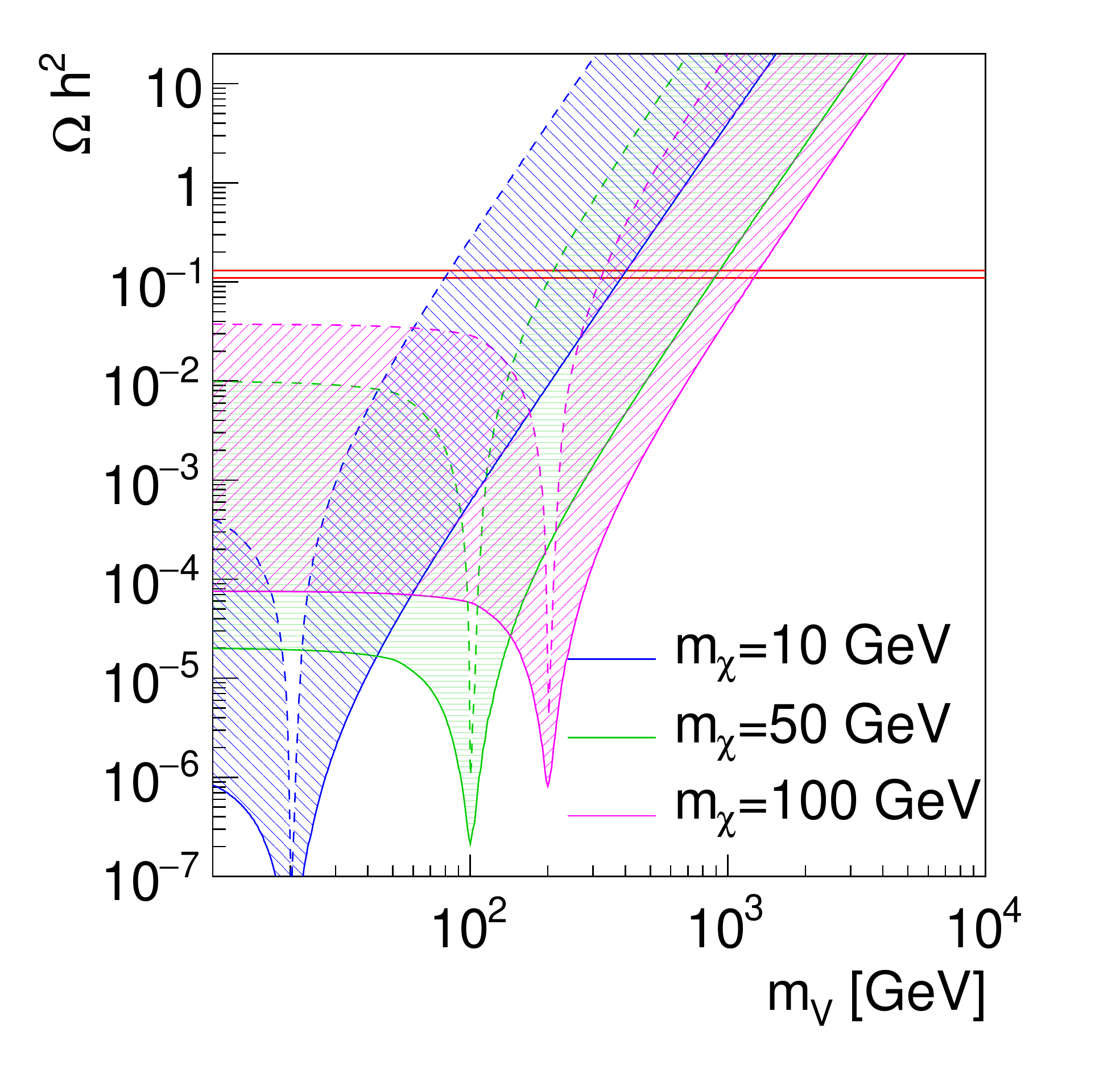}\hspace*{-0.4cm}
\includegraphics[width=0.35\textwidth]{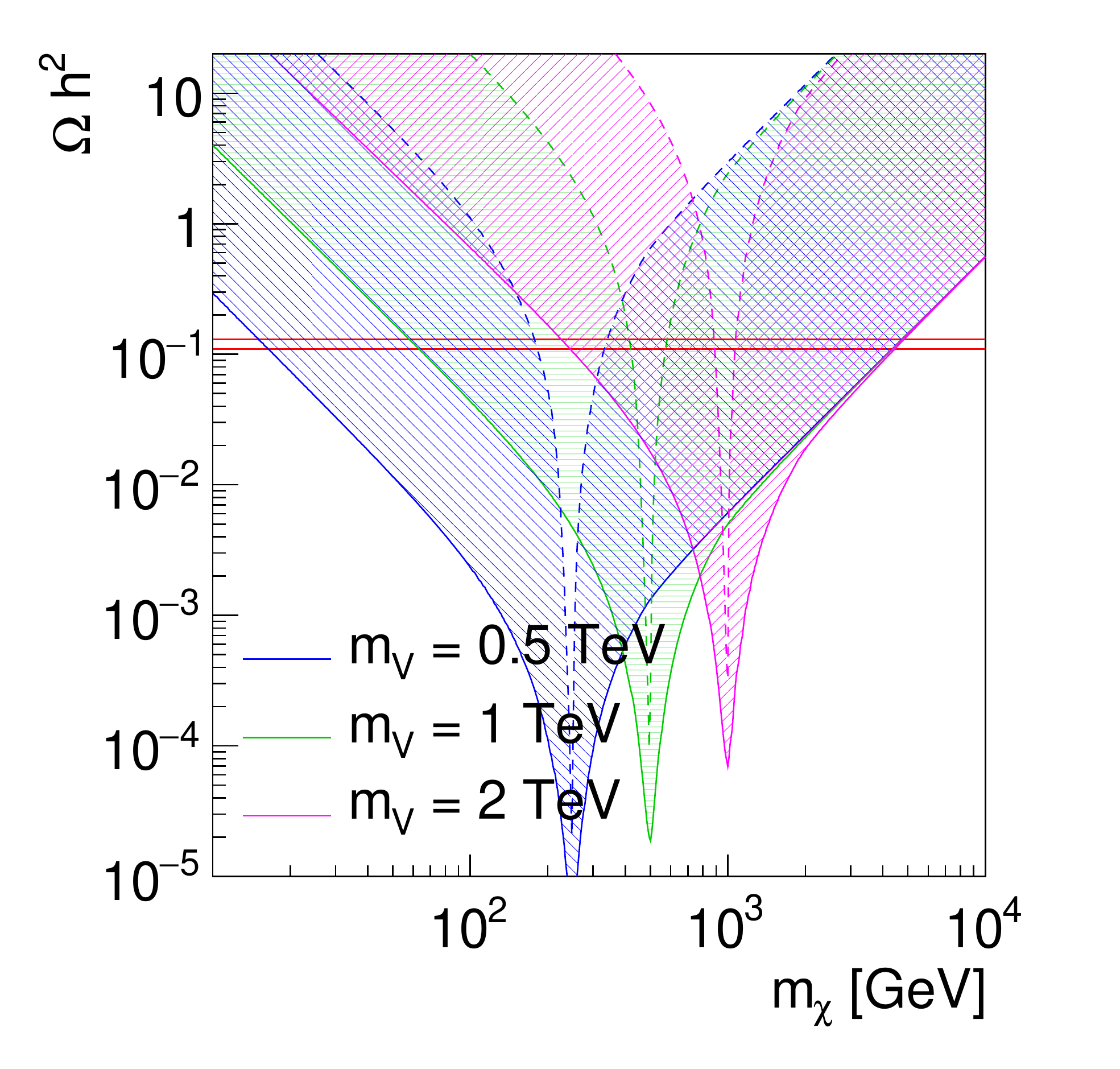}\hspace*{-0.4cm}
\includegraphics[width=0.35\textwidth]{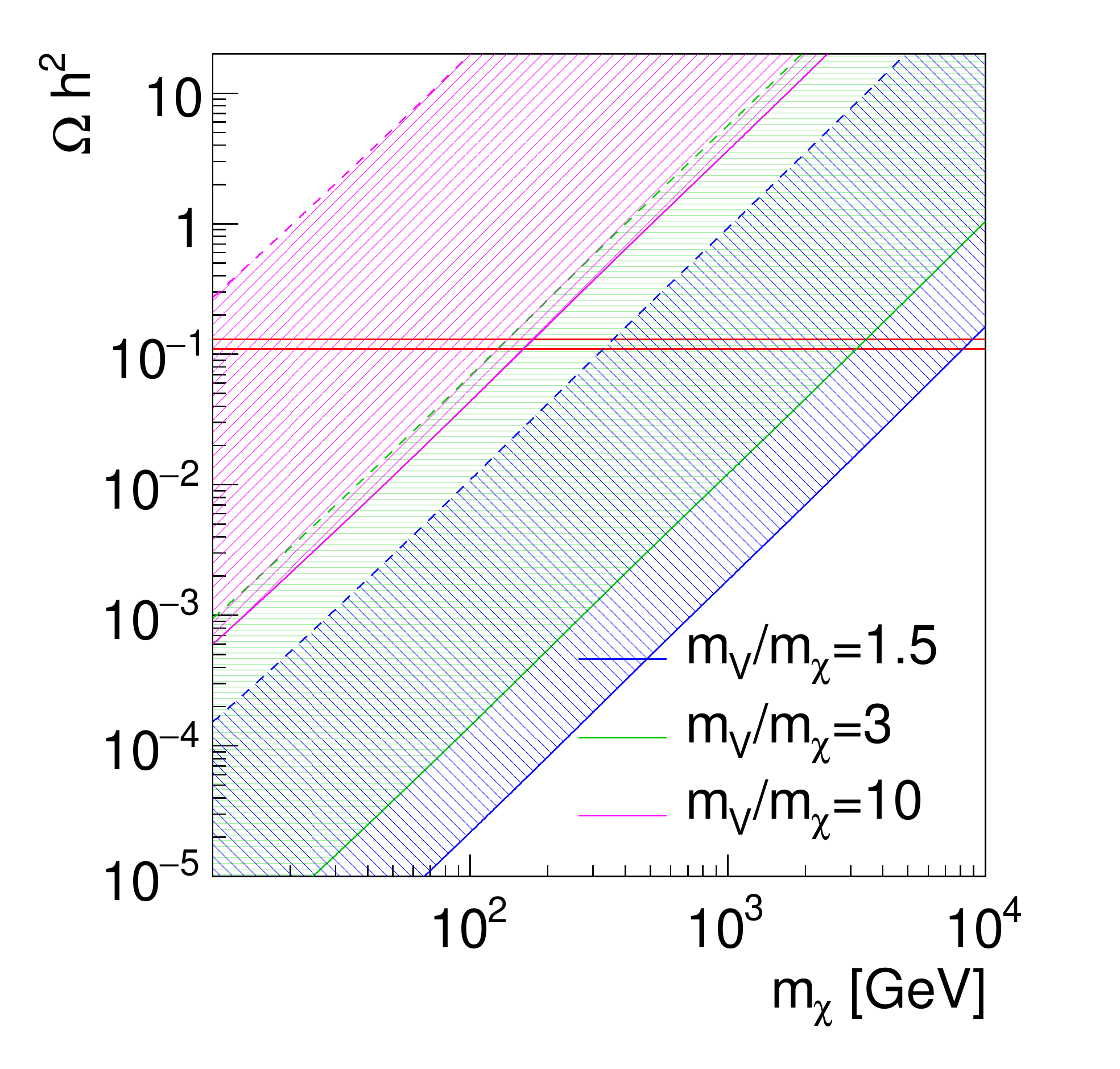}
\caption{Relic density for the $s$-channel vector mediator model as a
  function of the mediator mass for constant dark matter mass (left),
  as a function of the dark matter mass for constant mediator mass
  (center) and as a function of the dark matter mass for a constant
  ratio of mediator to dark matter mass. Over the shaded bands we vary
  the couplings $g_u=g_\chi=0.2~...~1$; large relic densities
  correspond to small coupling.}
\label{fig:relic_schannel}
\end{figure}

As before, we can use the Lagrangian of Eq.\eqref{eq:s_model} to
compute the predicted relic density with the help of
\textsc{Micromegas}.  In all three panels of
Fig.~\ref{fig:relic_schannel} we observe that for light dark matter
the annihilation rate is typically large. Unlike in the $t$-channel
model the predicted relic density easily matches the observed value.
In the left panel we can identify the three kinematic regimes defined
in Eq.\eqref{eq:s_regimes}: for small mediator masses, $m_V < 2
m_\chi$, the annihilation is a $2 \to 2$ process, but the dependence
on the light, off-shell mediator mass is small. For a global analysis
of the light-mediator case the non-relativistic annihilation is
essentially insensitive to the mediator mass.  Around the on-shell
condition we can reach the measured relic density with very small
couplings, effectively turning off any LHC signature; for heavy
mediators the $2 \to 2$ annihilation process rapidly decouples with
large mediator masses. As a side remark, the $\chi-\chi-V$ interaction
also introduces a $t$-channel annihilation $\chi \chi \to V^* V^*$,
but for the vector mediator its contribution is always strongly
suppressed by its 4-body phase space.

In the center panel we observe two, almost symmetric solutions in the
dark matter mass for a given, large mediator mass. The solution to the
right of the pole is hardly consistent with an EFT
description, but the solution with $m_\chi < m_V/2$ shows no problems
in generating the observed relic density. The limiting factor towards
small dark matter masses for example for $m_V = 2$~TeV can be seen in
Eq.\eqref{eq:eft_sigann}, where the annihilation rate scales with
$m_\chi^2$ and should not be too small to predict the correct relic
density.  In the right panel we focus on mass ratios consistent with
an effective theory interpretation and for example find a broad band
with $m_V/m_\chi =10$ and $m_\chi = 10~...~100$~GeV with a valid relic
density prediction. In this range we also expect a small mediator
width $\Gamma_V/m_V \lesssim 10\%$. As for the $t$-channel model,
additional couplings to the full Standard Model fermion spectrum are
likely to increase the dark matter annihilation rate by up to an order
of magnitude.

\subsubsection*{Total rate}

\begin{figure}[t]
\includegraphics[width=0.33\textwidth]{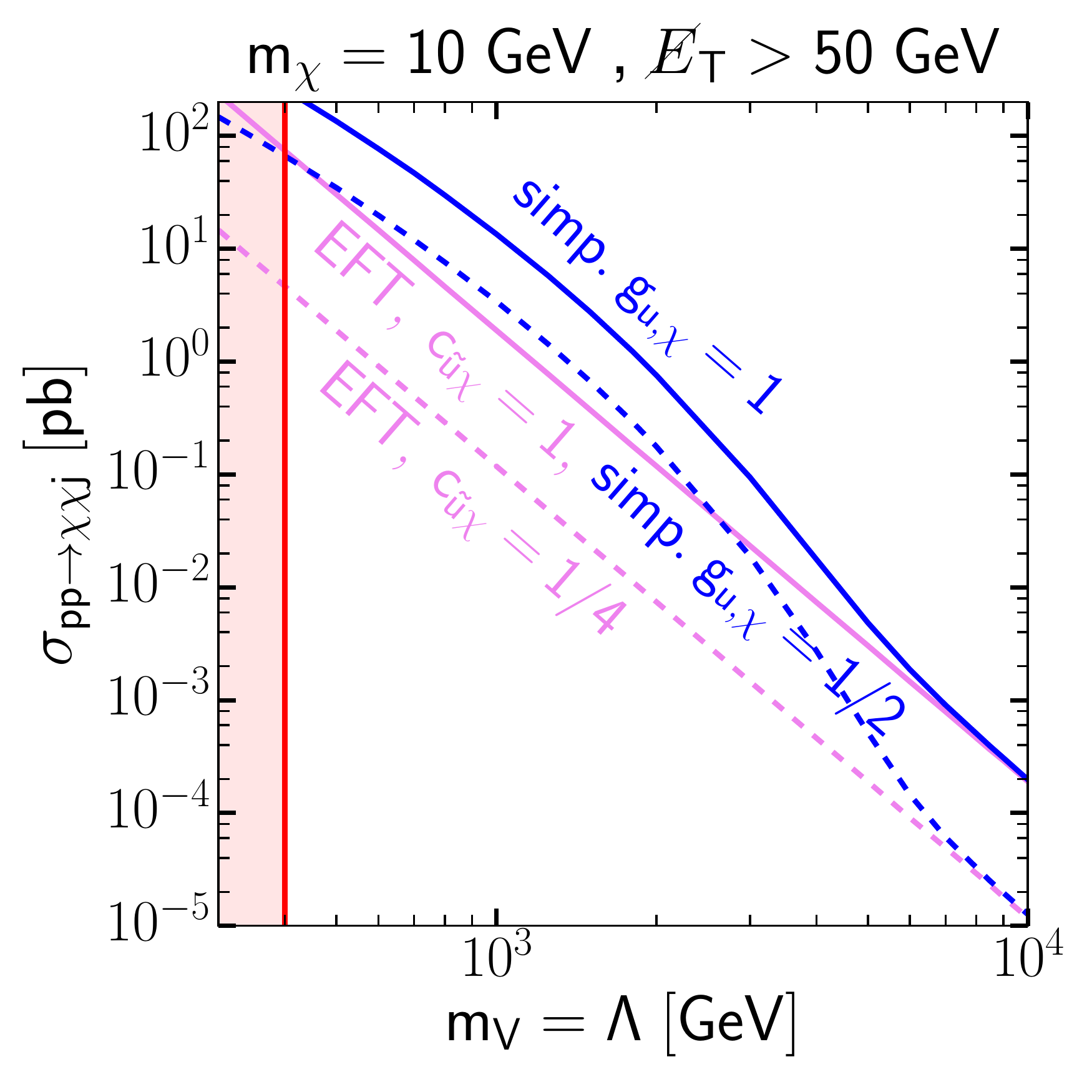}\hspace*{-0.2cm}
\includegraphics[width=0.33\textwidth]{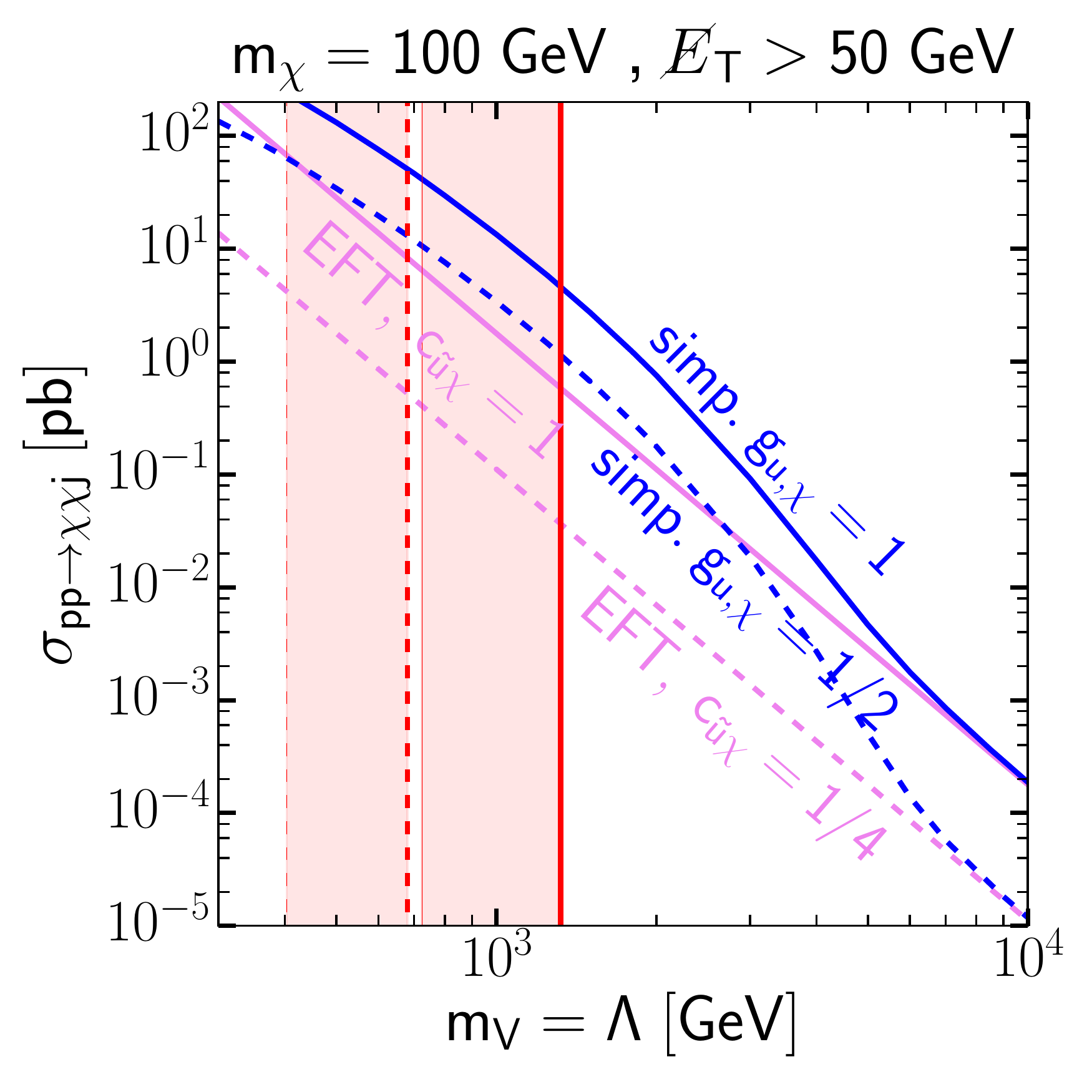}\hspace*{-0.2cm}
\includegraphics[width=0.33\textwidth]{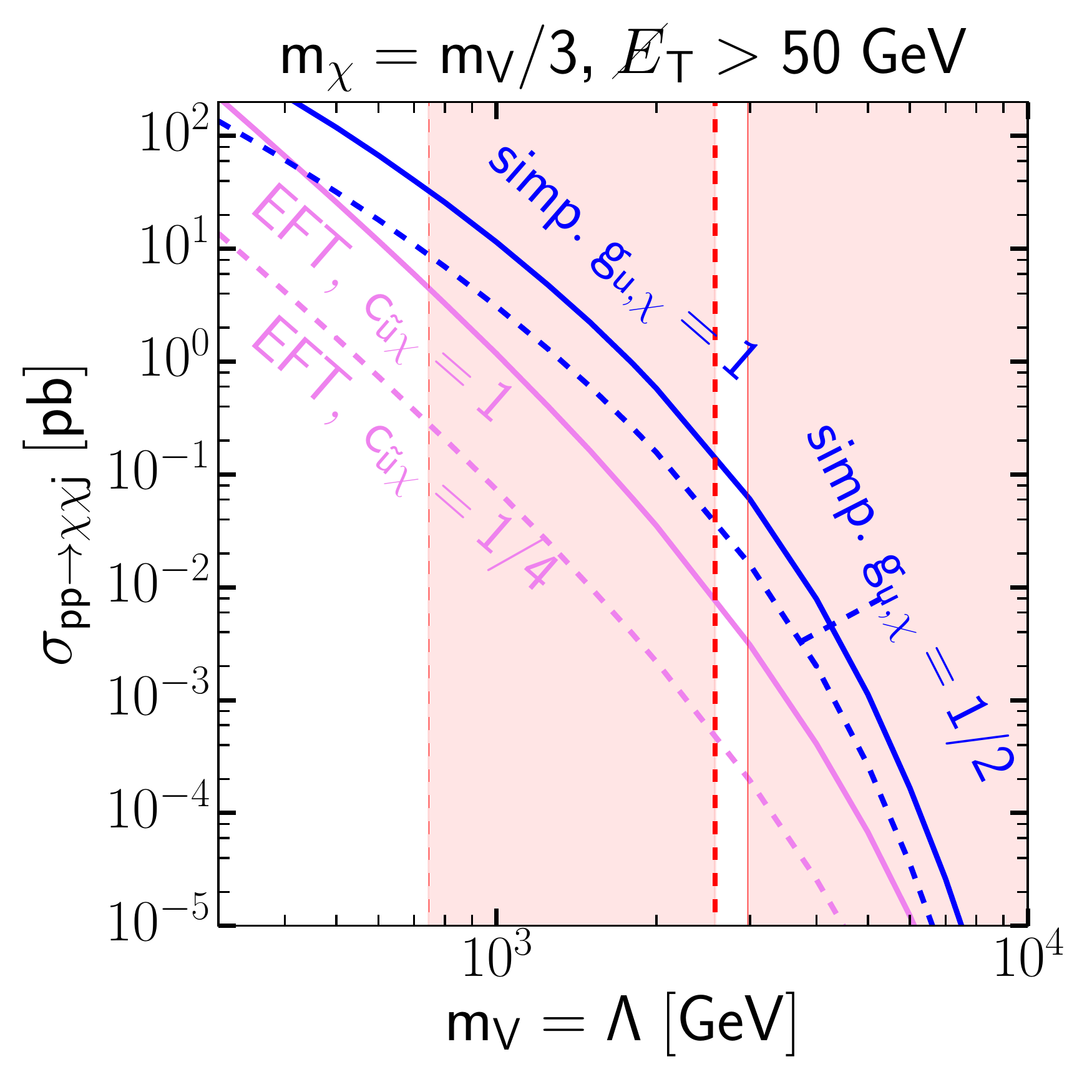} \\
\includegraphics[width=0.33\textwidth]{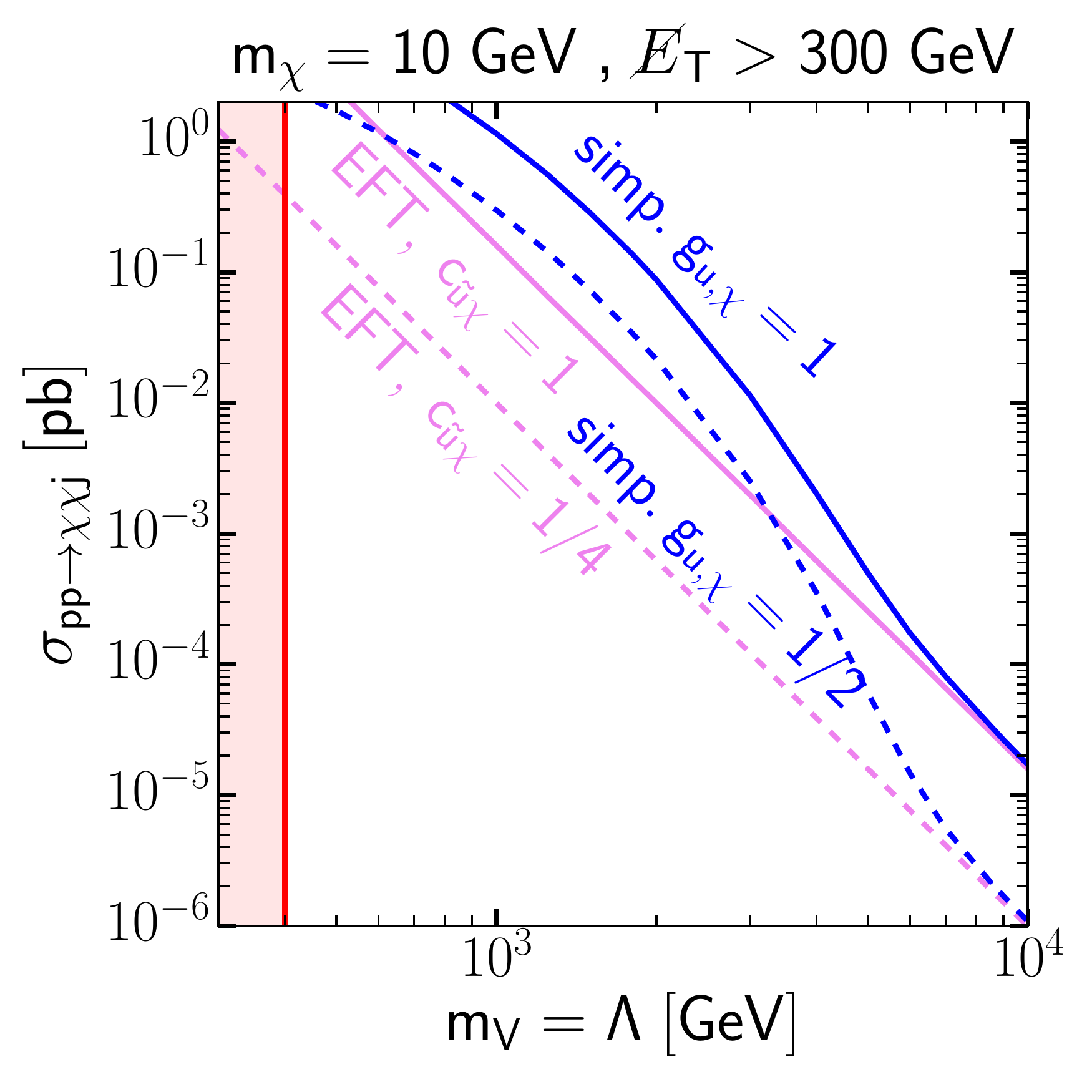}\hspace*{-0.2cm}
\includegraphics[width=0.33\textwidth]{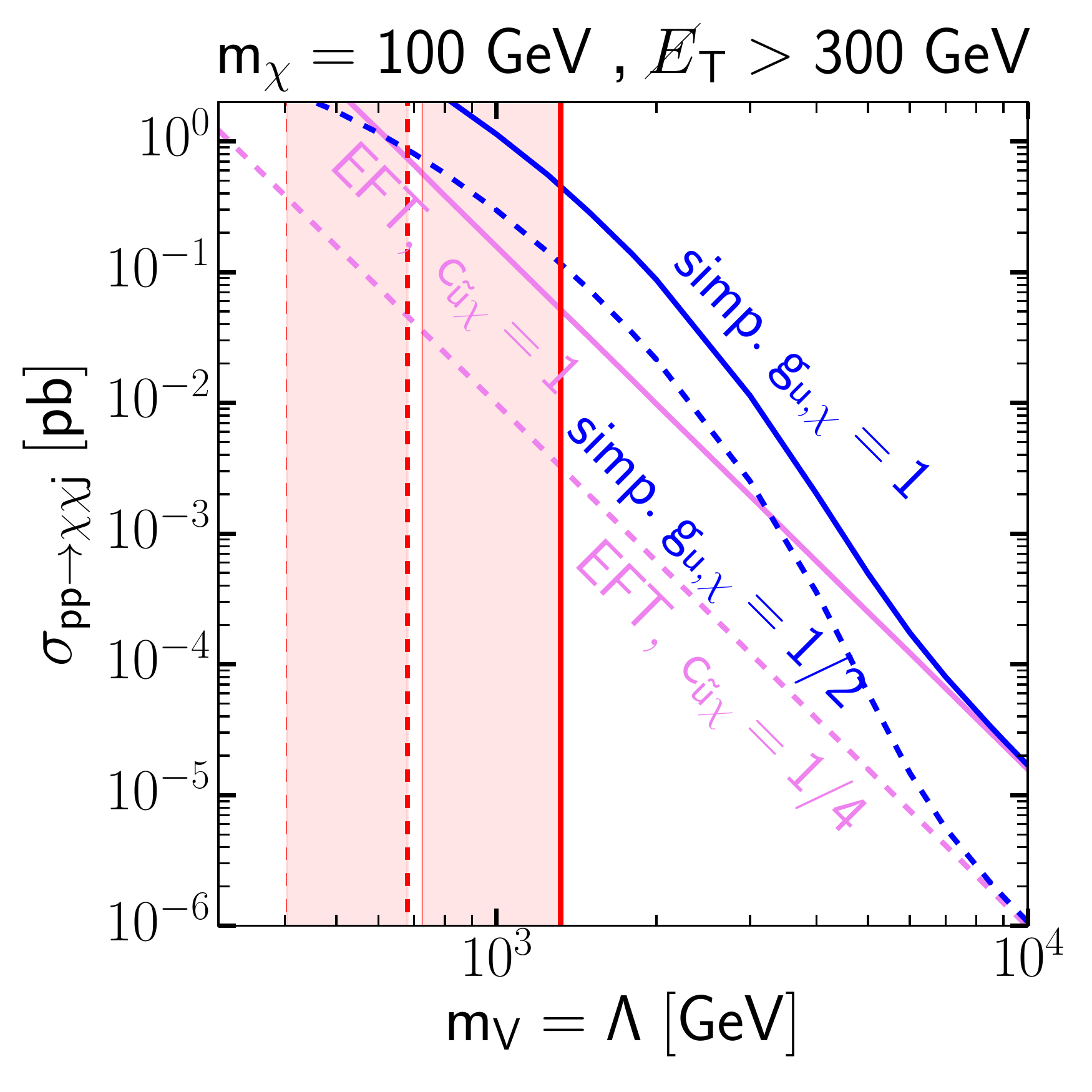}\hspace*{-0.2cm}
\includegraphics[width=0.33\textwidth]{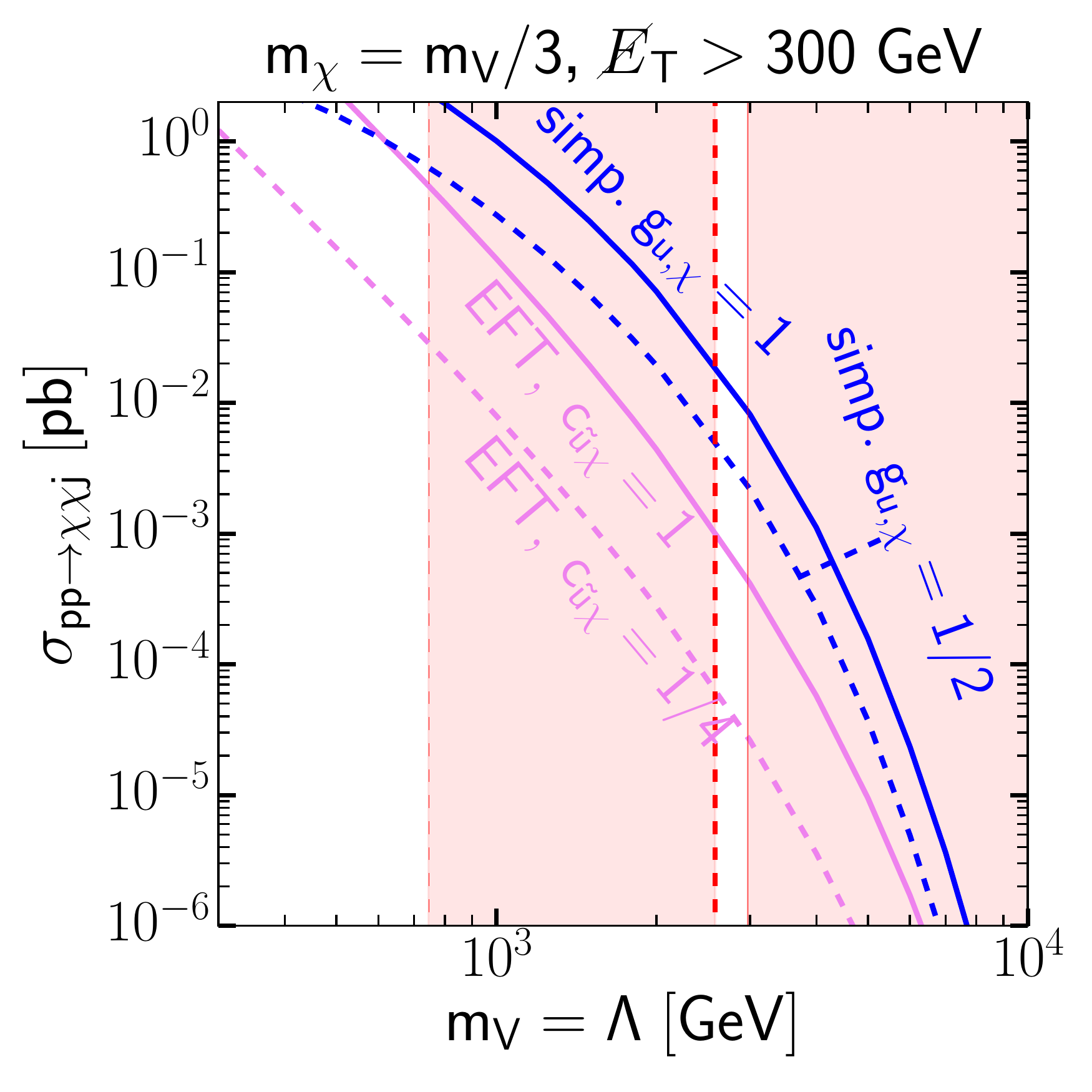} \\
\caption{Total production rate in the $s$-channel vector model as a
  function of the mediator mass.  The cut on $\met$ corresponds to a
  cut on the leading jet at parton level.  The vertical bands show the
  mediator masses predicting the observed relic density: upper edge
  for $\Omega_\chi^\text{obs}+10\%$ and lower edge for
  $\Omega_\chi^\text{obs}/10$.}
\label{fig:s_cross}
\end{figure}

As the first LHC observable we show the total mono-jet production rate
for the $s$-channel vector mediator in Fig.~\ref{fig:s_cross},
\begin{align}
\sigma_{\met+j}(m_\chi, m_V, g_u = g_\chi) \; .
\end{align}
As for the $t$-channel model we start with a low acceptance cut $\met
> 50$~GeV, where at parton level the hard jet just recoils against
the missing momentum. The cut on $p_{T,j}$, correlated with $\met$,
always regularizes and dials the relative size of
$\sigma_{V+j}/\sigma_V$. Because of the final state kinematics, a too
stiff cut on $p_{T,j}$ will not allow the mediator to be produced
on-shell. This correlation can have a large effect on the cross
section after cuts.

The three upper panels of Fig.~\ref{fig:s_cross} cover three different
dark matter mass values, for mediator masses up to 10~TeV. Following
Fig.~\ref{fig:relic_schannel}, the heavy mediator regime is consistent
with the observed relic density for $m_V/m_\chi \approx 3~...~10$ and
$g_\chi = g_u = 1$. For a light dark matter candidate with $m_\chi =
10$~GeV the mediator mass corresponding to the observed relic density
would be significantly below 1~TeV and likely ruled out by current LHC
searches. Heavier mediators are allowed, if there is another dark
matter candidate; lighter mediators need another annihilation
channel. In addition to the generic case, we also show an alternative,
more weakly interacting scenario with $g_\chi = g_u = 1/2$, for which
the mediator is clearly a narrow resonance. As expected, the relic
density constraint points towards half the mediators mass of the case
$g_\chi = g_u = 1$.

The main feature in all curves is that for fixed dark matter masses we
observe a change in the scaling of the total rate around $m_V \sim
5$~TeV. At this point the partonic energy scale of the LHC changes the
production kinematics from an on-shell mediator production with
$\sigma_{\met+j} = \sigma_{V+j} \times \br_{\chi\chi}$ to a $2 \to
2$ description of the hard sub-process $u \bar{u} \to \chi \chi$ with
a decoupled mediator. This change is given by a combination of the
proton--proton energy of 13~TeV and the typical momentum fractions of
the incoming valence-sea quark pair, and it does not depend on the
dark matter model. For example, when we include a stiffer transverse
momentum cut of $\met > 300$~GeV this turn-over point does not
move. The main change in the lower panels of Fig.~\ref{fig:s_cross} is
that the signal rate is suppressed.  This suppression is more enhanced
for the small mediator masses. Typical mono-jet rates in the
heavy-mediator regime can still reach femto-barn cross sections
for light dark matter.\medskip

\begin{figure}[t]
\includegraphics[width=0.33\textwidth]{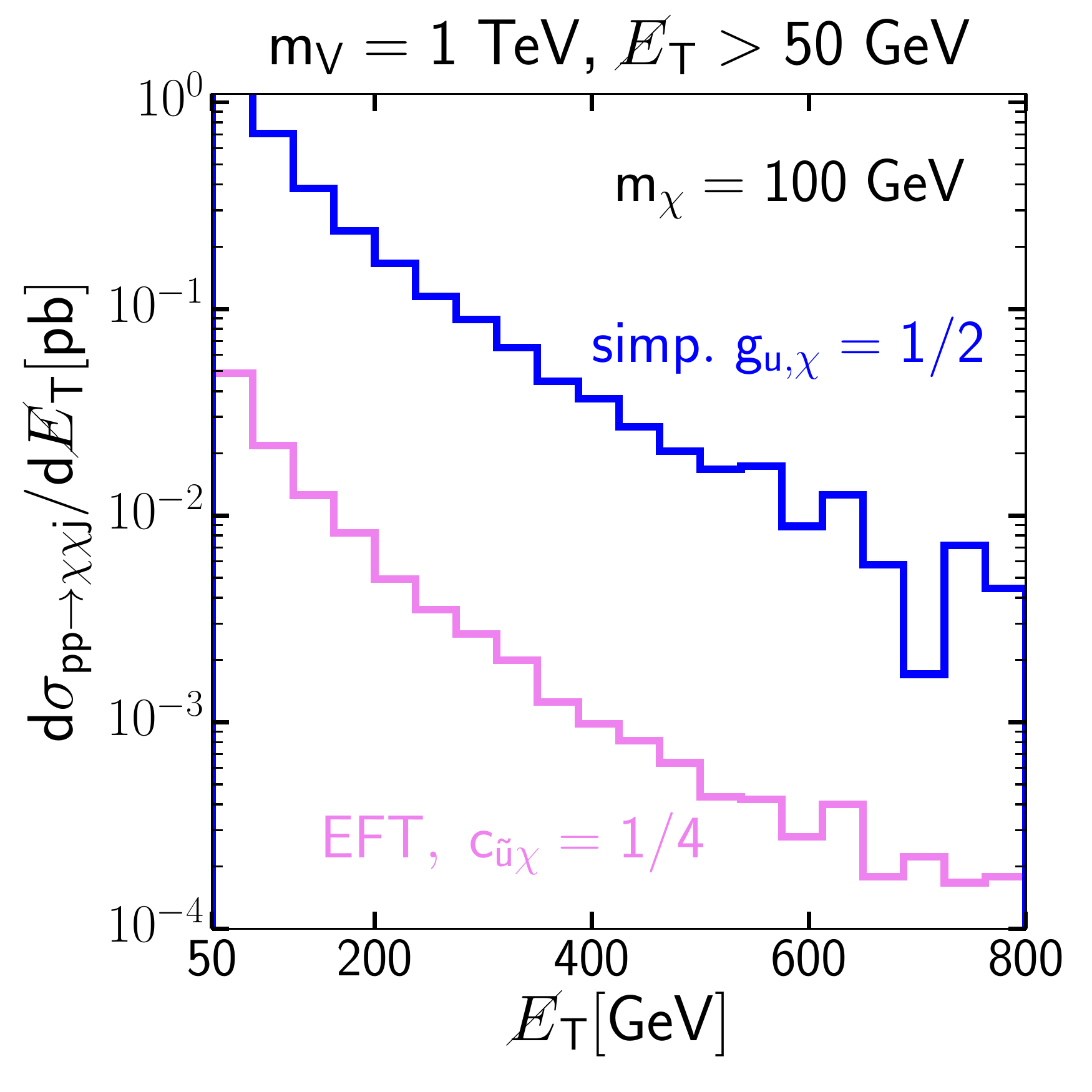}\hspace*{-0.2cm}
\includegraphics[width=0.33\textwidth]{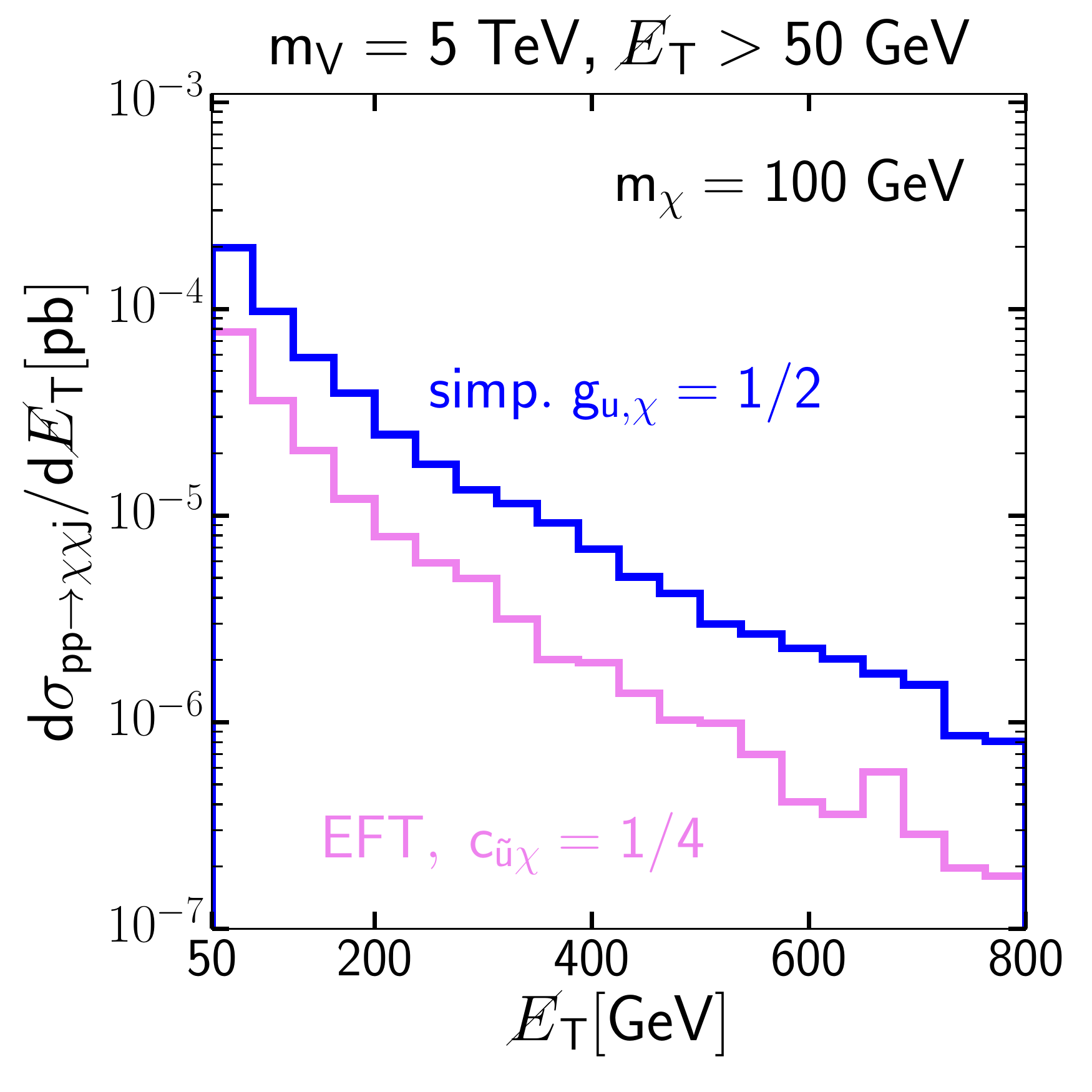}\hspace*{-0.2cm}
\includegraphics[width=0.33\textwidth]{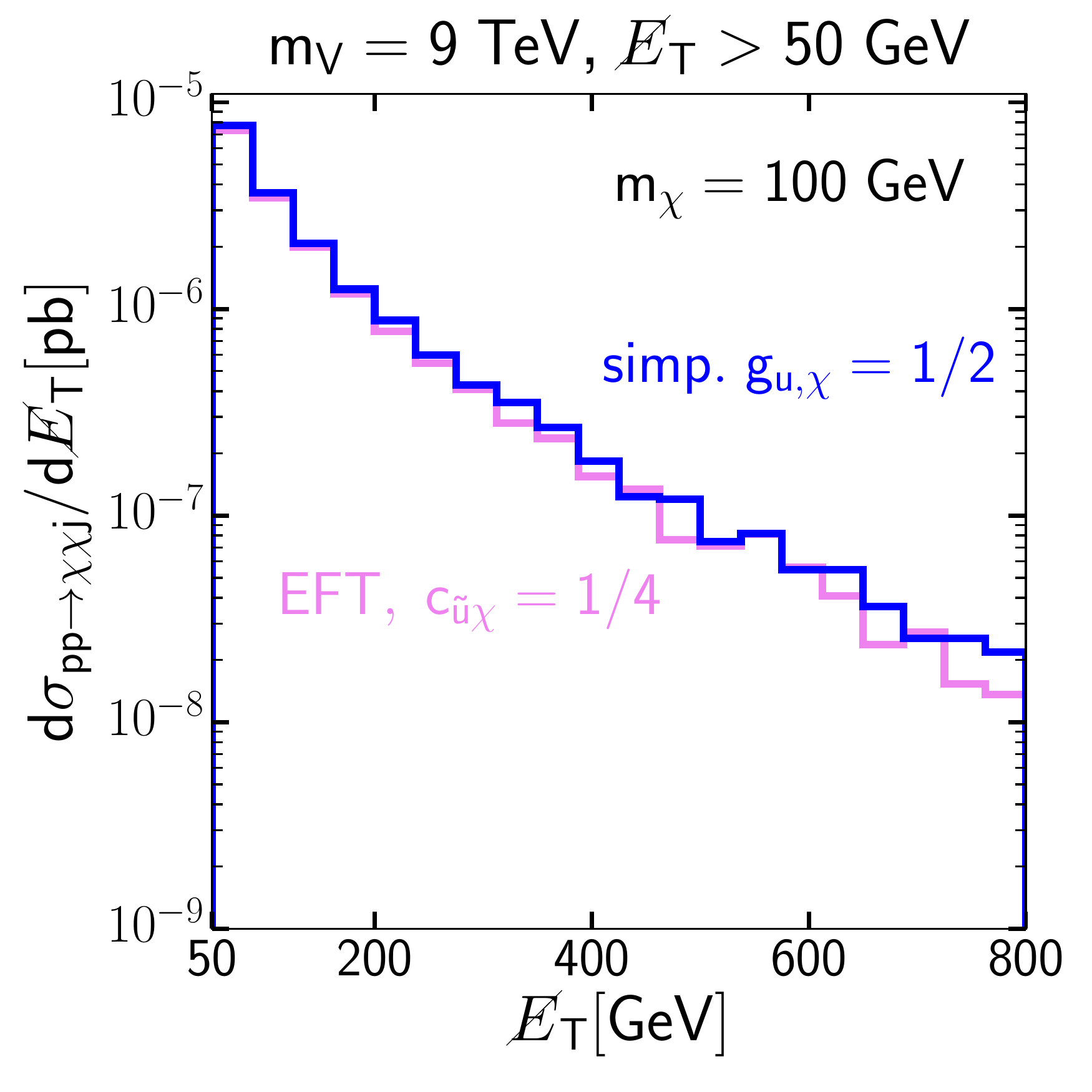} \\
\includegraphics[width=0.33\textwidth]{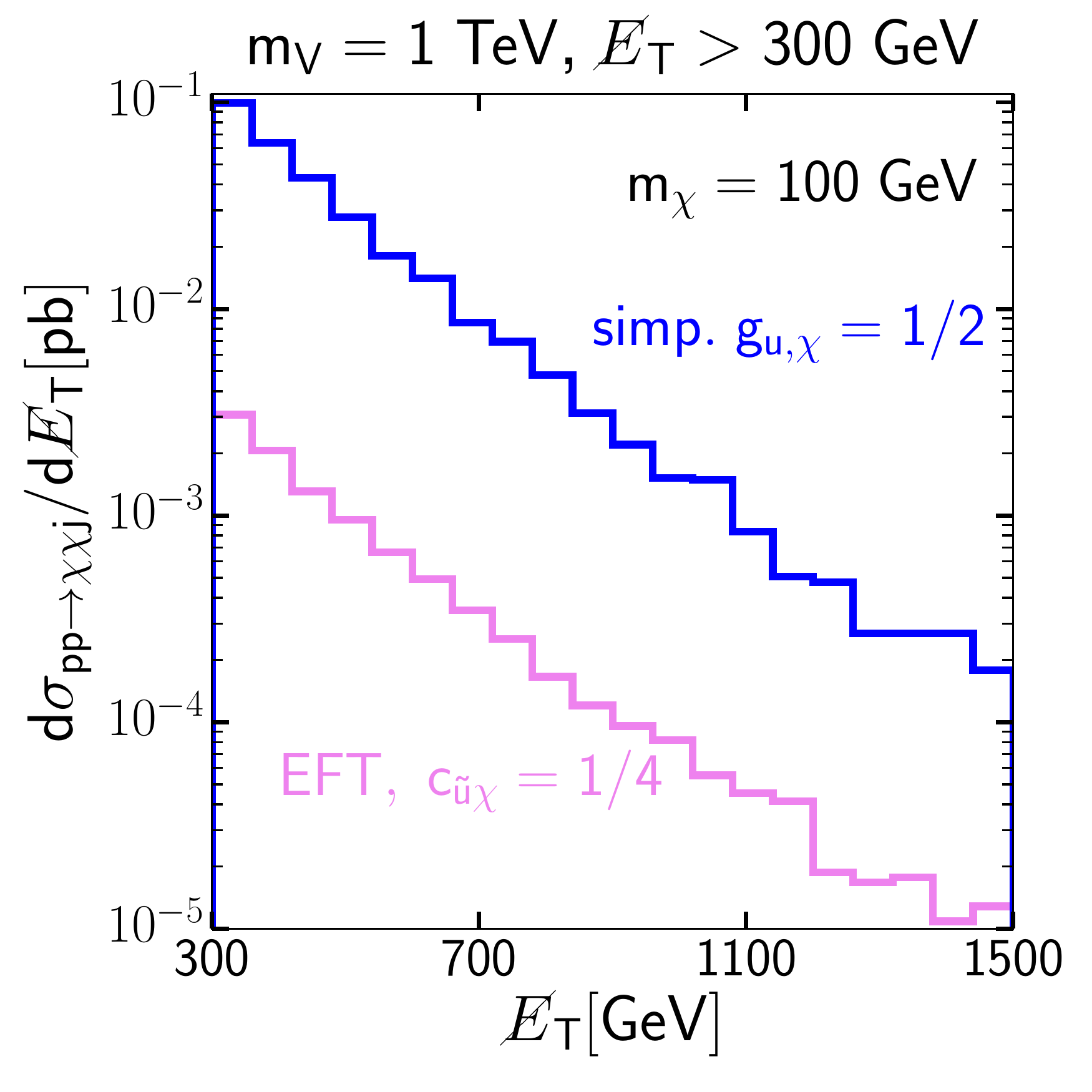}\hspace*{-0.2cm}
\includegraphics[width=0.33\textwidth]{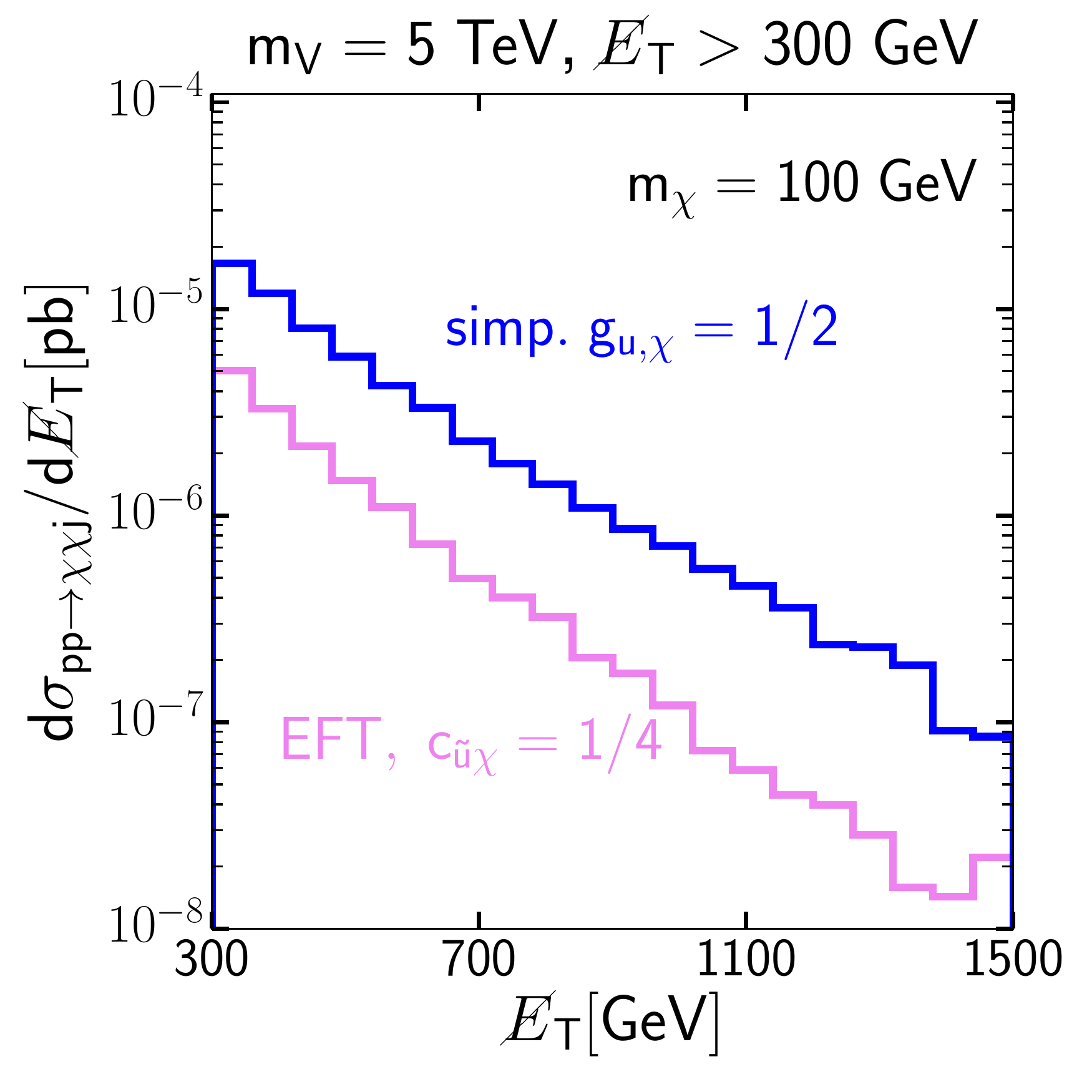}\hspace*{-0.2cm}
\includegraphics[width=0.33\textwidth]{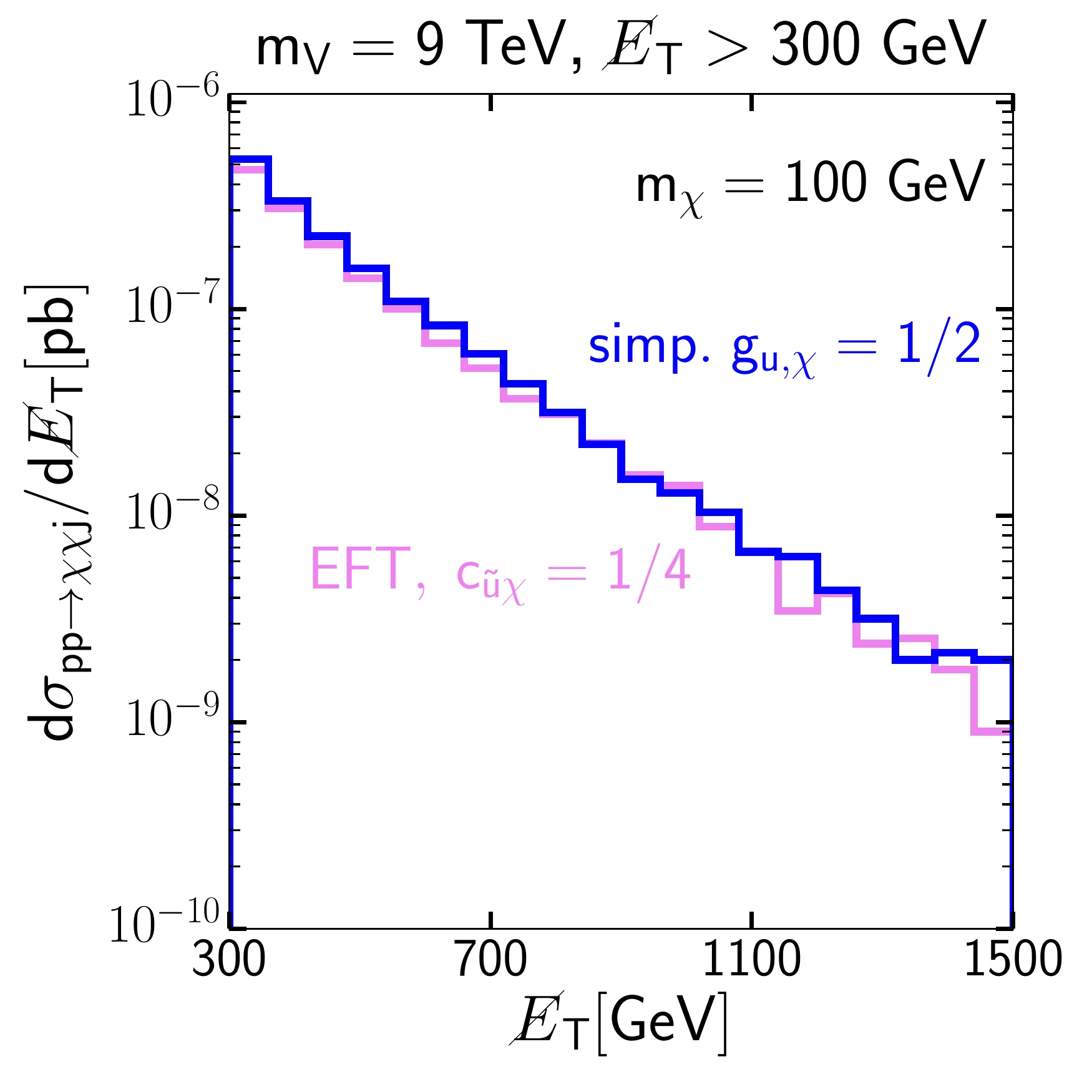}
\caption{$\met$ distributions in the $s$-channel vector mediator
    model.}
\label{fig:s_ptj}
\end{figure}

Once the $s$-channel mediator decouples from the mono-jet production
process we can describe the model in terms of an effective Lagrangian
with the dimension-6 four-fermion operator
\begin{align}
\lag_\text{eff} \supset \frac{c_{u\chi}}{\Lambda^2} 
  \left(\bar{u} \gamma^\mu u \right) \;
  \left( \bar{\chi} \gamma_\mu \chi \right)\; .
\label{eq:s_eft}
\end{align}
Matching at $\Lambda=m_V$ gives us a Wilson coefficient $c_{u\chi}=1$
for $g_\chi = g_u = 1$ and $c_{u\chi}=1/4$ for $g_\chi = g_u =
1/2$. In Fig.~\ref{fig:s_cross} we show the effective Lagrangian
predictions for the total LHC rates. As expected, it agrees with the
full model predictions in the decoupling region $m_V \gtrsim
5$~TeV. To predict the correct relic density the dark matter candidate
should be comparably light, in line with the effective theory
requirement.  For a constant ratio $m_V/m_\chi=3$, the agreement of
the full model with the effective theory is reached slowly, and it
only happens for masses where the LHC rates are heavily suppressed.
In this case, the effective Lagrangian description does never
approximate the mono-jet cross section rate well.

\subsubsection*{Kinematic distributions}

In Fig.~\ref{fig:s_ptj} we show the differential $\met$ cross section
for a selection of dark matter and mediator mass values and for two
choices on the $\met$ cut in the upper and lower panels.  We describe
the simple, small width scenario with $g_\chi = g_u = 1/2$. The $\met$
distributions are peaked towards the minimum allowed $\met$ values and
all shown mass values show the same behavior, up to the rate
normalization. To allow for an on-shell mediator production we always
choose a low acceptance cut on $\met$~\cite{olli}.

The normalized distributions for the matched effective operator
approximate the simplified model well. The EFT gives a slightly harder
distribution for small mediator masses, while it agrees with the full
model for heavier mediators. For small mediator masses there exists a
difference in the normalization of the total rate, \ie an EFT
interpretation would lead to a wrong measurement of $g_{\chi,
  u}^2/m_V^2$. This effect is essentially independent of the dark
matter mass or the $\met$ cut, and it poses a major problem for a
global EFT analysis. For $m_V \gtrsim 5$~TeV both, the cross section
and the differential distributions are in excellent agreement between
the full model and the EFT.

\subsubsection*{Effective Lagrangian vs model}

The $s$-channel vector mediator model can, in principle, cover the
entire $m_\chi - m_V$ mass plane. This plane can be separated in to
two preferred regimes by the relic density constraints and by the pole
condition $m_V/m_\chi =2$.  The half plane with $m_V \lesssim m_\chi$
will not be described by an effective Lagrangian approximation, for
$m_V = m_\chi~...~2 m_\chi$ the mediator will stay below its mass
shell, but the regime $m_V \gg m_\chi$ can in principle work fine.  In
the light-mediator case the non-relativistic dark matter
annihilation is essentially insensitive to the mediator mass.

The mediator width will be small, as long as we choose the couplings
$g_\chi = g_u < 1$. Towards a strongly interacting model it can reach
$\Gamma_V/m_V \sim 10\%$, slowly starting to deviate from a narrow
resonance description.  When we separate the two couplings $g_u$ and
$g_\chi$, a smaller value of $g_u$ will suppress the production rate,
penalizing the potentially observable LHC mono-jet signatures.  The
effect of the suppressed cross section is tamed by the increased
invisible branching ratio. This way, scenarios with $g_\chi = g_u =
1/2$ will lead to equivalent cross sections as scenarios with $g_\chi
=1$ and $g_u = 1/4$. Conversely, larger values of $g_u$ will allow us
to include two-jet resonance searches in a global analysis.  In the
effective theory's region of validity at the LHC, the mono-jet
production rate factorizes into the mediator production rate and the
invisible branching ratio. The three ingredients $\sigma_V(m_V,g_u)$,
$\Gamma_{\chi\chi}(m_V,m_\chi,g_\chi)$, and
$\Gamma_{uu}(m_V,m_\chi,g_\chi)$ have to be combined in a global fit.

The key feature of the effective theory description is that the LHC
kinematics introduce another energy scale, namely the maximum amount
of partonic energy available to produce a resonance. We find that the
mediator turns into a non-propagating state for $m_V \gtrsim
5$~TeV. This value is largely independent of the $\met$ cut, as well
as the specific dark matter mass, as long as $m_V > m_\chi$. This
limits the region of validity for a proper effective theory
description to a region with very small LHC cross sections. The effect
of our relic density constraint is that for very heavy mediators the
dark matter particle cannot be too light, either. In essence, the EFT
description does not hold in the two regions we are interested in at
the LHC: light mediators for $m_V < 2 m_\chi$ and moderately heavy
mediators with $m_V > 2 m_\chi$ but $m_V < 5$~TeV.

\subsubsection*{Axial-vector case}

\begin{figure}[t]
\includegraphics[width=0.35\textwidth]{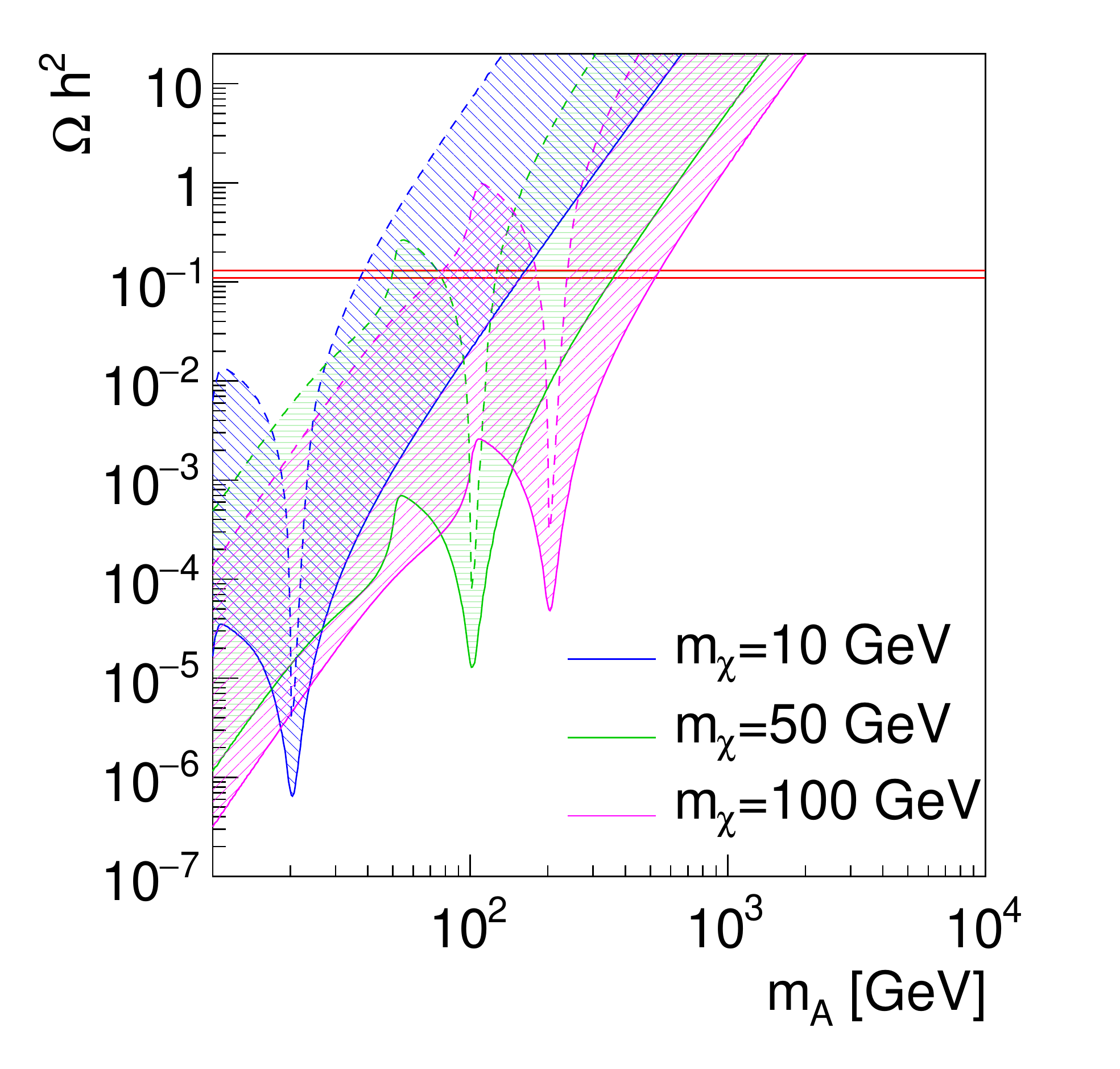}\hspace*{-0.4cm}
\includegraphics[width=0.35\textwidth]{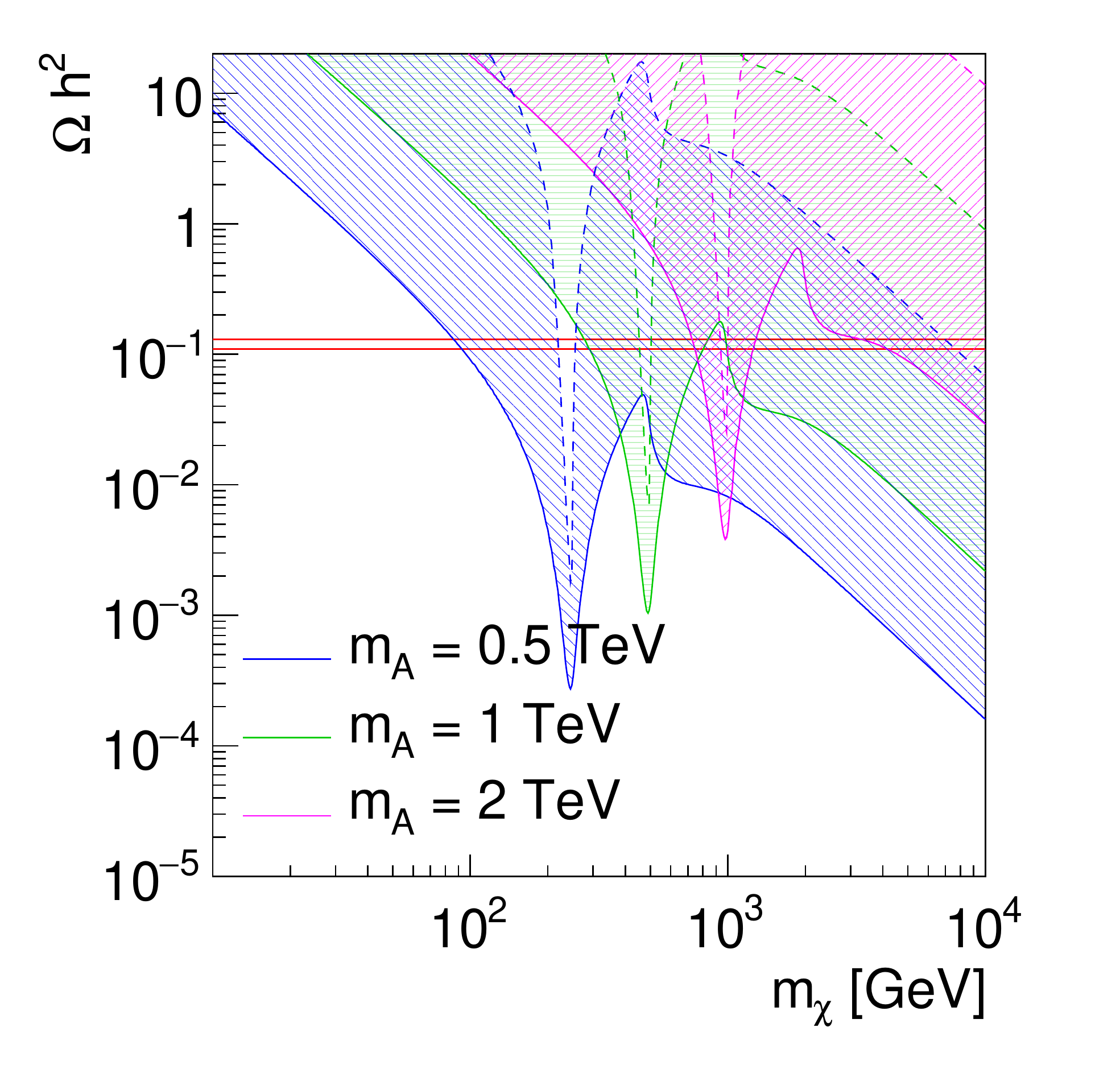}\hspace*{-0.4cm}
\includegraphics[width=0.35\textwidth]{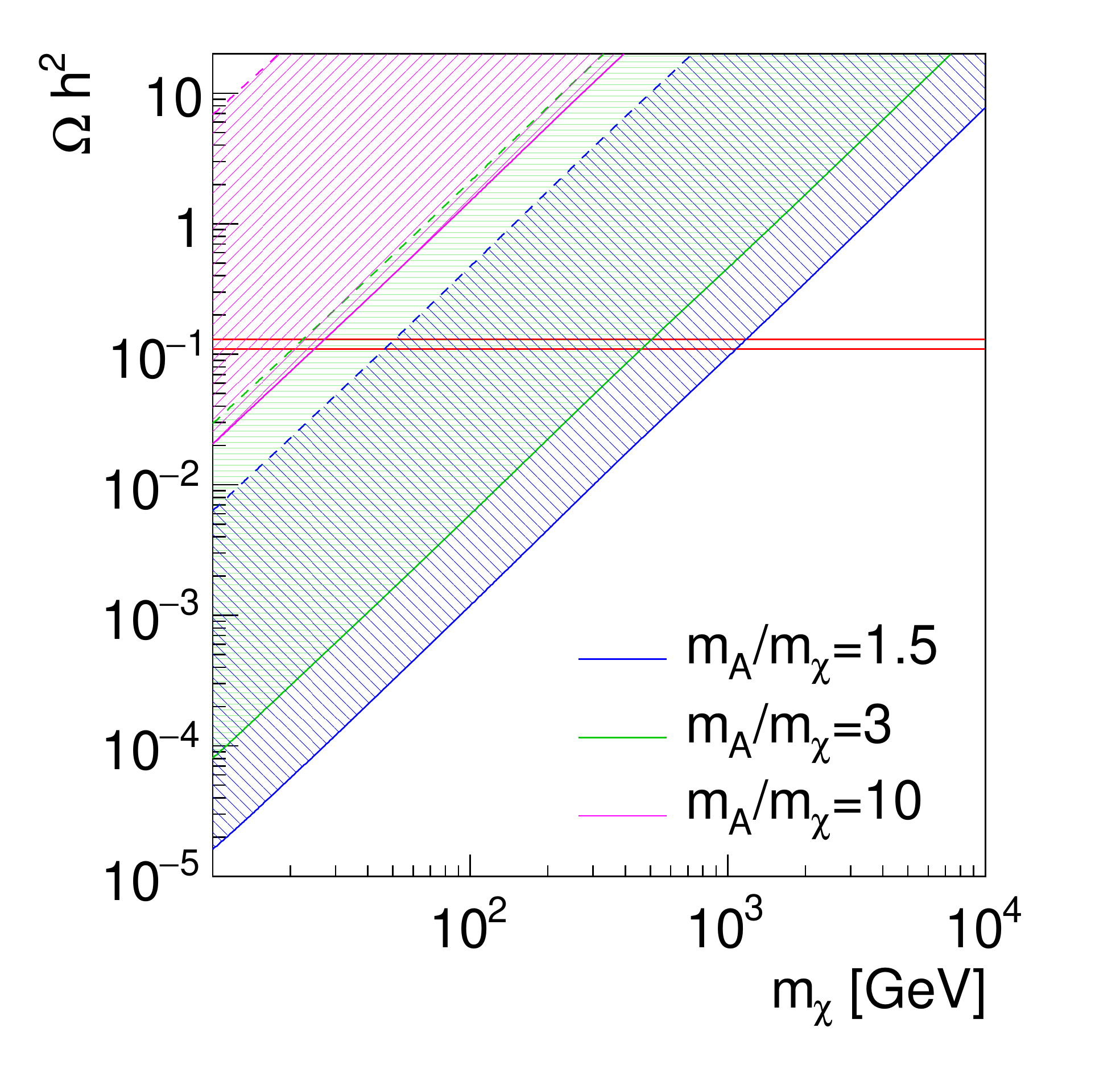}
\caption{Relic density for the $s$-channel axial-vector mediator model
  as a function of the mediator mass for constant dark matter mass
  (left), as a function of the dark matter mass for constant mediator
  mass (center) and as a function of the dark matter mass for a
  constant ratio of mediator to dark matter mass. We assume
  $g_u=g_\chi=0.2~...~1$; large relic densities correspond to small
  coupling.}
\label{fig:relic_ax}
\end{figure}

We can also assign axial-vector interactions rather than vector
interactions to the $s$-channel mediator,
\begin{align}
\lag \supset g_u\ \bar{u}\ \gamma^\mu\gamma^5 A_\mu u  
          + g_\chi\ \bar{\chi}\ \gamma^\mu\gamma^5 A_\mu \chi\ \; .
\label{eq:sax_model}
\end{align}
The preferred region by the observed relic density is given in
Fig.~\ref{fig:relic_ax}, indicating that the model can easily
reproduce the observed relic density for a wide range of parameters,
not just around the pole $m_A \approx 2 m_\chi$.\medskip

The total and differential rates are shown in
Fig.~\ref{fig:sa_cross}. We see literally no change compared to the
vector case in Fig.~\ref{fig:s_cross} and Fig.~\ref{fig:s_ptj}. The
only difference is that the preferred parameter choices for the relic
density are shifted. Moreover, the different combinations of vector
and axial-vector couplings make a difference in direct detection
(constraints), because of different coherent vs incoherent scattering
on the nucleus. The axial-vector coupling couples to the spin of the
nucleon, there is no coherent enhancement for large nuclei, and the
direct detection constraints are significantly weakened. In other
words, vector mediator models which are ruled out by direct detection
constraints can survive with an axial-vector mediator. For the LHC
this assignment makes no difference.\medskip

The effective Lagrangian corresponding to Eq.\eqref{eq:sax_model}
again includes a four-fermion operator,
\begin{align}
\lag_\text{eff} \supset \frac{c_{u\chi}}{\Lambda^2} \; ( \bar{u}\ \gamma^\mu\gamma^5 u) \ ( \bar{\chi} \gamma_\mu\gamma^5 \chi)  \;.
\label{eq:sax_eft}
\end{align}
As for the vector mediator case the crucial question for the effective
theory is if the mediator can be produced as a propagating state, \ie
if $m_V < 5$~TeV.

\begin{figure}[t]
\includegraphics[width=0.33\textwidth]{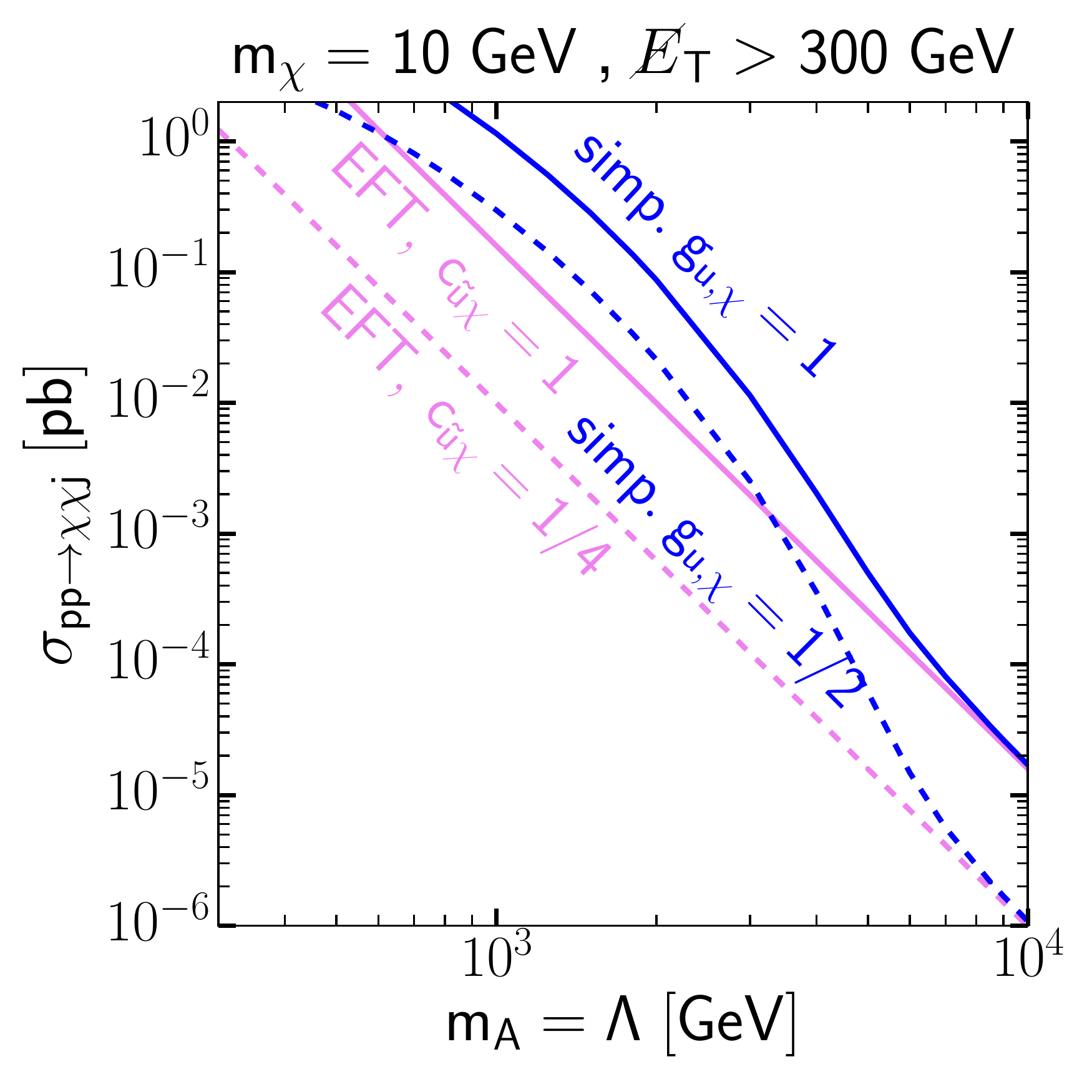}\hspace*{-0.2cm}
\includegraphics[width=0.33\textwidth]{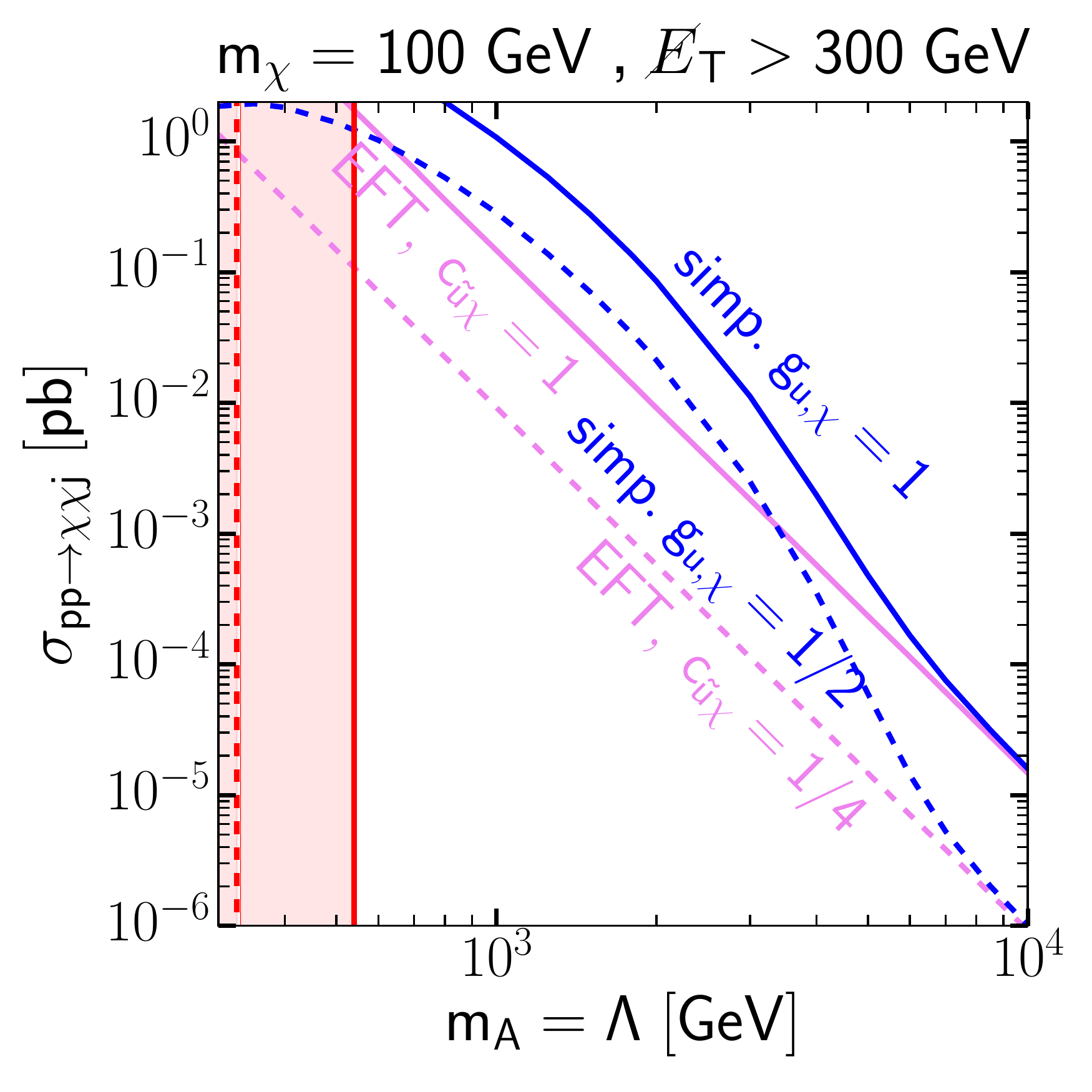}\hspace*{-0.2cm}
\includegraphics[width=0.33\textwidth]{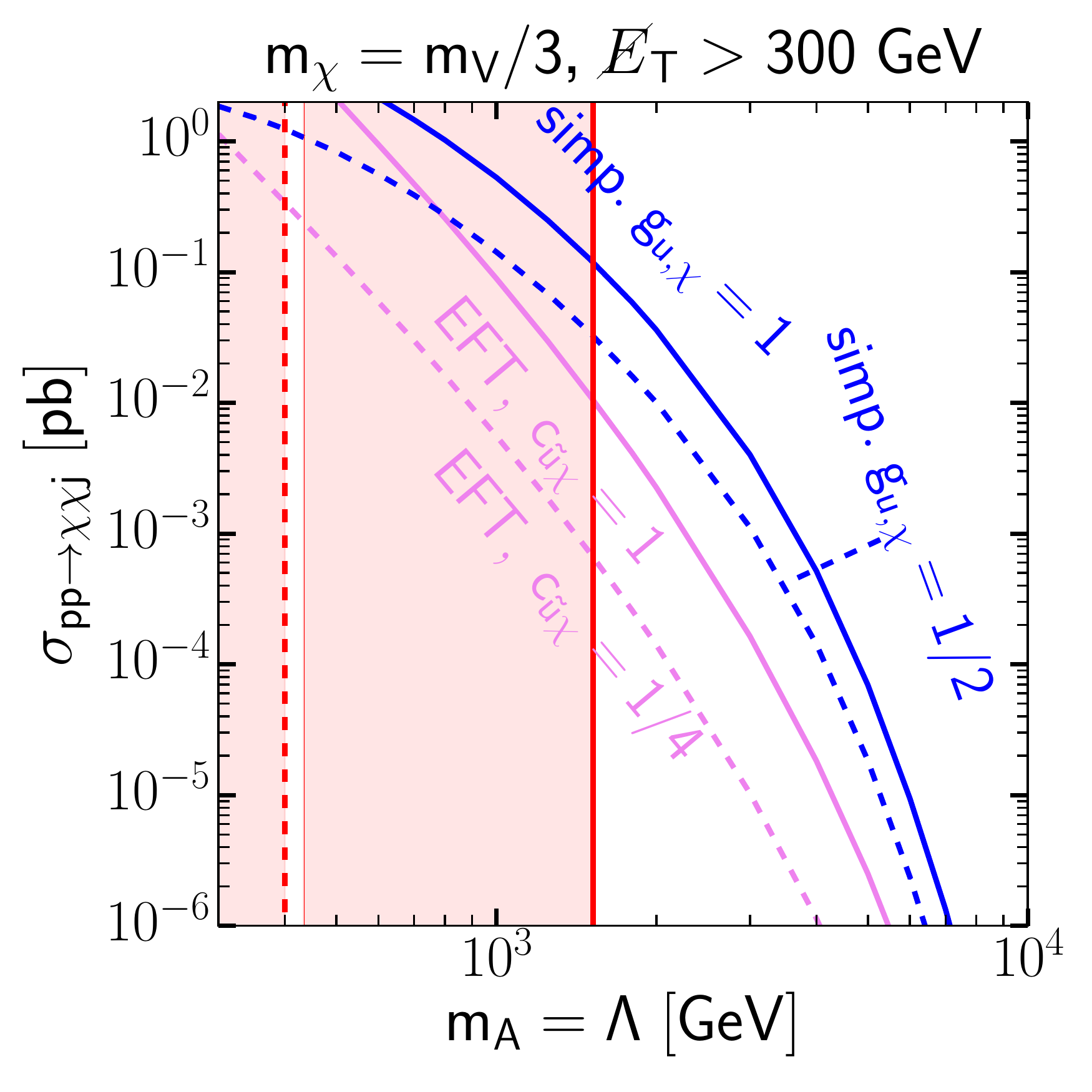} \\
\includegraphics[width=0.33\textwidth]{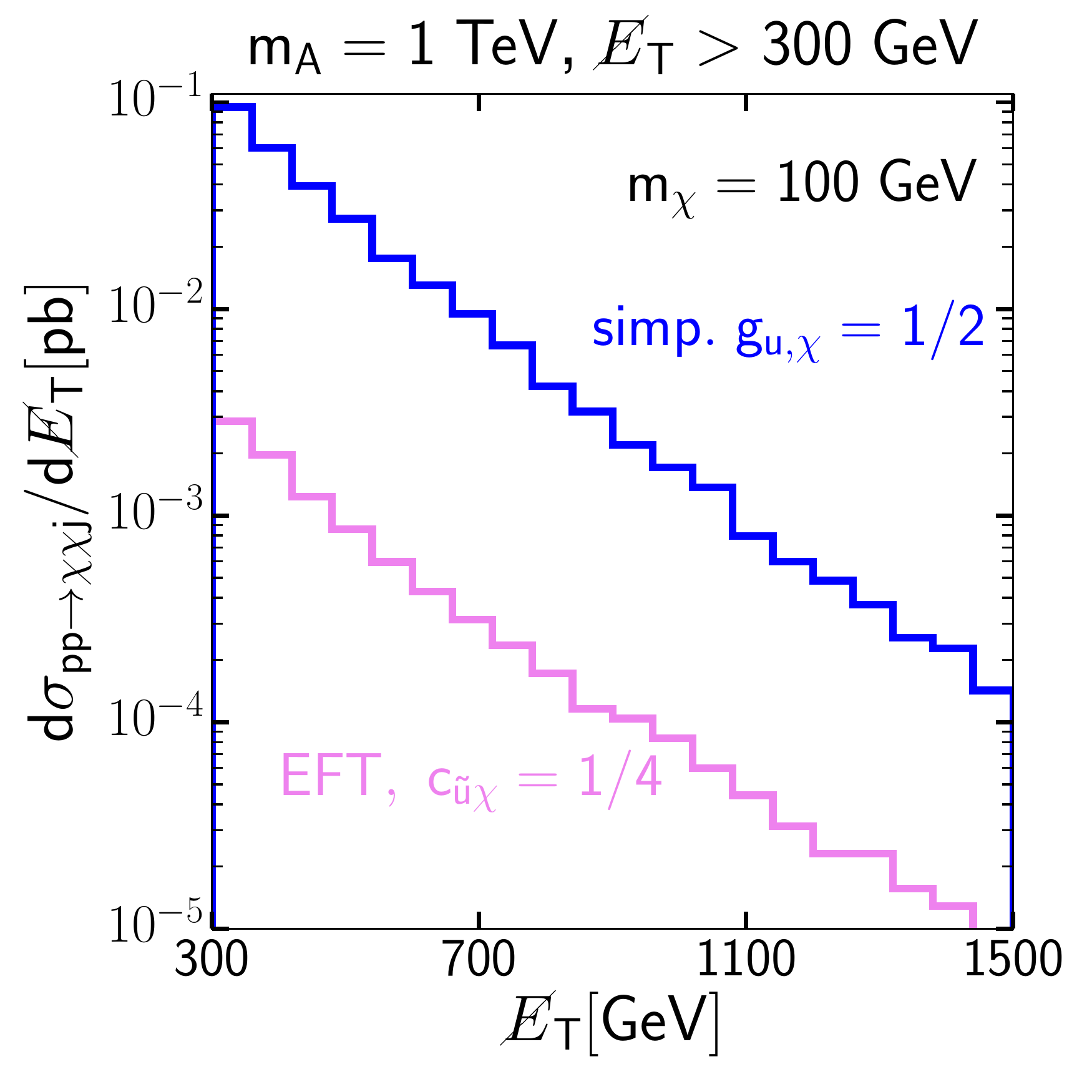}\hspace*{-0.2cm}
\includegraphics[width=0.33\textwidth]{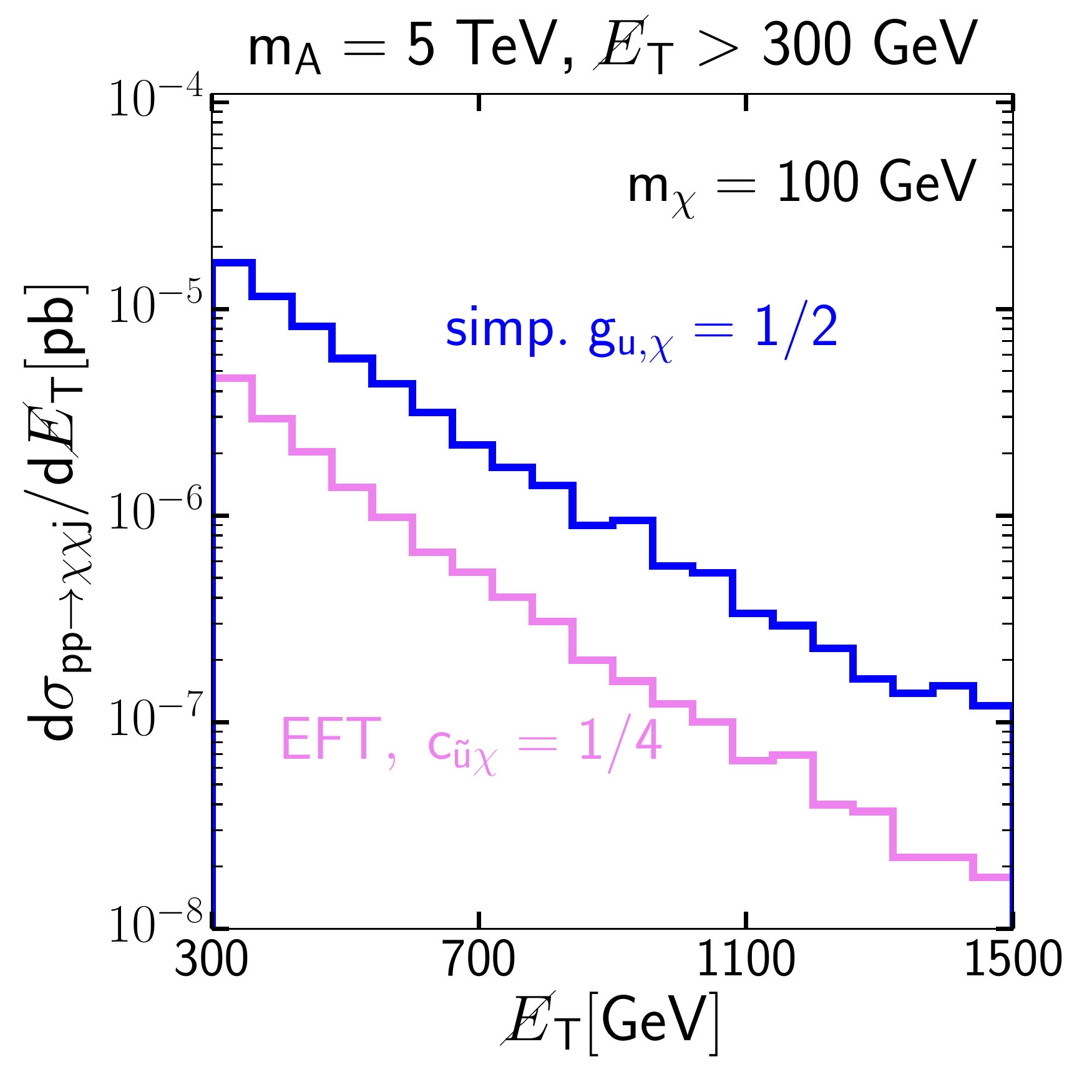}\hspace*{-0.2cm}
\includegraphics[width=0.33\textwidth]{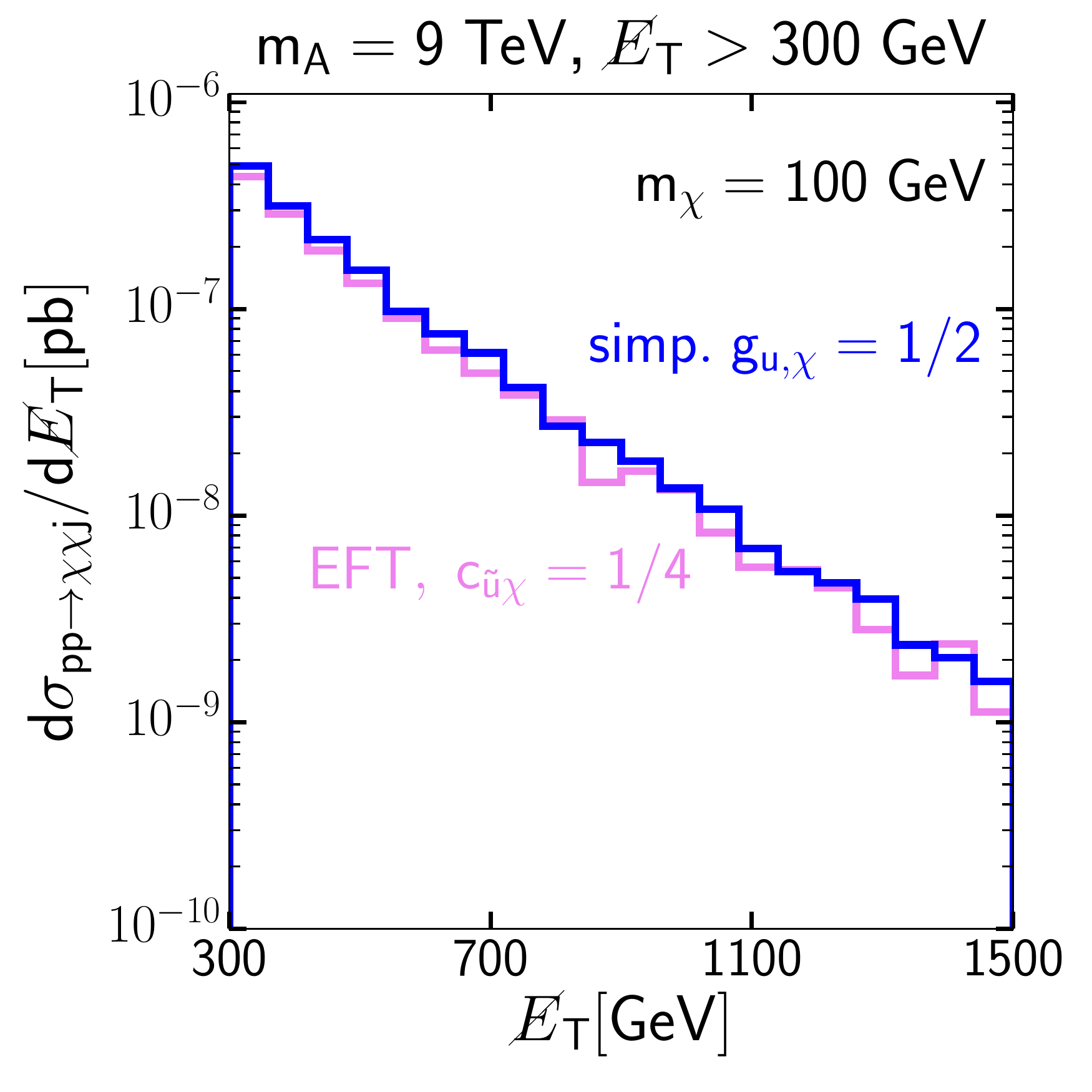}
\caption{Total production rate (upper panels) and $\met$ distributions
  (lower panels) in the $s$-channel vector mediator model. The
  vertical lines indicate the mediator masses predicting the observed
  relic density. The vertical bands show the mediator masses
  predicting the observed relic density: upper edge for
  $\Omega_\chi^\text{obs}+10\%$ and lower edge for
  $\Omega_\chi^\text{obs}/10$.}
\label{fig:sa_cross}
\end{figure}

\clearpage
\section{Loop-mediated scalar in s-channel}
\label{sec:s_loop}

For a (pseudo-)scalar $s$-channel mediator the situation becomes more
complicated. In this case, flavor bounds require minimal flavor
violating couplings, resulting in dominant couplings to heavy
quarks. A supersymmetric UV-completion includes the heavy
(pseudo-)scalar Higgs and neutralino dark matter. As usual, we need to
postulate two interactions,
\begin{align}
\chi-\chi-S \qquad \text{dark matter}
\qqqquad 
t-t-S \qquad \text{top quarks} \; .
\end{align}
This assumption already reflects the fact that for many reasons we
want to avoid introducing very large Yukawa couplings of the new
scalar to light quarks. Couplings which are not aligned with the SM
Yukawa couplings induce FCNCs, which are lead to strong bounds from
flavor
observables~\cite{top_scalar,matt_dorival,Mattelaer:2015haa,sloop,Dolan:2014ska}. We
therefore allow also for a $c-c-S$ coupling; we will see that it
induces the leading dark matter annihilation channel, but have no
effect on the LHC signatures. Minimally flavor-violating couplings are
automatically realized if we assume a mixing of the SM singlet
mediator with the SM Higgs through a portal
coupling~\cite{Bauer:2016gys}. Alternative realizations include a
singlet coupling to the Standard Model through dimension-5
operators~\cite{Bauer:2016hcu}, or a $SU(2)_L$ doublet mediator in the
case of singlet-doublet dark matter. For the mass spectrum we can
distinguish the same three cases as for the vector mediator,
Eq.\eqref{eq:s_regimes}. For our EFT analysis at the LHC we will
eventually focus on the parameter region
\begin{align}
m_S > 2 m_\chi 
\qquad \text{and} \qquad 
m_S > 2 m_t \; .
\label{eq:spectrum_sloop}
\end{align}
If we remain agnostic about the underlying theory leading to minimal
flavor violation, we can define our toy Lagrangian with the Yukawa
interaction
\begin{align}
\lag \supset 
- \frac{m_t}{v}\,S\,\bar t (g_{S,t}+i\,\gamma_5 g_{P,t}) t 
+ S\,\bar \chi (g_{S,\chi}+i\,\gamma_5 g_{P,\chi}) \chi 
+ \text{h.c.}
\label{eq:s_loop_model}
\end{align}
We only write out the mediator coupling to top quarks, but we will
also study the impact of light-quark couplings to the mediator in a
minimal flavor violation structure.  The ansatz of the top Yukawa
coupling proportional to $m_t/v$ bears no physical relevance, in
general it is likely to be suppressed by a new physics scale $\Lambda
> v$.  The scalar coupling to the dark matter fermions can be linked
to $m_\chi$, but does not have to.  The partial width of the scalar
mediator decaying to top quarks increases with the scalar mass, just
like for the Higgs case,
%
\begin{align}
\frac{\Gamma_{S \to tt}}{m_S} 
= \frac{3 G_F m_t^2 g_{S,t}^2}{4 \sqrt{2} \pi} \; \left( 1 - \frac{4m_t^2}{m_S^2} \right)^{3/2} 
< \frac{3 G_F m_t^2 g_{S,t}^2}{4 \sqrt{2} \pi}  
< 5\% \; ,
\end{align}
assuming $g_{S,t} <1$. A lighter mediator just decaying to bottom
quarks and dark matter would be significantly more narrow, unless we
introduce even larger Yukawa couplings than to the top quark, or we
couple the scalar to the gauge sector.

Just as for the vector mediator, the mono-jet rate in our parameter
region will factorize into $\sigma_{S+j} \times \br_{\chi\chi}$.  Like the
Higgs, a light scalar mediator will dominantly be produced through a
top-loop-induced coupling to gluons, with an additional gluon jet
giving the mono-jet signature.  For the Higgs, the corresponding
dimension-6 operator does not decouple with the top mass, but is
instead suppressed by the electroweak VEV.\medskip

\begin{figure}[b!]
\begin{center}
\begin{fmffile}{feyn1}
\begin{fmfgraph*}(100,60)
\fmfset{arrow_len}{2mm}
\fmfleft{i1,i2}
\fmfright{o1,o2}
\fmf{fermion,width=0.6,tension=1.7}{i1,v1}
\fmf{fermion,width=0.6,tension=1.7}{v1,i2}
\fmf{dashes,width=0.6,tension=1.7,label=$S$}{v1,v2}
\fmf{gluon,width=0.6,tension=1.7}{o2,v4}
\fmf{gluon,width=0.6,tension=1.7}{v3,o1}
\fmf{fermion,width=0.6,label=$t$}{v3,v4}
\fmf{fermion,width=0.6}{v2,v3}
\fmf{fermion,width=0.6}{v4,v2}
\fmflabel{$\chi$}{i1}
\fmflabel{$\chi$}{i2}
\fmflabel{$g$}{o1}
\fmflabel{$g$}{o2}
\fmfv{decor.shape=circle,decor.filled=full,decor.size=2thick}{v1}
\fmfv{decor.shape=circle,decor.filled=full,decor.size=2thick}{v2}
\end{fmfgraph*}
\hspace{1.5cm}
\begin{fmfgraph*}(100,60)
\fmfset{arrow_len}{2mm}
\fmfleft{i1,i2}
\fmfright{o1,o2}
\fmf{fermion,width=0.6}{i1,v1}
\fmf{fermion,width=0.6}{v1,i2}
\fmf{dashes,width=0.6,label=$S$}{v1,v2}
\fmf{fermion,width=0.6}{o2,v2}
\fmf{fermion,width=0.6}{v2,o1}
\fmflabel{$\chi$}{i1}
\fmflabel{$\chi$}{i2}
\fmflabel{$c,t$}{o1}
\fmflabel{$c,t$}{o2}
\fmfv{decor.shape=circle,decor.filled=full,decor.size=2thick}{v1}
\fmfv{decor.shape=circle,decor.filled=full,decor.size=2thick}{v2}
\end{fmfgraph*}
\hspace{1.5cm}
\begin{fmfgraph*}(100,60)
\fmfset{arrow_len}{2mm}
\fmfleft{i1,i2}
\fmfright{o1,o2,o3,o4}
\fmf{fermion,width=0.6,tension=1.7}{i1,v1}
\fmf{fermion,width=0.6,tension=1.7}{v2,i2}
\fmf{fermion,width=0.6,tension=1.0,label=$\chi$}{v1,v2}
\fmf{dashes,width=0.6,label=$S$}{v1,v3}
\fmf{dashes,width=0.6,label=$S$}{v2,v4}
\fmf{fermion,width=0.6}{o1,v3}
\fmf{fermion,width=0.6}{v3,o2}
\fmf{fermion,width=0.6}{o3,v4}
\fmf{fermion,width=0.6}{v4,o4}
\fmflabel{$\chi$}{i1}
\fmflabel{$\chi$}{i2}
\fmflabel{$c$}{o1}
\fmflabel{$c$}{o2}
\fmflabel{$c$}{o3}
\fmflabel{$c$}{o4}
\fmfv{decor.shape=circle,decor.filled=full,decor.size=2thick}{v1}
\fmfv{decor.shape=circle,decor.filled=full,decor.size=2thick}{v2}
\fmfv{decor.shape=circle,decor.filled=full,decor.size=2thick}{v3}
\fmfv{decor.shape=circle,decor.filled=full,decor.size=2thick}{v4}
\end{fmfgraph*}
\end{fmffile}
\end{center}
\caption{Feynman diagrams describing dark matter annihilation in the
  scalar $s$-channel mediator model defined in
  Eq.\eqref{eq:s_loop_model}.}
\label{fig:feyn_schannel}
\end{figure}
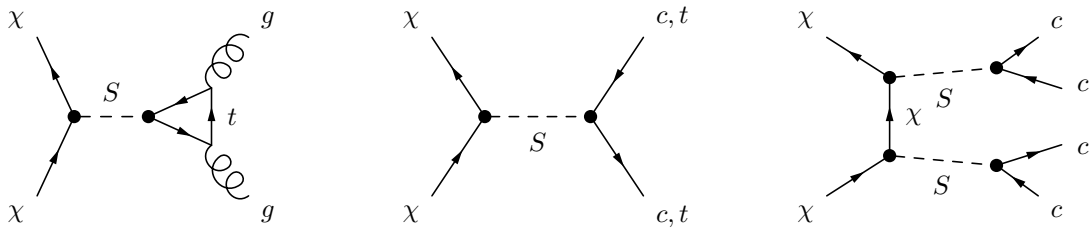

\begin{figure}[t]
\includegraphics[width=0.35\textwidth]{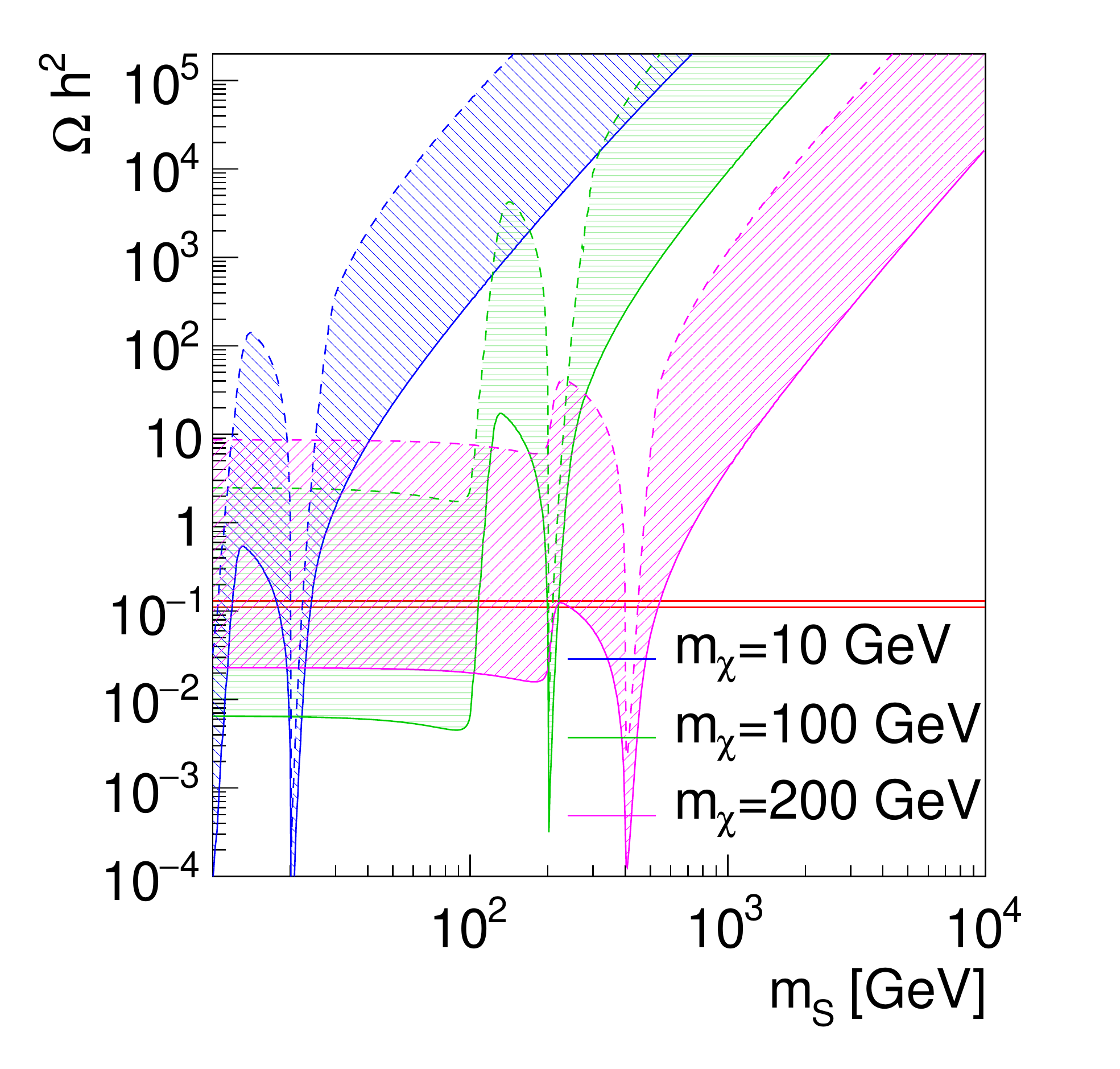}\hspace*{-0.6cm}
\includegraphics[width=0.35\textwidth]{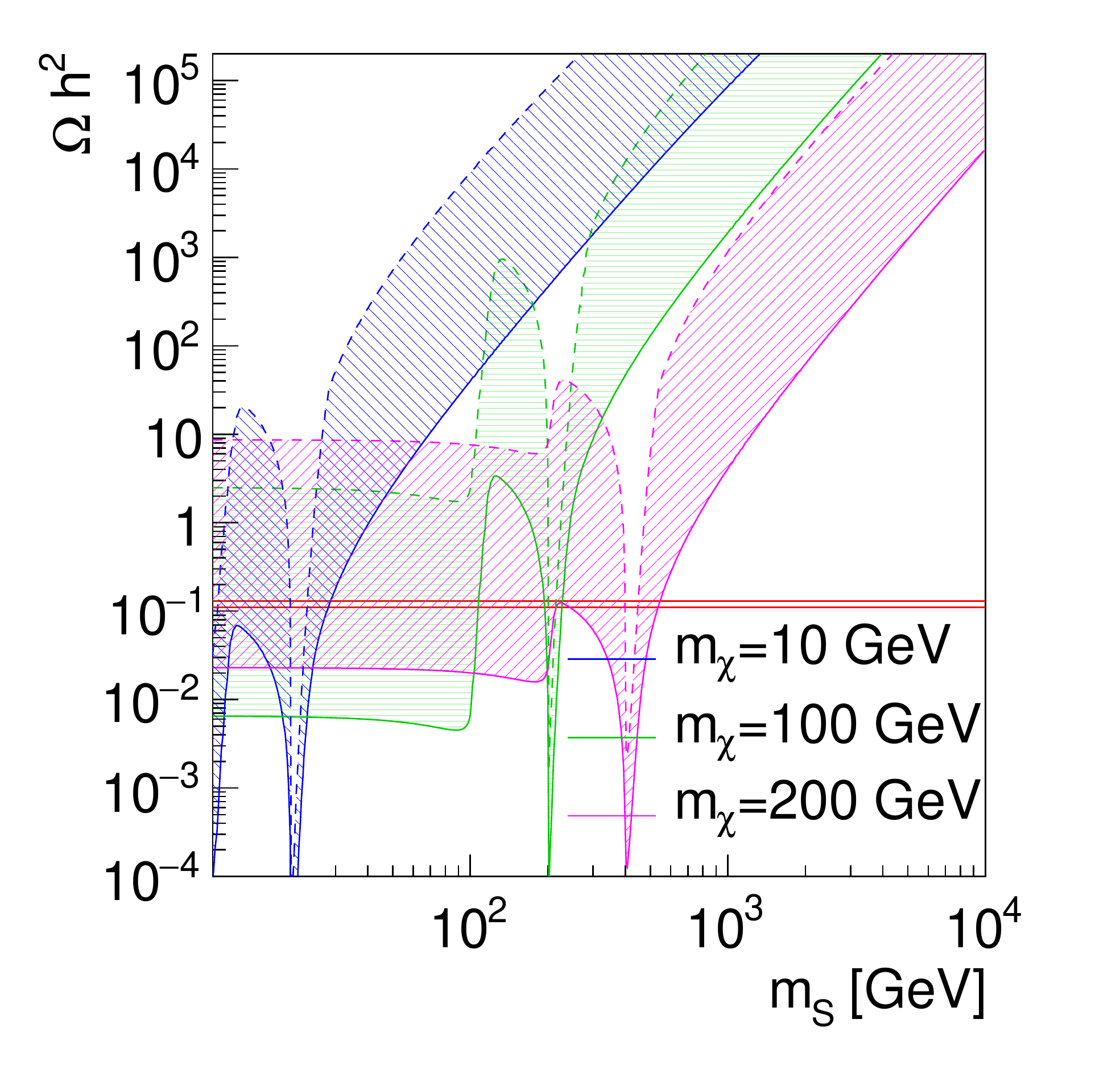}\hspace*{-0.6cm}
\includegraphics[width=0.35\textwidth]{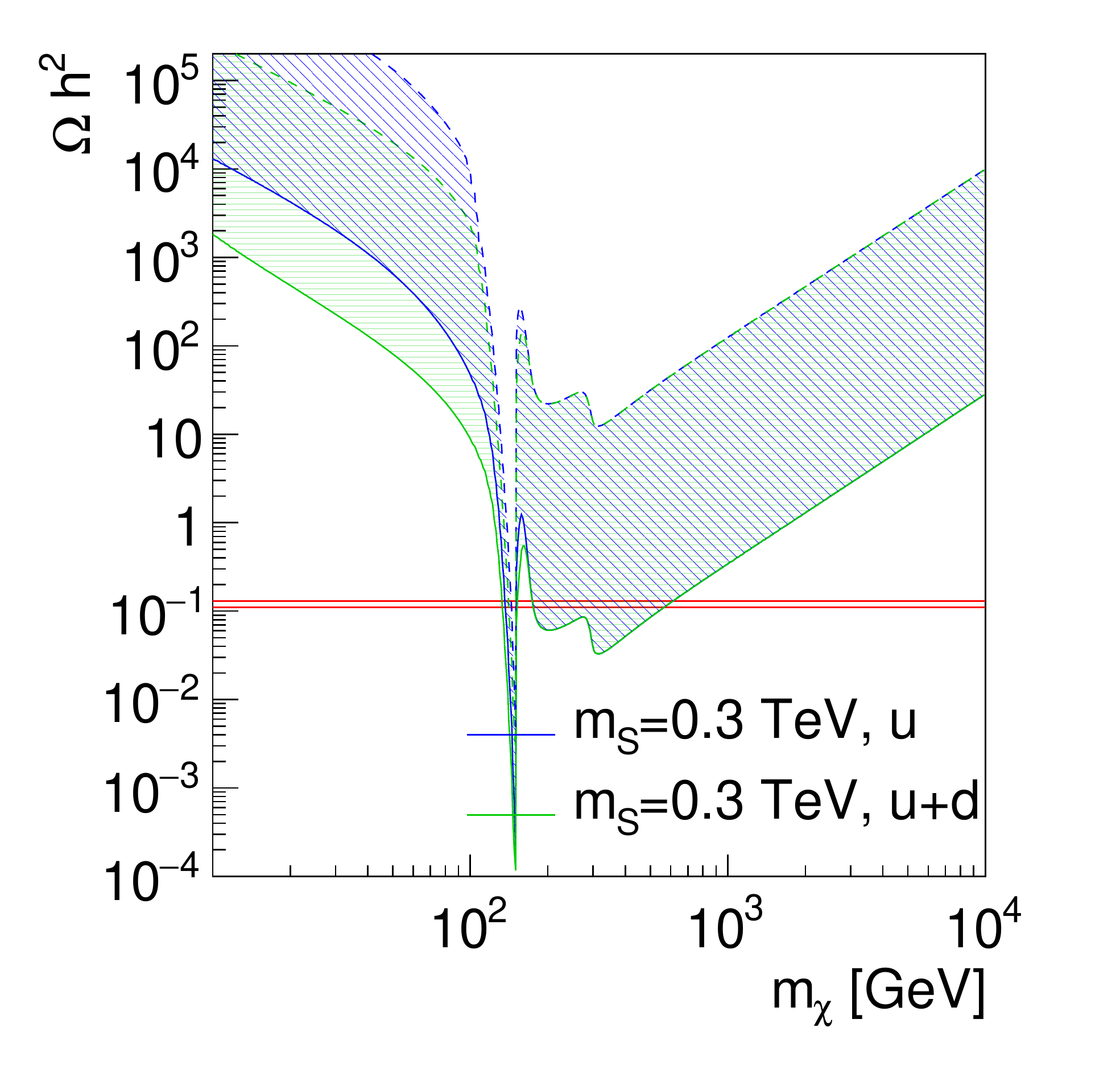} \\
\includegraphics[width=0.35\textwidth]{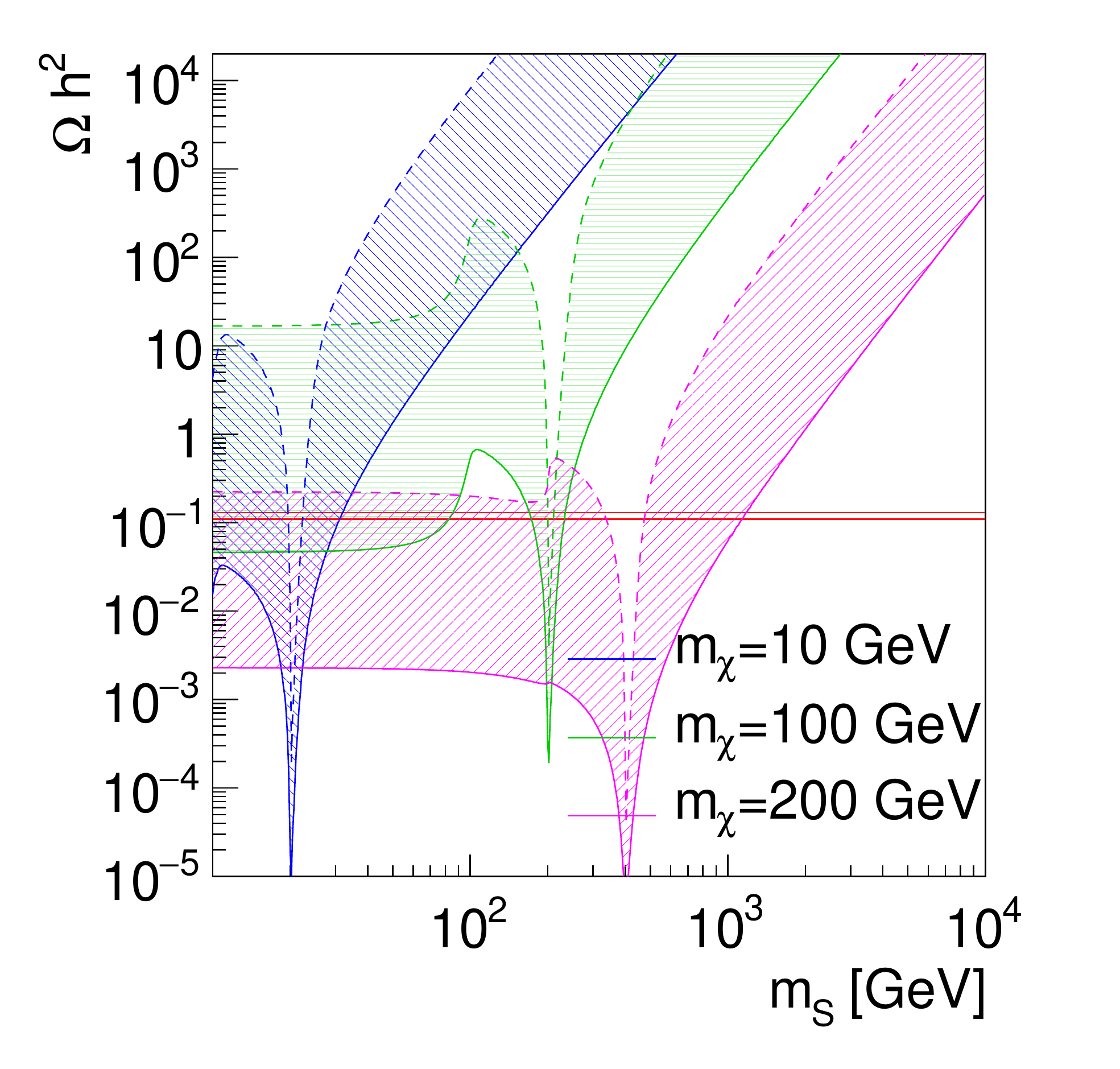}
\hspace*{-0.6cm}
\includegraphics[width=0.35\textwidth]{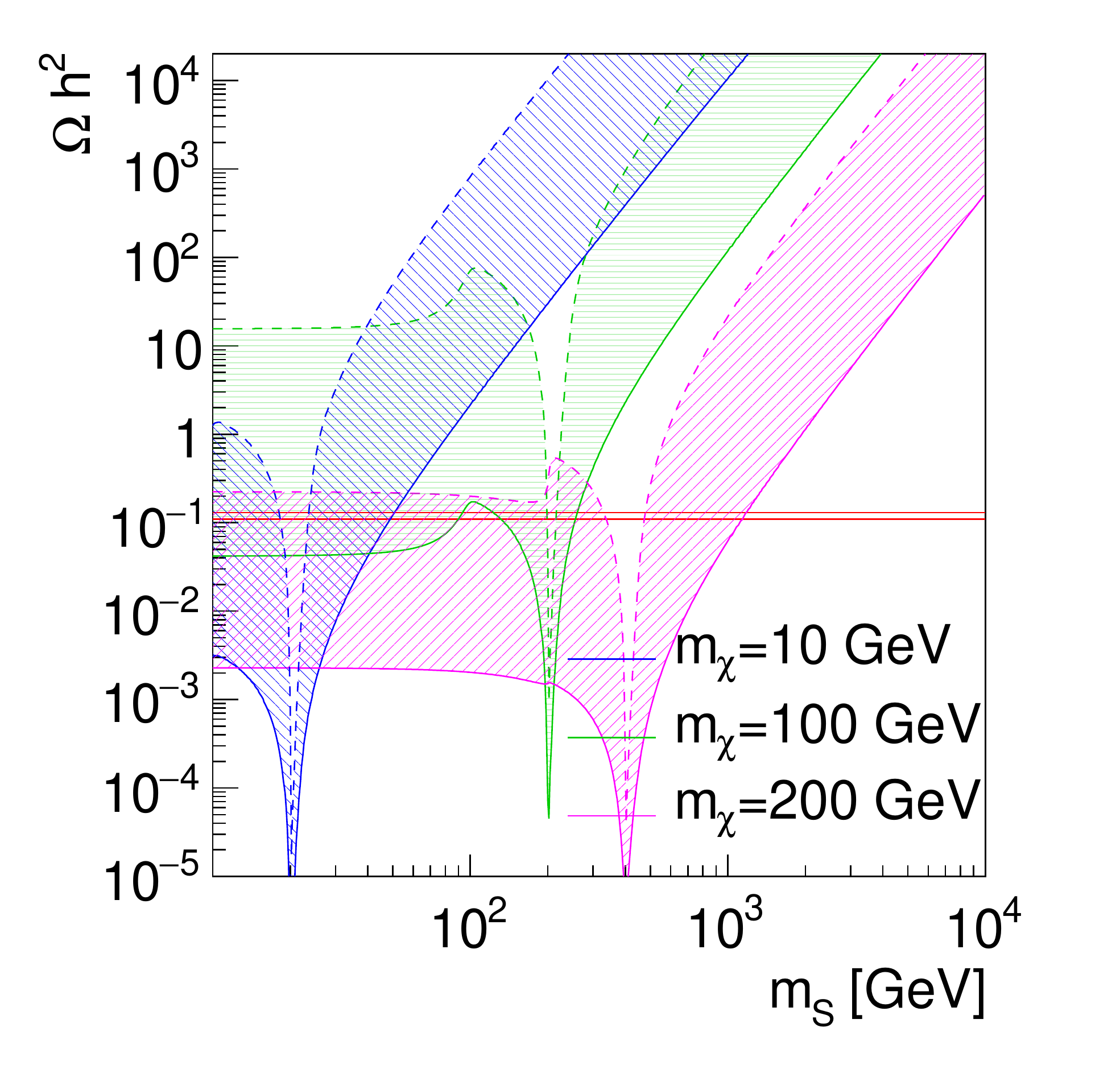}\hspace*{-0.6cm}
\includegraphics[width=0.35\textwidth]{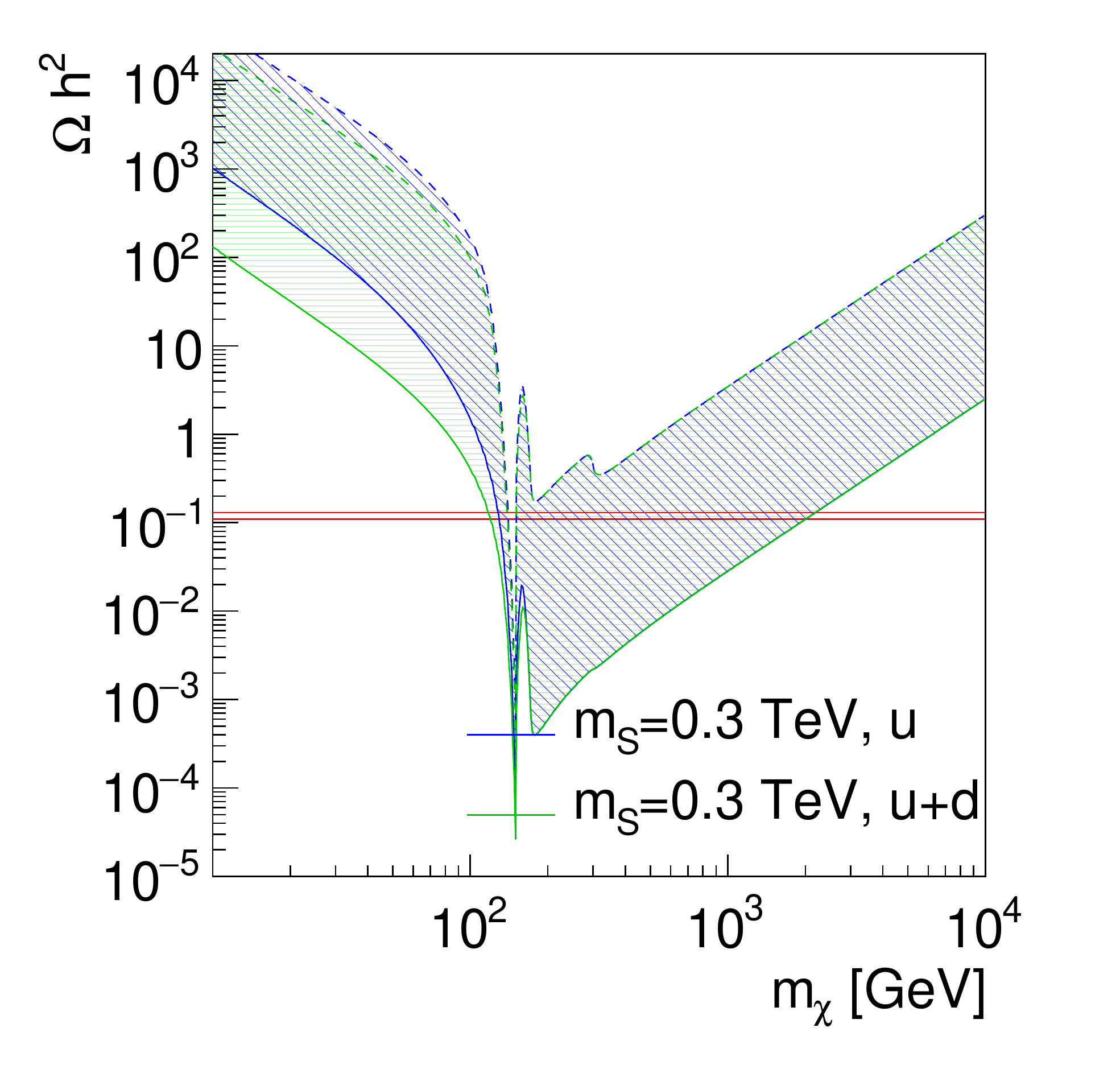} 
\caption{Relic density for the loop-induced $s$-channel scalar (upper
  panels) and pseudo-scalar (lower panels) mediator model as a
  function of the mediator mass for constant dark matter mass with
  only couplings to up-type quarks (left), with couplings to up- and
  down-type quarks (center) and as a function of the dark matter mass
  for a constant mediator mass (right). Over the shaded bands we vary
  the couplings $g_{S,t}=g_{S,\chi}=0.2~...~1$; large relic densities
  correspond to small coupling.}
\label{fig:relic_sloop}
\end{figure}

As usual, we use the Lagrangian of Eq.\eqref{eq:s_loop_model} to
compute the predicted relic density with the help of
\textsc{FeynRules}\cite{feynrules} and \textsc{Micromegas}\cite{micromegas}.  What
distinguishes this model from the two models discussed before is that
the dark matter annihilation process is not directly related to the
LHC production process. Three dark matter annihilation channels
are illustrated in Fig.~\ref{fig:feyn_schannel}. A very light mediator
will decay to two gluons through a top loop. To avoid strong flavor
constraints, a tree-level amplitude $\chi \chi \to c\bar{c}$ will
dominate for slightly heavier dark matter and a light mediator. Based
on the same couplings, there is a $t$-channel diagram with a decay to
four fermions. If there also exists a Yukawa coupling to bottom
quarks, the annihilation of a slightly heavier mediator will then take
over as $\chi \bar{\chi} \to b\bar{b}$.  An even heavier mediator will
annihilate into off-shell top quarks, $\chi \bar{\chi} \to (W^+ b)(W^-
\bar{b})$, and for $m_\chi > m_t$ the simple annihilation process
$\chi \bar{\chi} \to t\bar{t}$ will lead to a very efficient
annihilation. In the latter two cases, the link between the LHC
production rate and the dark matter annihilation rate becomes very
model dependent.

In Fig.~\ref{fig:relic_sloop} we show the predicted relic density for
a range of dark matter and mediator masses for scalar and pseudoscalar mediators. The coupling
$c-c-S$ with the $s$-channel (pseudo-)scalar and $t$-channel dark matter
exchange dominates over the process $\chi \bar{\chi} \to gg$, which we
nevertheless include in our numerical analysis. We see two major
structures in the mediator mass dependence of the relic
density. First, there is the usual peak for non-relativistic
annihilation at $m_S = 2 m_\chi$, which appears exactly as for a
tree-level $s$-channel vector. In addition, the predicted relic
density rapidly drops around $m_S = m_\chi$. The reason for this rapid
increase in the annihilation rate is the $t$-channel diagram $\chi
\chi \to S^* S^*$. It allows for a mass insertion, covering all
possible helicity combinations and therefore dominating over the
velocity-suppressed $s$-channel diagram. For the pseudoscalar this velocity suppression is absent, and the peak is less pronounced. This additional diagram makes
it easy to match the observed relic density over wide range of dark
matter and mediator masses, easier than for example for the
$s$-channel vector case.\medskip

The variety of dark matter annihilation channels in our model is not
reflected at the LHC. First, the tree-level couplings to charm and top
quarks are only relevant for alternative mediator decays. The dominant
production process 
\begin{align}
gg \to S +\text{jets} \to \chi \bar{\chi} + \text{jets}
\end{align}
will introduce the mono-jet signature through initial-state
radiation. The corresponding Feynman diagram is shown in the left
panel of Fig.\ref{fig:feyn_schannel}. This process is very well
known from Higgs physics, including the phase space region with
sizeable jet recoil momentum~\cite{baur}.

\subsubsection*{Total rate}

We implement the loop-mediated $s$-channel mediator model, as well as
its EFT approximations, in \textsc{FeynRules}~\cite{feynrules} and
simulate the mono-jet signal with
\textsc{MadGraph5}~\cite{madgraph}. For our toy model we show the
total rates for $\chi \bar{\chi} + j$ at parton level in
Fig.~\ref{fig:sjet1}. We choose a range of mediator masses of $m_S =
10~...~10^5$~GeV, dark matter masses of $m_\chi =10,\ 100$ and 200~GeV, and an
acceptance cut $\met > 100$~GeV, translating at parton level into $p_{T,j} >
100$~GeV. Just as for the tree-level $s$-channel mediator we observe a
distinct change in the mediator mass dependence around $m_S =
5$~TeV. Below this transition point the mediator is produced as a
propagating degree of freedom and the mono-jet rate factorizes into
$\sigma_{S+j} \times \br_{\chi \chi}$. Heavier mediators cannot be produced
on-shell, leading to the simple scaling $\sigma_{\met+j} \propto
1/m_S^4$. Interestingly, this transition point is almost identical for
the $q\bar{q}$-induced processes and $gg$-induced processes at the
LHC.\medskip

\begin{figure}[t]
\begin{center}
\includegraphics[width=1\textwidth]{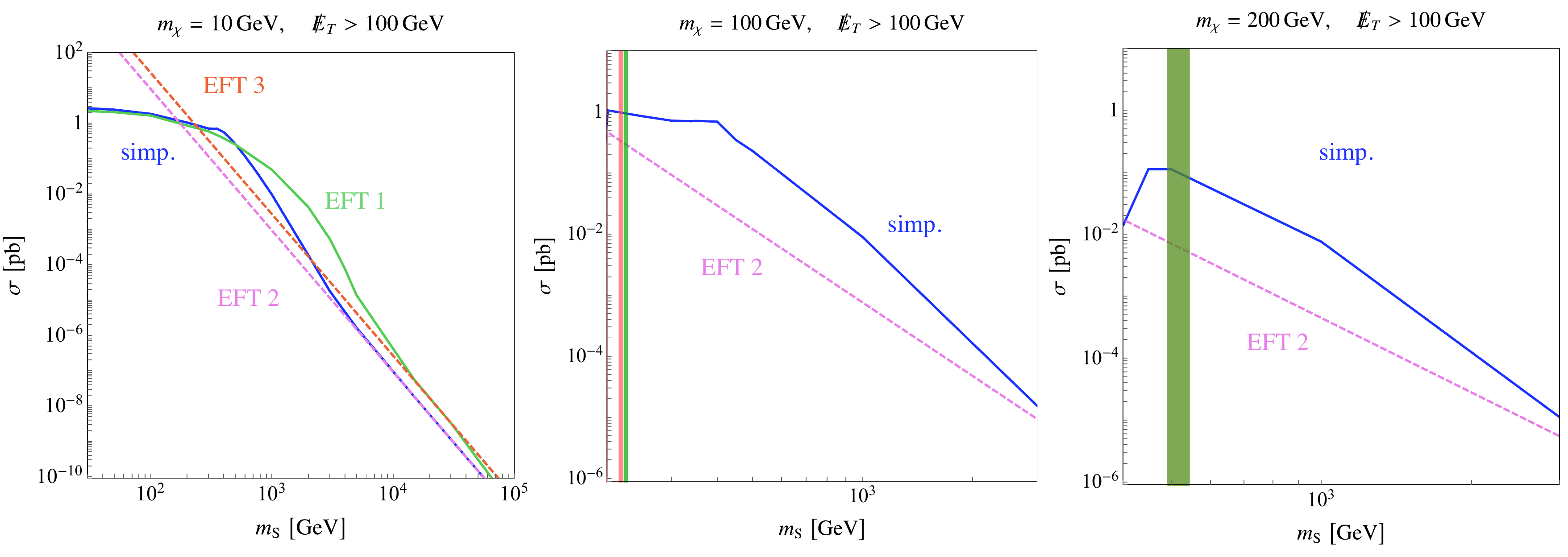}
\includegraphics[width=1\textwidth]{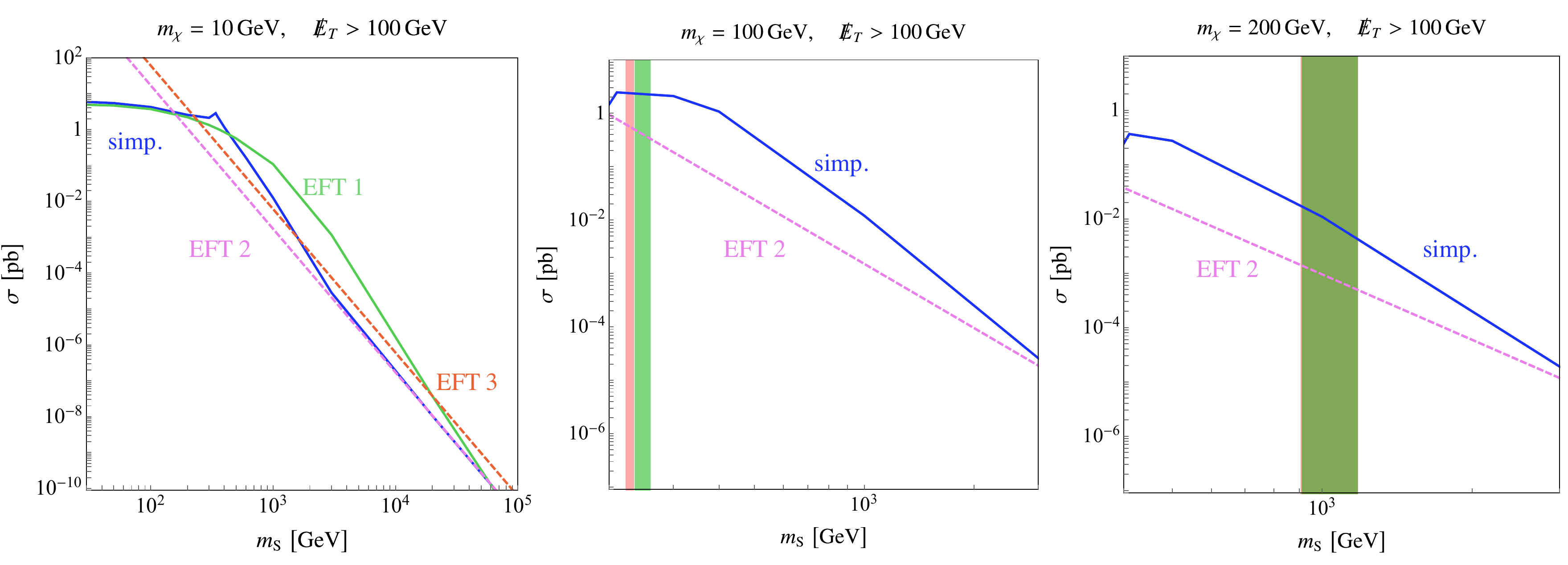}
\end{center}
\vspace{-.7cm}
\caption{Total production rate in the loop-mediated $s$-channel scalar
  model as function of the mediator mass for scalar (upper panels) and
  pseudoscalar couplings (lower panels). We show all three different
  effective Lagrangians motivated by different parameter ranges for
  $m_\chi=10$ GeV and the most relevant models for $m_\chi=100$~GeV
  and 200~GeV. For the red (green) shaded regions the annihilation
  cross section reproduces the observed relic density within
  $\Omega_\chi^\text{obs}/3$ and $\Omega_\chi^\text{obs}+10\%$. }
\label{fig:sjet1}
\end{figure}

Depending on the mass spectrum of our model we can define and match
different effective Lagrangians, which can describe the LHC signal
more or less reliably~\cite{matching_matters}. The light dark matter
obviously has to remain a propagating degree of freedom. For mediator
masses
\begin{align}
m_t> m_S > 2m_\chi 
\end{align}
we can --- in analogy to the Higgs case --- decouple the top quark
only, leading to the Lagrangian
\begin{align}
\lag_\text{eff,1} \supset 
  \frac{c_S^{g}}{\Lambda}S \, G_{\mu\nu}G^{\mu\nu}
+ \frac{c_P^{g}}{\Lambda}S \, G_{\mu\nu}\tilde G^{\mu\nu}
- \left[ S\,\bar \chi (g^\chi_S+i\,\gamma_5g^\chi_P) \chi + \text{h.c.} \right] \; .
\label{eq:eff_lag1}
\end{align}
The Wilson coefficients can be determined at $\Lambda = m_t$
\begin{alignat}{7}
\frac{c_S^g}{\Lambda}   
&=\frac{\alpha_s}{8\pi}\frac{g_S^t}{v}\tau \left[ 1+(1-\tau)f(\tau) \right]
\quad &&\stackrel{\tau\rightarrow \infty}{\longrightarrow}\quad 
\frac{\alpha_s}{12\pi}\frac{g_S^t}{v} \notag \\
\frac{\tilde c_P^g}{\Lambda}   
&=\frac{\alpha_s}{8\pi}\frac{g_P^t}{v}\tau\,f(\tau)
\quad &&\stackrel{\tau\rightarrow \infty}{\longrightarrow}\quad 
\frac{\alpha_s}{8\pi}\frac{g_P^t}{v} \; ,
\end{alignat}
where $f(\tau)=\text{arcsin}^2(1/\sqrt{\tau})$ and $\tau = 4
m_t^2/m_S^2$.  The only difference between this model and the Higgs
case is that the effective theory in the Higgs sector is formulated in
terms of doublets, leading to dimension-6 operators. We know that in
this effective theory the transverse momentum spectra will fail to
reproduce large logarithms of the type $\log
(p_T/m_t)$~\cite{baur,matt_dorival}, limiting the agreement between
the full model and the EFT approximation.\medskip

Alternatively, we can follow our original assumption in
Eq.\eqref{eq:spectrum_sloop} and decouple the mediator assuming 
\begin{align}
m_S> m_t, 2m_\chi \; .
\end{align}
We are left with dimension-6 four-fermion operators coupling to the
resolved top loop,
\begin{align}
\lag_\text{eff,2} \supset 
 \frac{c_S^t}{\Lambda^2}(\bar t t) \; (\bar\chi\chi) 
+\frac{c_P^t}{\Lambda^2}(\bar t \gamma_5t) \; (\bar\chi\gamma_5\chi)  \; .
\label{eq:eff_lag2}
\end{align}
The Wilson coefficients after matching at $\Lambda = m_S$ are by
\begin{align}
\frac{c_S^t}{\Lambda^2}=\frac{g_S^tg_S^\chi}{m_S^2}\frac{m_t}{v}
\qquad \text{and}\qquad  
\frac{c_P^t}{\Lambda^2}=\frac{g_P^tg_P^\chi}{m_S^2}\frac{m_t}{v} \; .
\end{align}
By definition, this effective theory will retain all top mass effects
in the kinematic distributions. Moreover, it ensures that the
mediator--gluon coupling is only generated through the top loop and
this way allows for a firm link of the LHC observables to the dark
matter annihilation process.\medskip

Finally, we can brute-force decouple the top quark as well as the
scalar mediator
\begin{align}
m_S \sim m_t > 2m_\chi \; .
\end{align}
If we decouple both of them in one step we find
\begin{align}
\lag_\text{eff,3} \supset 
 \frac{c_\chi^{g}}{\Lambda^3}(\bar\chi\chi) \; G_{\mu\nu}G^{\mu\nu}
+\frac{\tilde c_\chi^{g}}{\Lambda^3}(\bar\chi\gamma_5\chi) \; G_{\mu\nu}\tilde G^{\mu\nu} \; .
\label{eq:eff_lag3}
\end{align}
The effective operators are dimension-7, \ie further
suppressed. Matching at $\Lambda = m_S \sim m_t$ gives us
\begin{alignat}{9}
\frac{c_\chi^g}{\Lambda^3}
&=\frac{\alpha_s}{8\pi}\frac{c_S^t}{\Lambda^2}\frac{1}{m_t}\tau \left[ 1+(1-\tau ) f(\tau) \right]
\quad &&\stackrel{\tau\rightarrow \infty}{\longrightarrow}\quad 
\frac{\alpha_s}{12\pi}\frac{c_S^t}{\Lambda^2}\frac{1}{m_t}
&=&\frac{\alpha_s}{12\pi}\frac{g_S^tg_S^\chi}{m_S^2}\frac{1}{v} \notag \\
\frac{\tilde c_\chi^g}{\Lambda^3}
&=\frac{\alpha_s}{8\pi}\frac{c_P^t}{\Lambda^2}\frac{1}{m_t}\tau\,f(\tau)
\quad &&\stackrel{\tau\rightarrow \infty}{\longrightarrow}\quad 
\frac{\alpha_s}{8\pi}\frac{c_P^t}{\Lambda^2}\frac{1}{m_t}
&=&\frac{\alpha_s}{8\pi}\frac{g_P^tg_P^\chi}{m_S^2}\frac{1}{v} \; .
\end{alignat}
In principle, this model allows for additional particles
contributing to the effective mediator--gluon coupling. In our case we
fix the limit of the loop-function such that only the top quark runs
in the loop.\medskip

We show the mono-jet production rates for all three effective
Lagrangians in Fig.~\ref{fig:sjet1}. As expected, the decoupled top
ansatz in Eq.\eqref{eq:eff_lag1} reproduces the total rate of the
simplified model only if $m_S < 2 m_t$.  Above that threshold, it
overestimates the simplified model cross section, because the $\tau
\rightarrow \infty$ limit of the loop function is always larger than
its actual value. The effective Lagrangian with the decoupled
mediator, Eq.\eqref{eq:eff_lag2}, reproduces the dynamic model for
$m_S \gtrsim 5$~TeV. Above this value the LHC energy is not sufficient
to produce the mediator on-shell. Finally, the effective Lagrangian of
Eq.\eqref{eq:eff_lag3} with a simultaneously decoupled top and
mediator does not reproduce the total production rate of the
simplified model anywhere\footnote{An observed agreement of the full
  model and this effective theory would correspond to a precise
  measurement of the mediator mass.}. In the center and right panels
of Fig.~\ref{fig:sjet1}, we show the total cross section for heavier
dark matter candidates as well as the windows in parameter space for
which the annihilation cross section reproduces between one third and
$110\%$ of the measured relic density. We show scenarios where the
mediator only couples to up-type quarks (shaded red) and where the
mediator also couples to down-type quarks (shaded green). For $m_\chi = 200$ GeV, these regions almost completely overlap. \medskip
  
As a side remark, for very large mediator masses the
effective Lagrangian with a decoupled top, Eq.\eqref{eq:eff_lag1},
fails to precisely reproduce the simplified model prediction. The
reason is the mass dependence of the mediator width in the heavy top
approximation,
\begin{align}
\Gamma_S 
> \Gamma_{S\rightarrow gg} 
= \frac{2m_S^3}{\pi} \frac{\big(c_{S}^g  \big)^2}{\Lambda^2}
= \frac{2m_S^3}{\pi} \left( \frac{\alpha_s}{12\pi} \, \frac{g_S^t}{v} \right)^2  \; .
\end{align}
Clearly, the large mediator mass regime violates the original
assumption $m_t \gg m_S$.

\begin{figure}[t]
\vspace{-.3cm}
\includegraphics[width=1.0\textwidth]{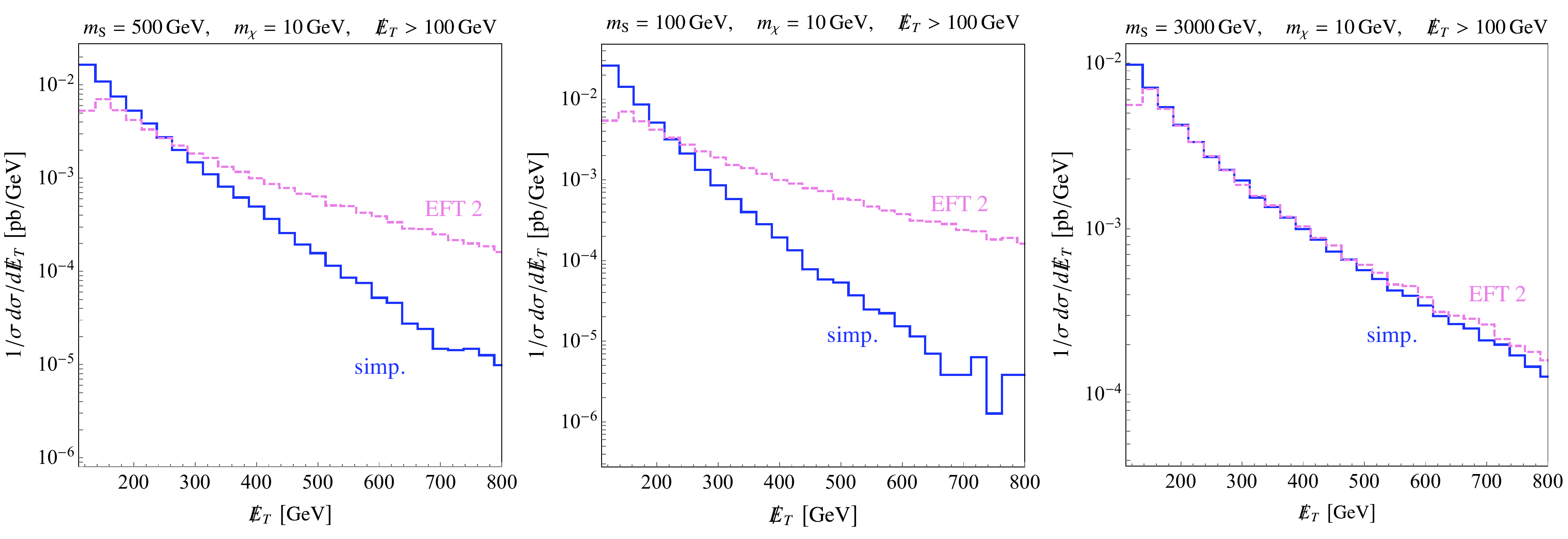}
\caption{$\met$ distributions in the loop-induced $s$-channel scalar
  mediator model.}
\label{fig:sjet3}
\end{figure}

\subsubsection*{Kinematic distributions}

In a second step, we consider the normalized $\met$ distributions.
Figure~\ref{fig:sjet3} shows this distribution for $m_\chi=10$ GeV
and three mediator
masses, $m_S=100,\ 500$ and $3000$~GeV, based on the
simplified model and the effective Lagrangian of
Eq.~\eqref{eq:eff_lag2} with a decoupled mediator.  The couplings are
chosen as $g_S^t=g_S^\chi=1$ in all cases, and we
require $\met>100$~GeV. Heavy mediators lead to harder $p_T$
spectra. Correspondingly, the EFT spectrum is harder than the
simplified model for light mediator masses, becoming indistinguishable
for large mediator masses. Since the dominant $Z_{\nu \nu}$+jets
background is softer than the signal, the signal over background ratio
for light mediators suffers from an aggressive $\met$ cut. For masses
$m_S > 2 m_t$, searches for a di-top resonance are a promising way to
identify an $s$-channel mediator with dominant couplings to top
quarks.

\subsubsection*{Effective Lagrangian vs model}

If a scalar $s$-channel mediator is predominantly coupled to
up-type quarks, the link between the LHC production rate and the
predicted relic density essentially vanishes. The two processes, and
direct detection limits, are only related if the mediator is very
light and hence decays through the one-loop diagram to a pair of
gluons. However, this is not the regime where an EFT description with
a decoupled mediator should be considered. In this most interesting
regime we should decouple the top quark in the loop and keep the
mediator and the dark matter agent as propagating degrees of
freedom. The features and eventually the failure of this effective
theory is well known from the Higgs sector~\cite{baur,matt_dorival}.

At the LHC, the assumption of a heavy mediator becomes numerically
accurate for $m_S > 5$~TeV, the same way as it does for $s$-channel
mediators produced in quark-antiquark scattering at tree level. This
is where an EFT approximation of the mono-jet rate and distributions
becomes accurate. The dark matter annihilation process in the early
universe is $\chi \bar{\chi} \to t \bar{t}$, and it can be large
around the pole $m_S \approx 2 m_\chi$. Because of this effective
$2\to 1$ annihilation topology, the mediator coupling or mediator
decay to Standard Model particles does not play a numerically relevant
role. A simultaneous decoupling of the top and the mediator leads to a
dimension-7 operators, which we find to be valid nowhere in model
parameter space for LHC physics or for dark matter
annihilation.\medskip

Independent of the EFT description we find that a loop-induced scalar
mediator is especially well-suited to explain the observed relic
density. The reason is that the $t$-channel annihilation $\chi
\bar{\chi} \to S S$ does not have the velocity suppression or the $2
\to 2$ annihilation process.

\clearpage
\section{Loop-mediated scalar in t-channel}
\label{sec:t_loop}
%
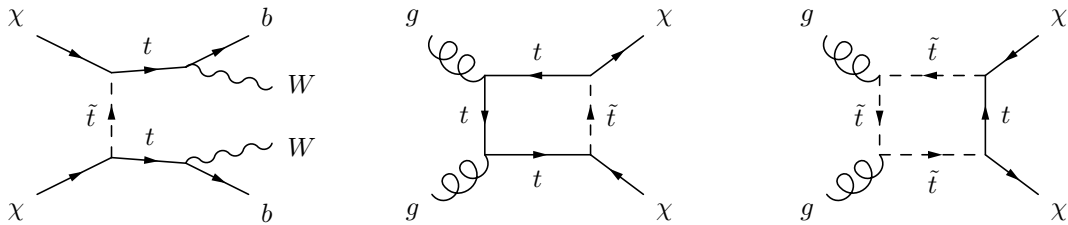
\begin{figure}[b!]
\vspace{.5cm}
\begin{fmffile}{feyn2}
\begin{fmfgraph*}(100,60) 
\fmfset{arrow_len}{2mm}
\fmfleft{i1,i2}
\fmfright{o1,o2,o3,o4}
\fmflabel{$\chi$}{i1}
\fmflabel{$\chi$}{i2}
\fmflabel{$b$}{o1}
\fmflabel{$W$}{o2}
\fmflabel{$b$}{o4}
\fmflabel{$W$}{o3}
\fmf{fermion,tension=2.0,width=0.6}{i1,v1}
\fmf{fermion,tension=2.0,width=0.6}{i2,v2}
\fmf{scalar,label=$\tilde{t}$,l.side=left,width=0.6}{v1,v2}
\fmf{fermion,label=$t$,l.side=left,tension=2.0,width=0.6}{v1,v3}
\fmf{fermion,label=$t$,l.side=left,tension=2.0,width=0.6}{v2,v4}
\fmf{fermion,width=0.6}{v3,o1}
\fmf{fermion,width=0.6}{v4,o4}
\fmf{photon,width=0.6}{v3,o2} 
\fmf{photon,width=0.6}{v4,o3} 
\end{fmfgraph*}
\hspace{1.5cm}
\begin{fmfgraph*}(100,60) 
\fmfset{arrow_len}{2mm}
\fmfleft{i1,i2}
\fmfright{o1,o2}
\fmflabel{$\chi$}{o1}
\fmflabel{$\chi$}{o2}
\fmflabel{$g$}{i1}
\fmflabel{$g$}{i2}
\fmf{gluon,tension=2.0,width=0.6}{i1,v1}
\fmf{gluon,tension=2.0,width=0.6}{i2,v2}
\fmf{fermion,tension=2.0,width=0.6}{v4,o2}
\fmf{fermion,tension=2.0,width=0.6}{o1,v3}
\fmf{fermion,label=$t$,l.side=right,width=0.6}{v2,v1}
\fmf{fermion,label=$t$,l.side=right,width=0.6}{v4,v2} 
\fmf{fermion,label=$t$,l.side=right,width=0.6}{v1,v3} 
\fmf{scalar,label=$\tilde{t}$,l.side=right,width=0.6}{v3,v4}
\end{fmfgraph*}
\hspace{1.5cm}
\begin{fmfgraph*}(100,60) 
\fmfset{arrow_len}{2mm}
\fmfleft{i1,i2}
\fmfright{o1,o2}
\fmflabel{$g$}{i1}
\fmflabel{$g$}{i2}
\fmflabel{$\chi$}{o1}
\fmflabel{$\chi$}{o2}
\fmf{curly,tension=2.0,width=0.6}{i1,v1}
\fmf{curly,tension=2.0,width=0.6}{i2,v2}
\fmf{scalar,label=$\tilde{t}$,l.side=right,width=0.6}{v4,v2,v1,v3} 
\fmf{fermion,label=$t$,l.side=right,width=0.6}{v3,v4}
\fmf{fermion,tension=2.0,width=0.6}{v3,o1}
\fmf{fermion,tension=2.0,width=0.6}{o2,v4}
\end{fmfgraph*}
\end{fmffile}\\[.5cm]
\caption{Feynman diagrams describing dark matter annihilation and LHC
  production (center and right) in the loop-induced scalar $t$-channel
  mediator model defined in Eq.\eqref{eq:t_model_1loop}.  Mono-jet
  production arises from attaching gluons to any of the colored
  particles.}
\label{fig:tchan-loop}
\end{figure} 

Finally, we consider a scalar $t$-channel mediator with couplings only
to the top quark. This combines some features discussed
in Sec.~\ref{sec:t_channel} with Sec.~\ref{sec:s_loop}.  Such a
scenario is for example realized by a light scalar top partner in
supersymmetry. The relevant interactions are
\begin{align}
t-\tilde{t}-\chi \qquad \text{dark matter annihilation}
\qqqquad 
\tilde{t}-\tilde{t}-g(-g) \qquad \text{QCD} \; ,
\label{eq:inttloop}
\end{align}
the first of which can lead to decay of $\chi \rightarrow \tilde t
t^{(*)}$.  The condition for stable dark matter, including multi-body
decays via an off-shell top and an off-shell $W$-boson, then reads
\begin{align}
m_\chi < \mst \; .
\end{align}
The second interaction in Eq.\eqref{eq:inttloop} occurs
through the QCD part of the covariant derivative. The mediator will
also have $Z$ and $\gamma$ interactions from the covariant derivative,
but we assume them to be sub-leading at the LHC.  In our simplified
model the dark matter interaction is described by 
\begin{align}
\lag \supset y_{\tilde t}\ \left( \bar{t}_R \chi \right) \tilde{t} + \text{h.c.}
\label{eq:t_model_1loop}
\end{align}
In contrast to the loop-induced $s$-channel scalar, couplings to a single quark
flavor do not lead to problems with flavor observables, so we
assume no interactions between dark matter and light
quark flavors. 

Dark matter annihilation in this last model proceeds through the
tree-level and one-loop diagrams shown in Fig.\ref{fig:tchan-loop}. In
the regime $m_\chi > m_t$ the annihilation rate is dominated by the
tree-level process $\chi \bar \chi \rightarrow t \bar t$.  For $m_\chi
< m_t$ annihilation can proceed through off-shell tops with up to a
$2\rightarrow 6$ topology, or the loop-mediated process $\chi \bar
\chi \rightarrow g g$.  In Fig.~\ref{fig:relic_tloop}, we present the
regions of parameter space for which this mediator can lead to the
correct relic density for mediator and dark matter masses above the
top threshold. We find that there is a sizeable parameter range for
sub-TeV dark matter and mediators where we recover the observed relic
density from annihilation into top quarks.

\begin{figure}[t]
\includegraphics[width=0.35\textwidth]{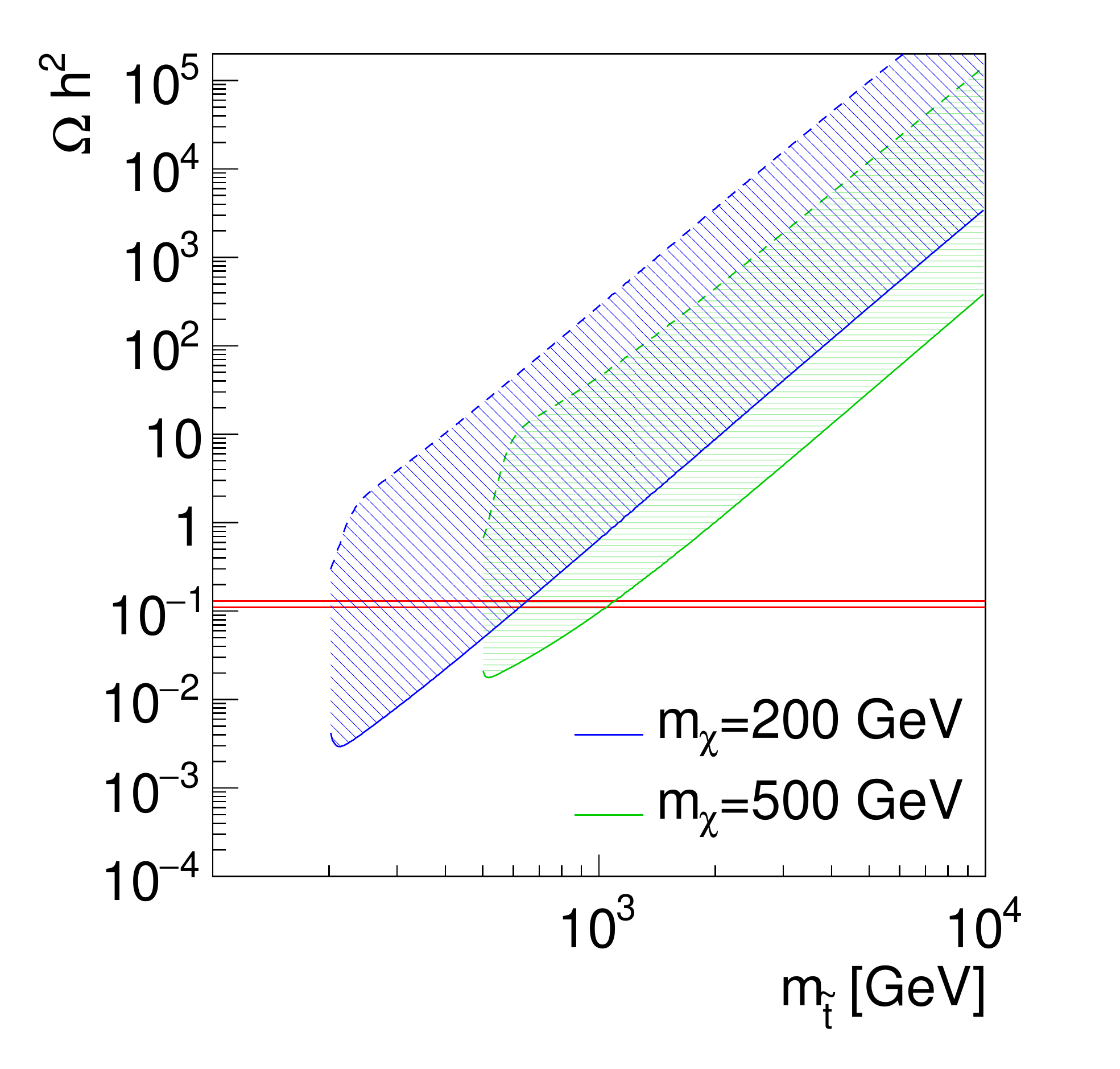}
\includegraphics[width=0.35\textwidth]{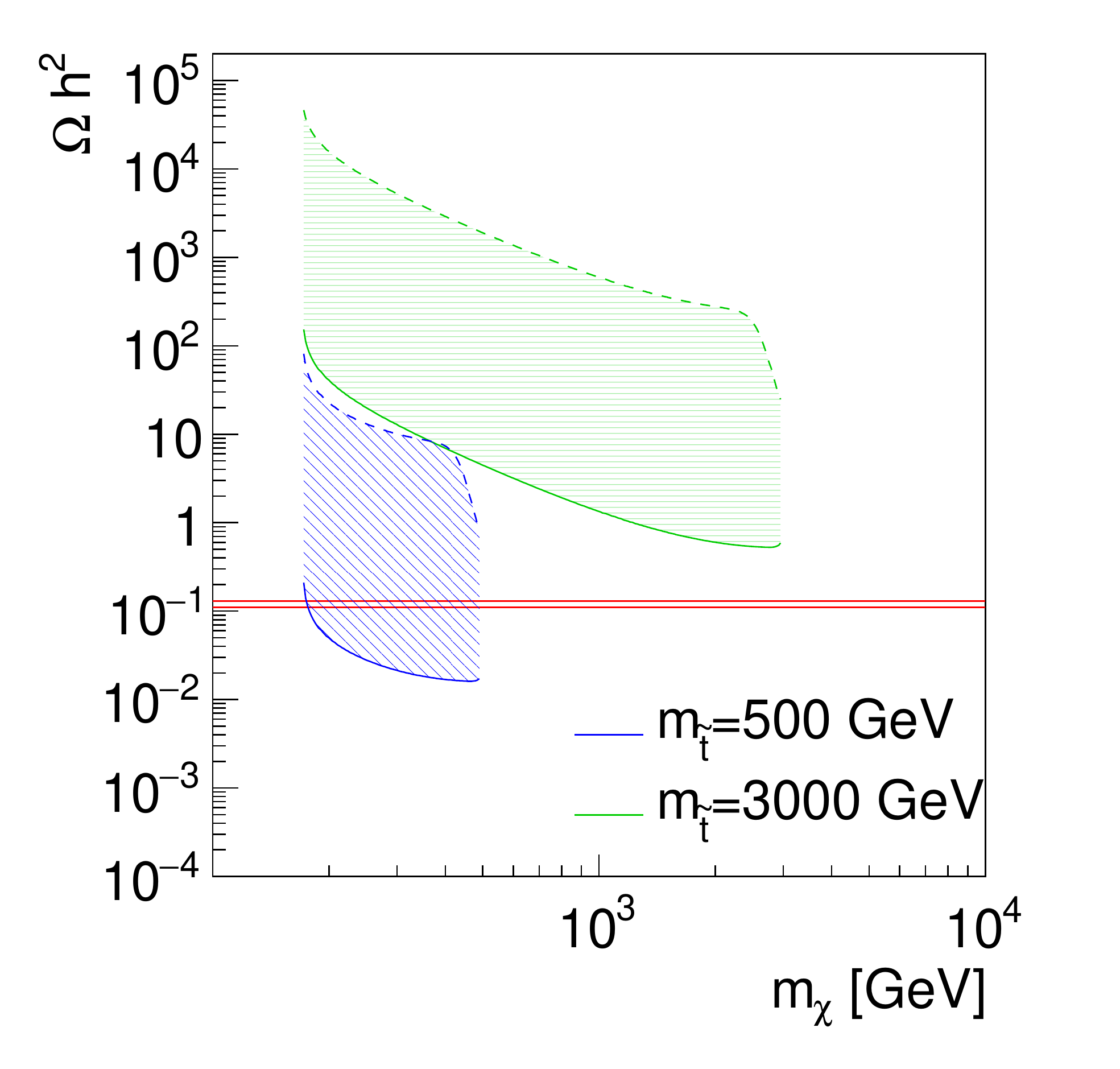}
\caption{Relic density for the loop-induced $t$-channel scalar
  mediator model as a function of the mediator mass for constant dark
  matter mass (left), as a function of the dark matter mass for
  constant mediator mass (center) and as a function of the dark matter
  mass for a constant ratio of mediator to dark matter mass. Over the
  shaded bands we vary the couplings $g=0.2\ldots 1$; large relic
  densities correspond to small coupling.}
\label{fig:relic_tloop}
\end{figure}

\subsubsection*{Total rate}

\begin{figure}[b!]
\begin{center}
\includegraphics[scale=.45]{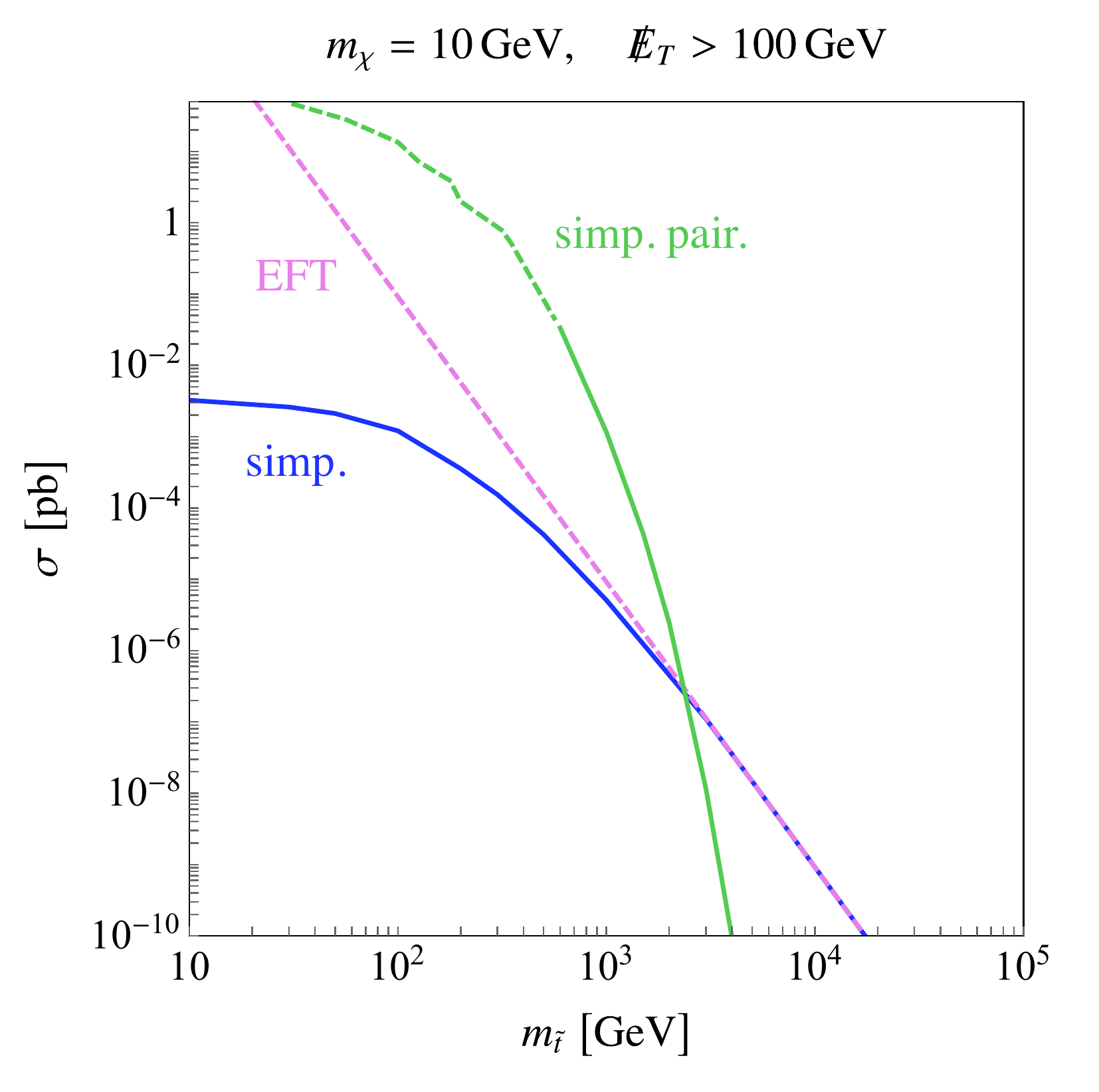}
\caption{Cross sections for the loop-mediated $t$-channel scalar model
  and its effective Lagrangian approximation for the mono-jet signals,
  and the pair production cross section with
  subsequent decay into $X + \slashed{E}_T$.}
\label{fig:tjet1}
\end{center}
\end{figure}

As above, we implement the simplified model with
\textsc{FeynRules}~\cite{feynrules} and produce mono-jet events with
\textsc{MadGraph5}~\cite{madgraph}.  At the one-loop level we evaluate
around 150 Feynman diagrams, because many internal and external lines
in Fig.~\ref{fig:tchan-loop} can radiate a gluon.  The total cross
section $\sigma_{\met + j}$ is shown in Fig.~\ref{fig:tjet1} for
mediator masses of $\mst = 10~...~10^5$~GeV, a dark matter mass of
$m_\chi =10$~GeV, and a cut $\met > 100$~GeV.
The effective field theory for heavy mediators $\mst \gg m_\chi (\sim
m_t)$ includes the dimension-6 four-fermion operator,
\begin{align}
\lag_\text{eff} \supset \frac{c_{\tilde{t}\chi}}{\Lambda^2}  \; 
                       \left(\bar{t}_R \chi \right) \; 
                       \left( \bar{\chi} t_R\right) \; ,
\label{eq:t_eft_1loop}
\end{align}
where $\Lambda=m_{\tilde t}$ and $c_{\tilde{t}\chi}=1$ after
matching to Eq.\eqref{eq:t_model_1loop}.\medskip

The mono-jet process for this specific model is mediated at one loop
in both the simplified model and the EFT.  Similar to tree-level
$t$-channel mediator, we again show the pair production rate for the
$t$-channel mediator, with a subsequent decay $\tilde{t} \tilde{t}^*
\to t\bar t +\met$. It occurs at tree level and cannot be described by
the same effective theory.  As a consequence, pair production
dominates the LHC signal up to very large mediator masses $\mst >
3$~TeV. Here, the mono-jet signal becomes competitive because of the
$1/\mst^8$ suppression of the pair production rate.

This defines the mass scale where the EFT in principle becomes
trustworthy, but where all LHC signals are tiny.  It is apparent that
the mono-jet cross section in the EFT always overestimates the
cross-section predicted by the simplified model, in contrast to the
tree-level $t$-channel mediator discussed in
Sec.~\ref{sec:t_channel}. The reason for this is the absence of an
$s$-channel-enhanced diagram like the one in the center of
Fig.~\ref{fig:feyn_tchannel}.

When the mediator is below the threshold to allow on-shell $t+\chi$ decays,
three body decays into $b+W+\chi$ are considered and similarly four body
decays below the W threshold. This region is shown as the
green-dashed line in Fig.~\ref{fig:tjet1}. The requirement of large missing energy
reduces the efficiency resulting in a plateau in the $\chi \bar \chi + X$
cross section as mediator mass becomes small even if the pair-production
cross section is large. ATLAS and CMS searches currently place upper limits
on $\mst$ in the range of 350..800 GeV depending on $m_\chi$ and $\mst - m_\chi$.
 
%

\subsubsection*{Effective Lagrangian vs model}

If we couple a scalar $t$-channel mediator only to the top quark, we
avoid the associated production which dominates the LHC signal for the
usual scalar $t$-channel mediator. Flavor constraints do not force us
to also include Yukawa couplings to light quarks.  Dark matter
annihilation then proceeds to a variety of channels, including
tree-level off-shell processes and loop-induced $2 \to 2$
topologies. The dominant annihilation mainly depends on the dark
matter mass and only slightly on the mediator mass.

In contrast, the decoupling pattern of the LHC rates is largely
described by an increasing mediator mass.  For relatively light
mediators LHC production is dominated by mediator pair production,
which is not linked to the leading four-fermion operator as the
mono-jet signature or most annihilation channels. On the other hand,
in the absence of an associated production channel the effective
theory replicates the cross section of the simplified model already
for $\mst > 3$~TeV. This regime is nevertheless not the most
interesting, because the predicted LHC rates are very small.

The mono-jet cross section for a $t$-channel mediator with couplings
to top quarks has the unique property that the EFT always
overestimates the prediction in the simplified model due to the
absence of any $s$-channel contribution with resonant enhancement.
It reproduces the behavior of the sub-leading $t$-channel diagrams shown
to the left in Fig.~\ref{fig:feyn_tchannel} for the 
tree-level $t$-channel mediator.

\clearpage
\section{Summary}
\label{sec:summary}

We have studied the performance of an effective field theory of dark
matter, focusing on the link between the mono-jet signature at the
LHC and the observed relic density. While a link between direct
detection and LHC physics is easier to establish~\cite{tim}, the relic
density is the most constraining ingredient of any global dark matter
fit; a correct description of the main features in the relevant
parameter space is crucial for global fits of dark matter in this
framework. Our analysis is based on four simple models: a $t$-channel
(darkoquark) mediator; an $s$-channel vector mediator with tree-level
couplings to a dark matter fermion and the Standard Model; and an
$s$-channel or $t$-channel scalar mediator, coupled to the Standard
Model at one loop.\medskip

A $t$-channel scalar mediator is problematic independent of its
effective Lagrangian. It is hard to explain the relic density, avoid
LHC constraints, and predict measurable LHC signals at the same
time. Because LHC prefer heavy mediators, the EFT approximation works
fine, including intermediate on-shell states, but it fails to describe
appreciable rates at the LHC. An interesting decoupling aspect that
the dominant LHC process switches from resonant mediator pair
production to single-resonant associated production, to dark matter
continuum production with initial state radiation. Note that for the
LHC analysis the dark matter mass plays hardly any rule, except for
degenerate dark matter and mediator scenarios which cannot be linked
to the relic density.

The $s$-channel vector mediator is interesting because of its
resonance structure in non-relativistic annihilation and at the
LHC. From a global fit perspective, the factorization into $2 \to 1$
production and decay matrix elements hurts the parameter analysis
hard. By definition, the EFT description works in a different region
of phase space. In this case the partonic LHC energy plays a key role;
mediator production is described by a propagating mediator up to $m_V
\approx 5$~TeV, above which the model merges into an EFT.

The loop-mediated $s$-channel scalar mediator can be matched it to
three effective Lagrangians. Obviously, we need to choose
the correct decoupling pattern for valid LHC predictions: a
decoupled top quark in the loop-induced gluon-mediator coupling is
essentially identical to the low-energy description of Higgs
production at the LHC, and it is well known that it should not be
applied to hard mono-jet production. A decoupled mediator scenario
works well at the LHC, provided the mediator mass exceeds
5~TeV. Decoupling the mediator and the top at the same time does not
yield a valid effective theory for the LHC.

For the loop-mediated $t$-channel scalar mediator there is no
associated production of the mediator with dark matter. This means
that the appropriate EFT provides a good description already for
mediator masses above 3~TeV. In contrast to the tree-level $t$-channel
and the $s$-channel models, the EFT always overestimates the
suppressed mono-jet rate predicted by the full model. This means that
the EFT again approximates the full model only in parameter regions
where the expected number of events is very small.

A side result of our study is that the relevant energy scales for an
EFT description at the LHC are the mediator mass and the partonic LHC
energy. The exact mass of the dark matter (away from a degenerate
spectrum $\mmed \approx m_\chi$) or the mediator width do not play a
noticeable role.\medskip

Based on our EFT comparison with four typical models we find that a
correct description of LHC observables like the total mono-jet rate
and the relevant range of the missing transverse momentum spectrum is
not actually the main challenge to the EFT framework. From a global
fit perspective it is crucial to link the predicted relic density to
LHC observables. Unlike the link between mono-jet production and
direct detection, this link between mono-jet production and the relic
density is not established in our models and does not suggest to
perform global fits to an effective theory of dark matter.\bigskip

\begin{center} \textit{Acknowledgments}\end{center}

We are very grateful to the MITP program \textsl{Understanding the
  First Results from LHC Run~II} for triggering the questions leading
to this paper and for their outstanding hospitality in a great town.
A.~B. is funded by the \textsl{Heidelberg Graduate School for
  Fundamental Interactions}.  All authors acknowledge support from the
German Research Foundation (DFG) through the Forschergruppe
\textsl{New Physics at the Large Hadron Collider} (FOR 2239).  ND also
acknowledges partial support from the OCEVU Labex (ANR-11-LABX-0060)
and the A*MIDEX project (ANR-11-IDEX-0001-02) funded by the
\textsl{Investissements d'Avenir} program.

\clearpage
\appendix
\section{Detector effects}

Throughout the main body of the paper we show parton level results,
\ie one hard jet recoils against the missing energy, $\met =
p_{T,j}$. We keep this choice to be able to illustrate all the
sub-channel contributions, their different behavior, and to allow for
a comparison between the different models. In this appendix we show
the robustness of these results when we include a proper event
simulation.  Aside from the implementation of the models in
\textsc{FeynRules}~\cite{feynrules} and the generation of events with
\textsc{Madgraph}\cite{madgraph}, we use \textsc{Pythia}~\cite{pythia}
for parton shower and hadronization, and the default ATLAS setup in 
\textsc{Delphes}~\cite{delphes} for the detector simulation. This
includes \textsc{FastJet}~\cite{fastjet} for jet clustering.  For the
event production we generate events including up to two hard jets,
plus soft and collinear jet radiation. As for the parton-level results
discussed before, we require at least one central hard jet,
$|\eta_j|<2.5$, consistent with the usual mono-jet searches. While
after the proper simulation $\met$ and $p_{T,j}$ are still highly
correlated, they are no longer strictly equal.  Therefore, we apply a
cut on $\met$ combined with $p_{T,j}^{\text{lead}}>50$~GeV for the
leading jet. This way, the reconstruction corresponds to the usual LHC
analyses. Finally, motivated by the mono-jet searches, we veto events
with more than two jets with $|\eta_j|<4.5$ and $p_{T,j}>30$~GeV.\medskip

\begin{figure}[b!]
  \includegraphics[width=0.33\textwidth]{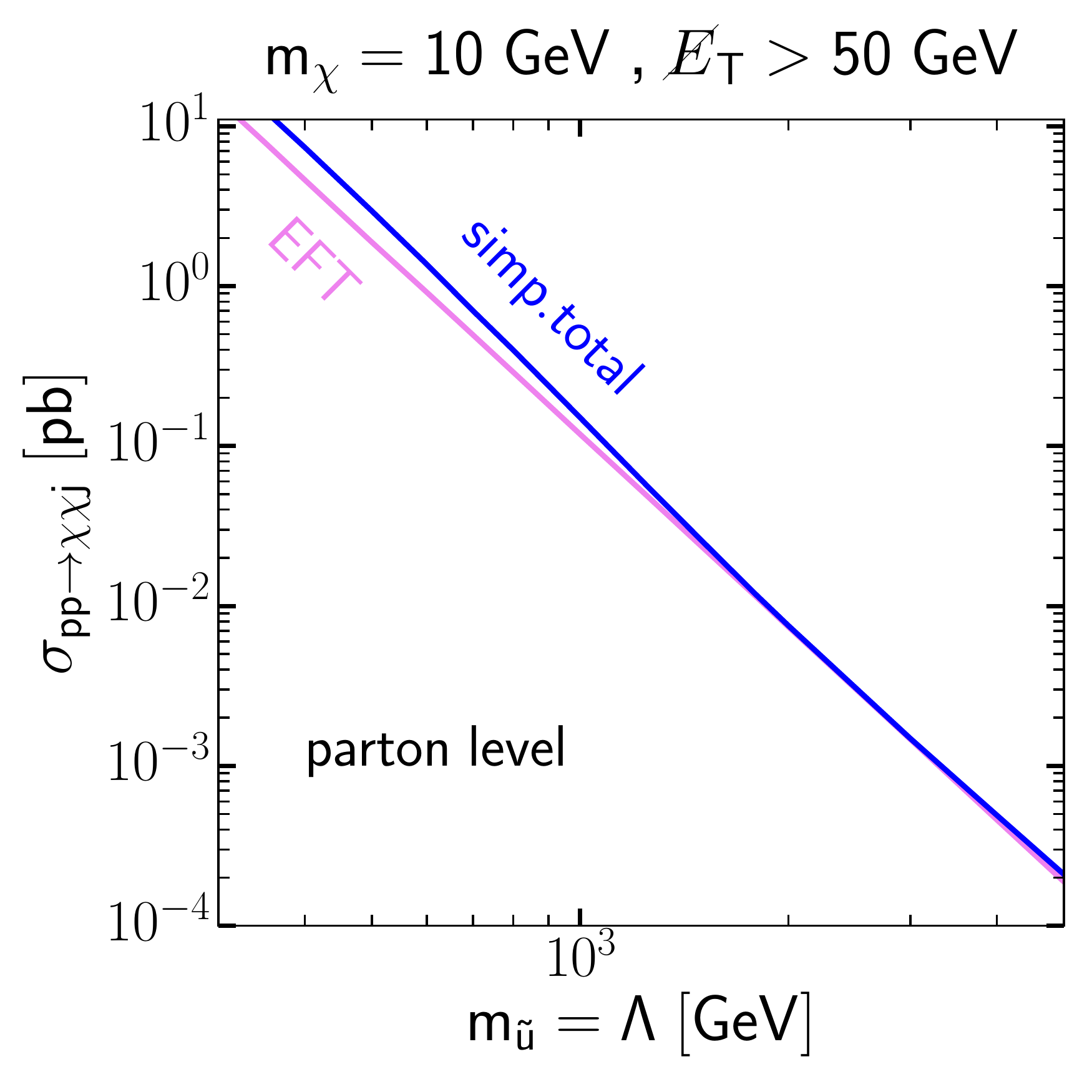}\hspace*{-0.2cm}
  \includegraphics[width=0.33\textwidth]{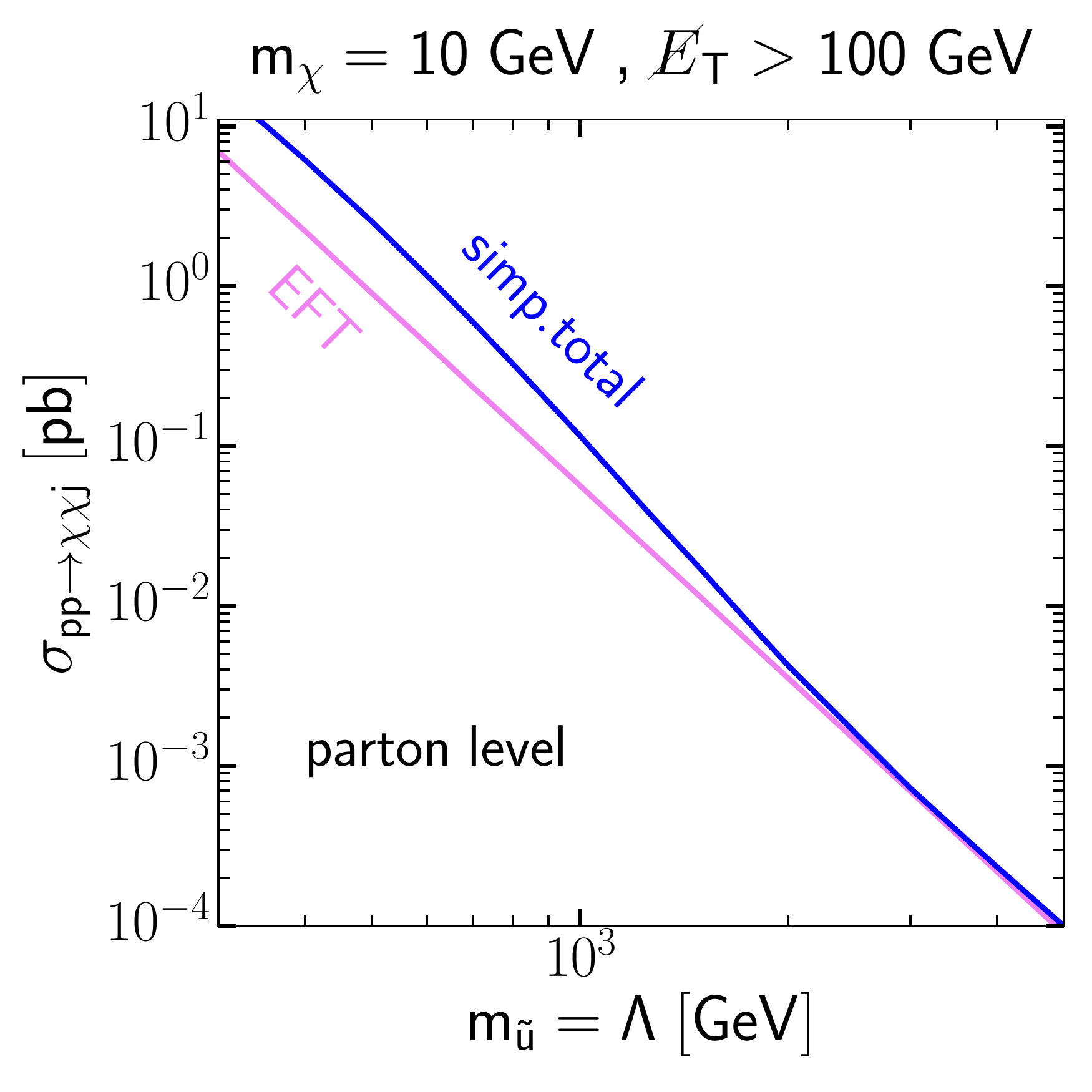}\hspace*{-0.2cm}
  \includegraphics[width=0.33\textwidth]{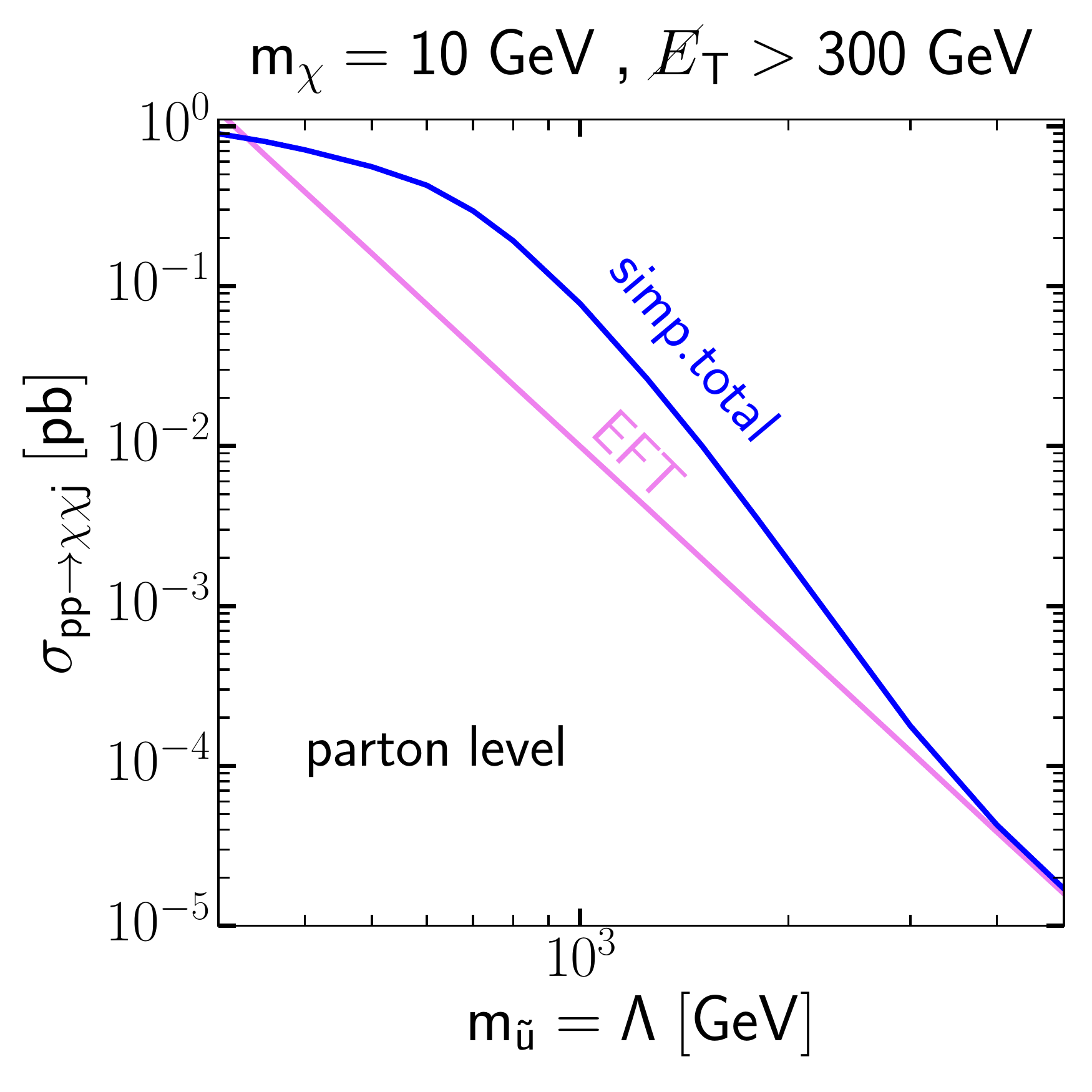} \\
  \includegraphics[width=0.33\textwidth]{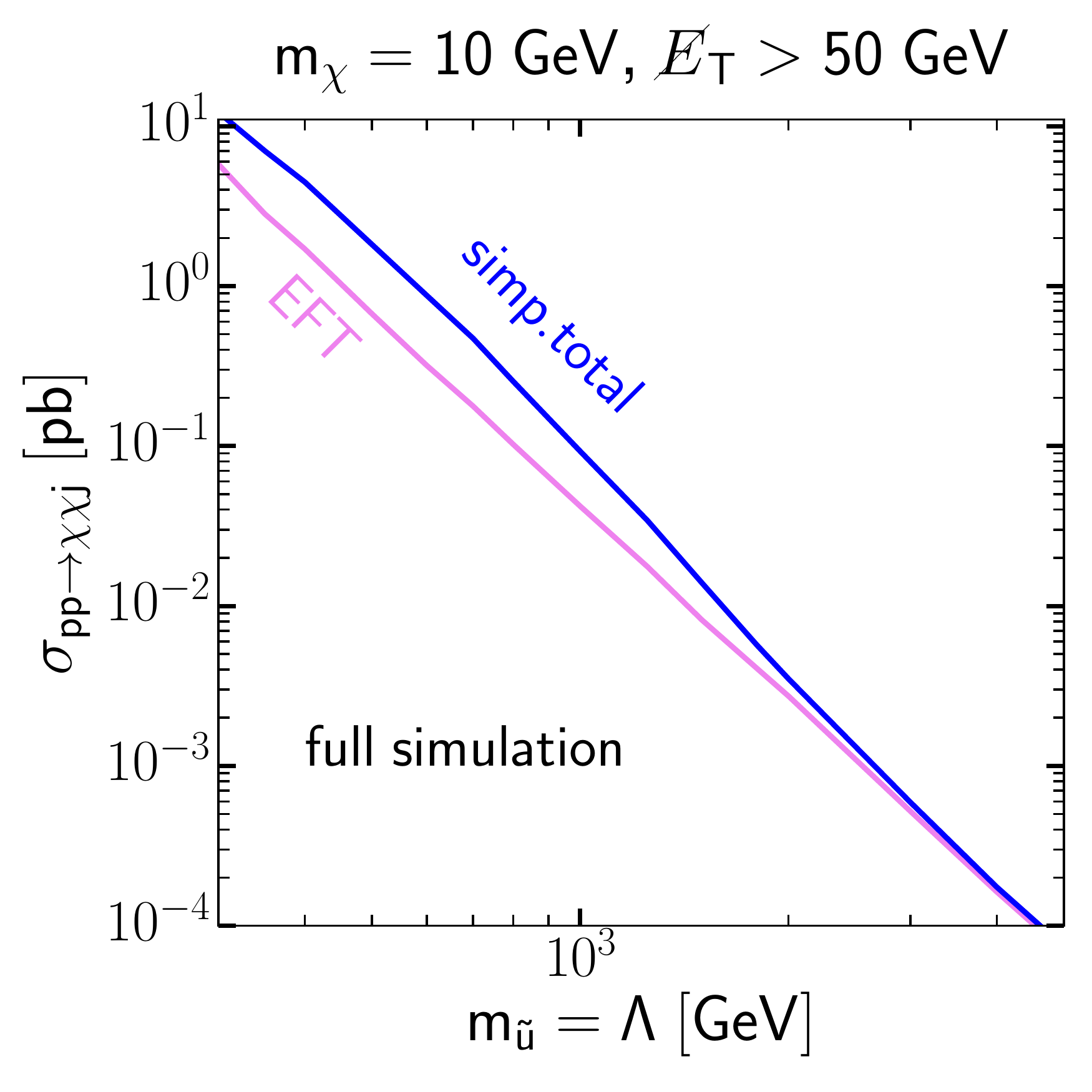}\hspace*{-0.2cm}
  \includegraphics[width=0.33\textwidth]{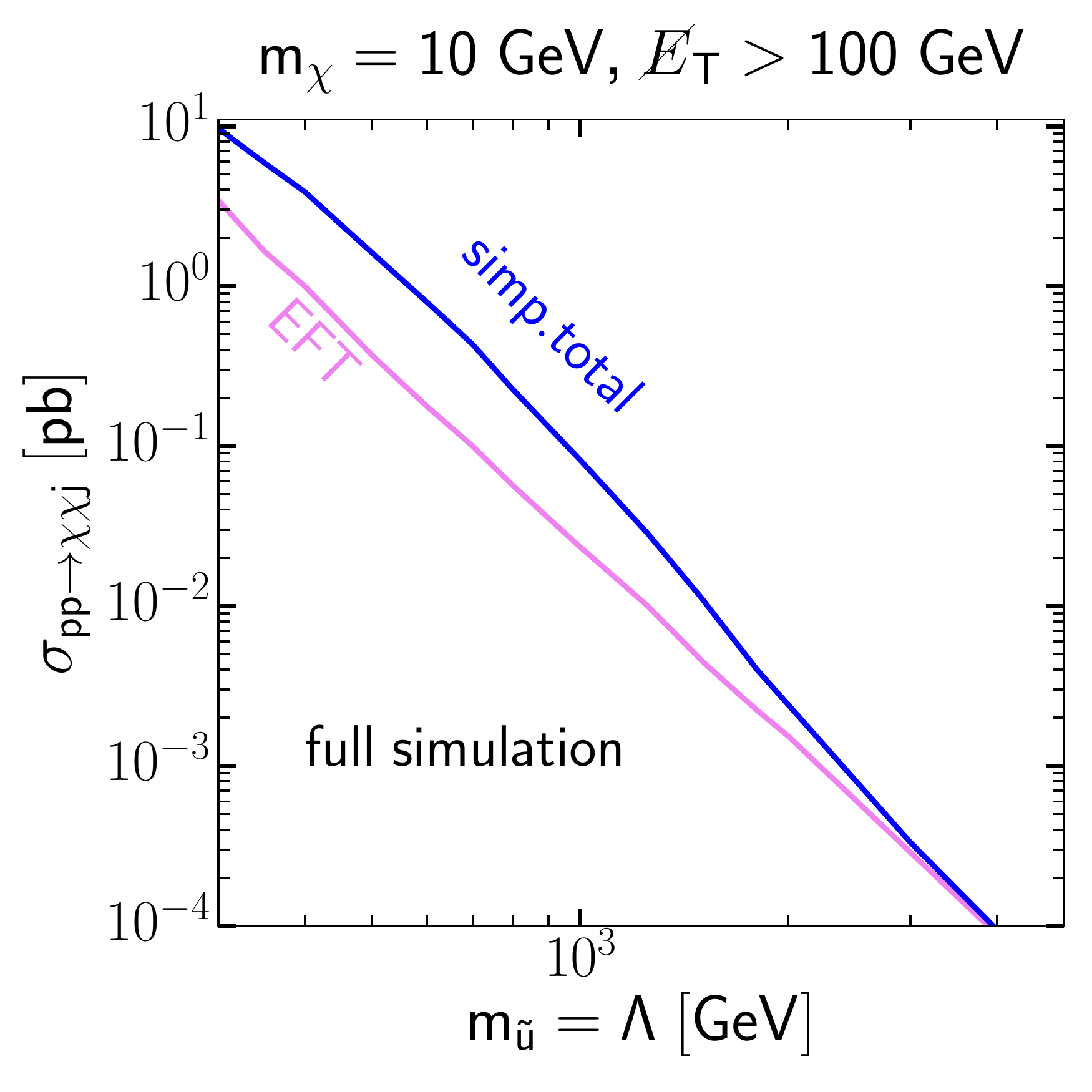}\hspace*{-0.2cm}
  \includegraphics[width=0.33\textwidth]{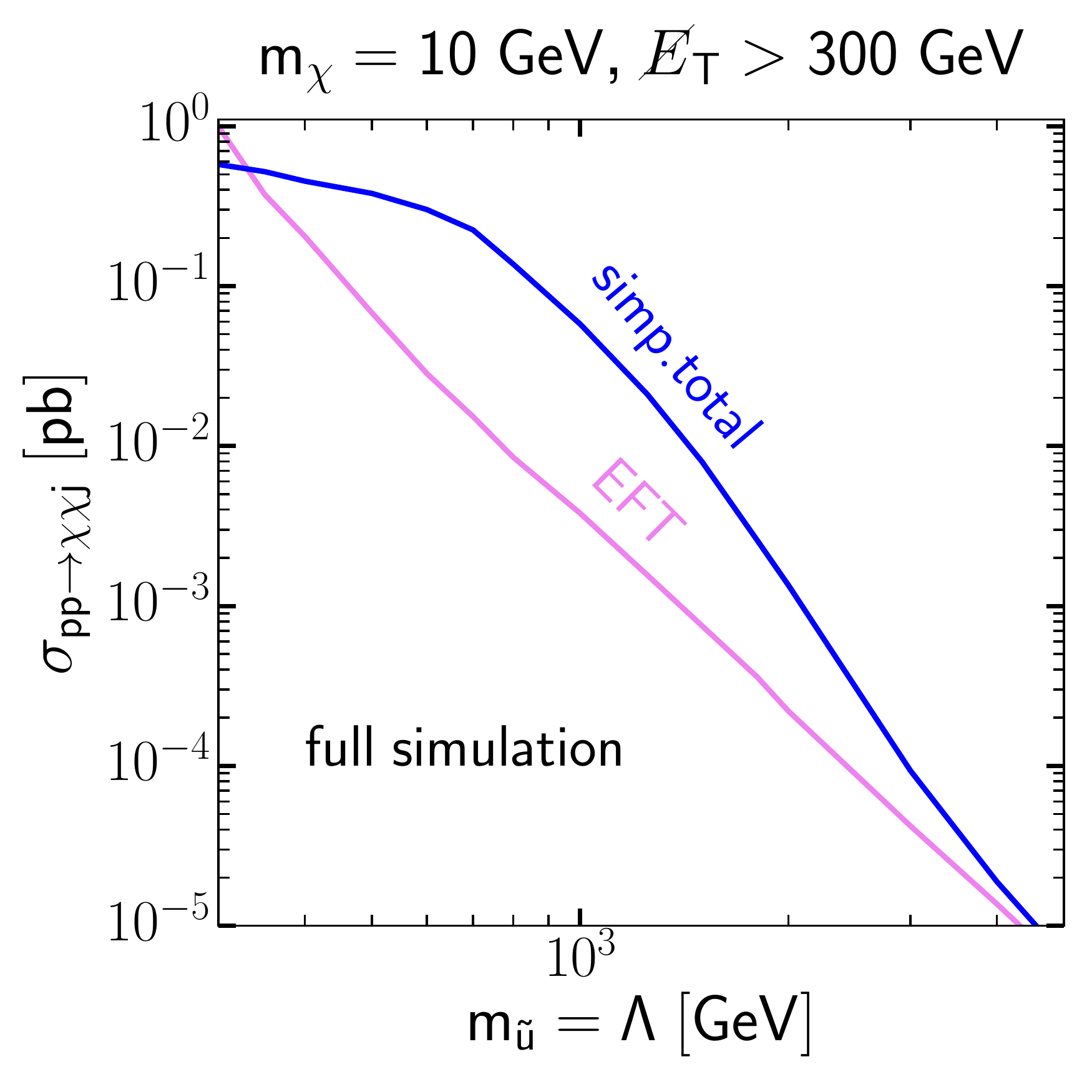} 
\caption{Total production rate in the $t$-channel model as a function
  of the mediator mass. In the upper panels we show again the results
  obtained from the parton level simulations. The lower plots include
  the full simulation procedure chain.}
\label{fig:t_cross_full}
\end{figure}

For illustration purposes it is sufficient to focus on the tree-level
$t$-channel mediator, described in Sec.~\ref{sec:t_channel}. We again
set $y_u=1$, or $c_{u\chi}=1$ for the EFT approximation. Following the
structure of the paper, we start showing total cross sections after
the above introduced selection cuts. Given
the mild dependence of the cross sections on the dark matter mass, we
use $m_\chi=10$~GeV.

As can be observed in Fig.~\ref{fig:t_cross_full}, all features we
describe in Sec.~\ref{sec:t_channel} remain after the proper
simulation. Specifically, the decoupling pattern for the simplified
model towards the effective Lagrangian is identical. For the smallest
and least realistic $\met$ cut, the agreement between the simplified
model cross section and the effective Lagrangian approximation is
already accomplished for mediator masses around 1~TeV. A larger cut on
$\met$ delays the decoupling limit to larger mediator masses,
rendering the effective Lagrangian approximation valid for multi-TeV
mediators. The only differences appearing in the proper simulation are
in the precise cross section values. They are reduced compared to
parton level; this is caused by the more sophisticated treatment of
the detection and reconstruction of the hard jet, as well as of the
missing energy.  We note here that the precise values for the
predicted cross sections will depend on the final requirements imposed
on the reconstructed jets, as well as on higher-order corrections.\medskip 

\begin{figure}[b!]
  \includegraphics[width=0.33\textwidth]{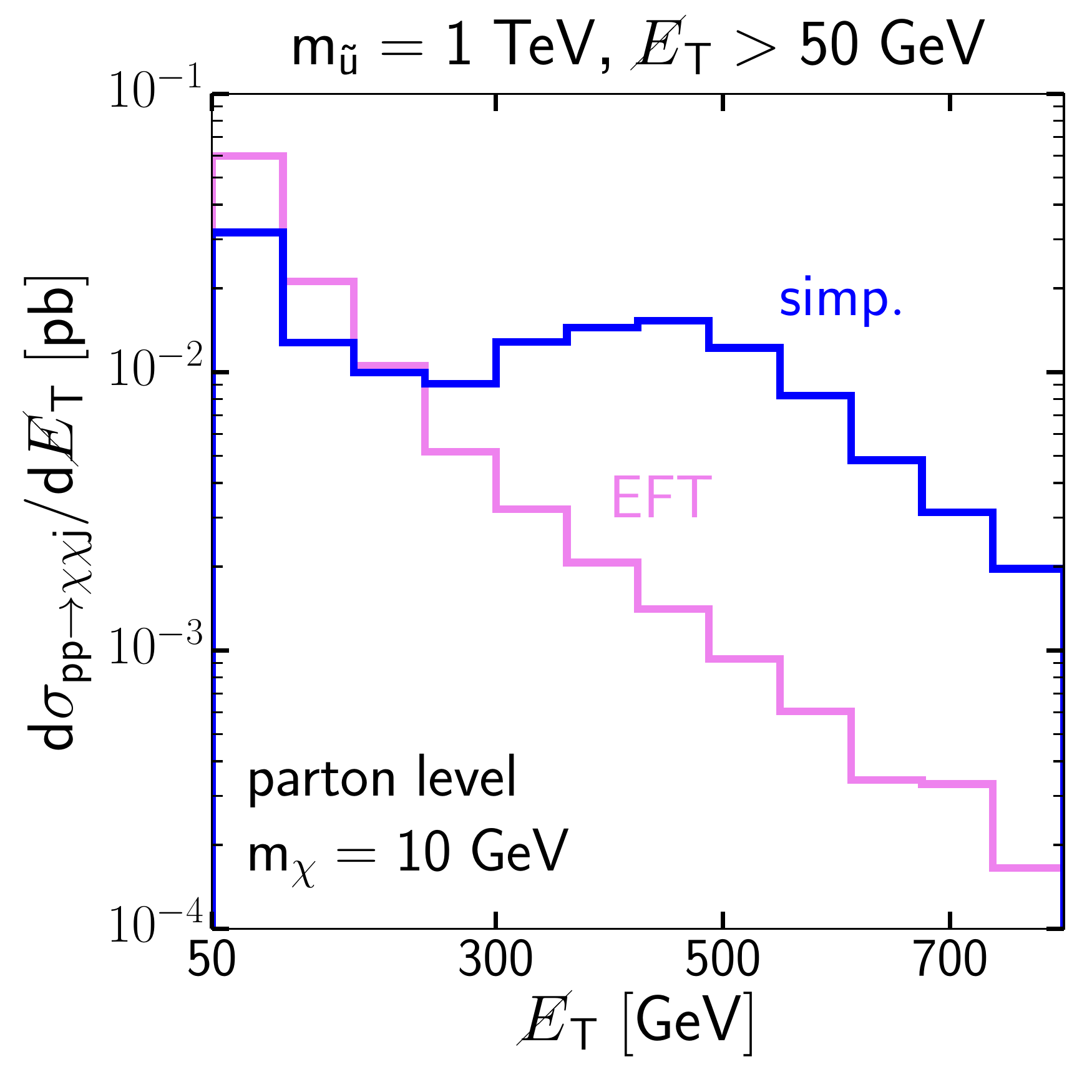}\hspace*{-0.2cm}
  \includegraphics[width=0.33\textwidth]{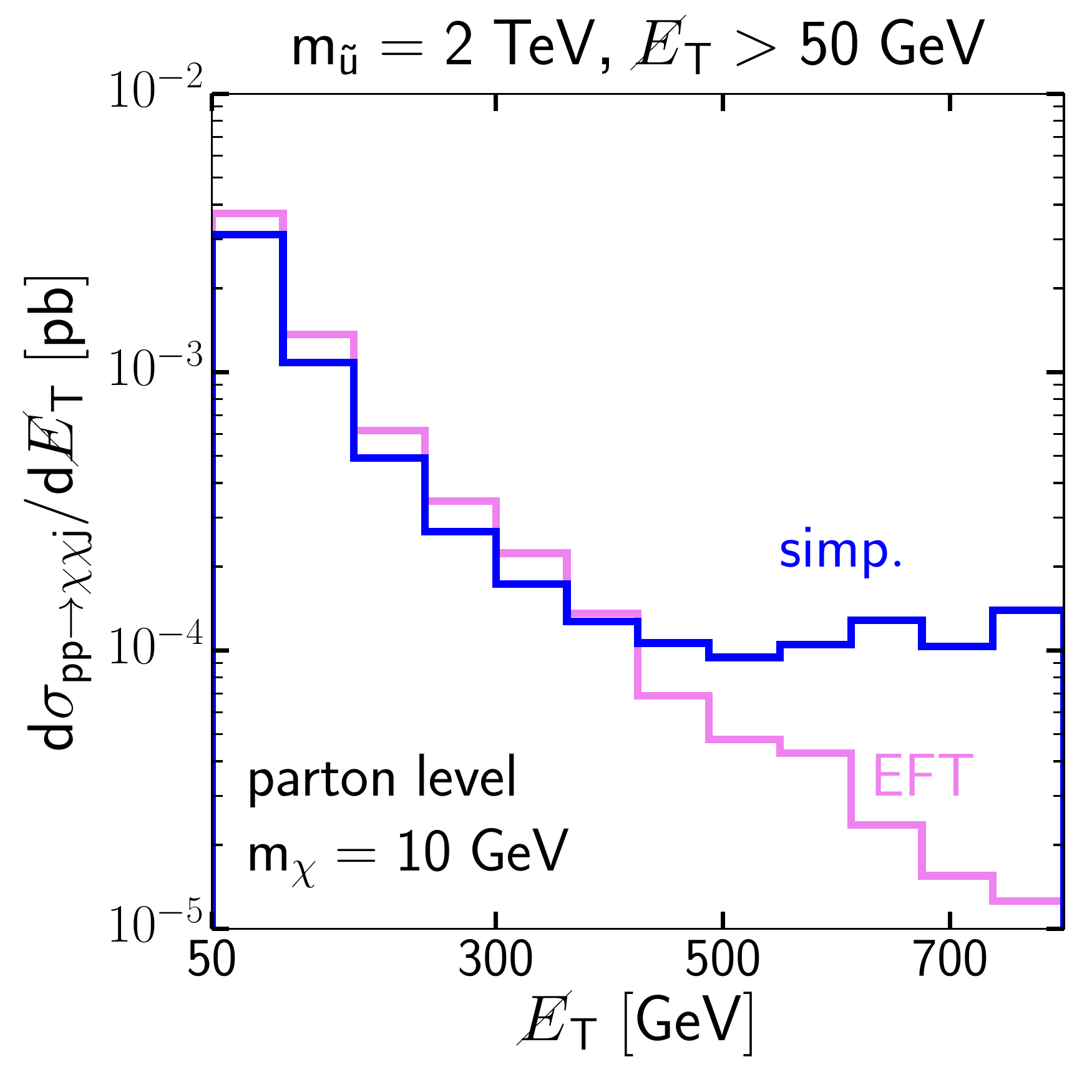}\hspace*{-0.12cm}
  \includegraphics[width=0.33\textwidth]{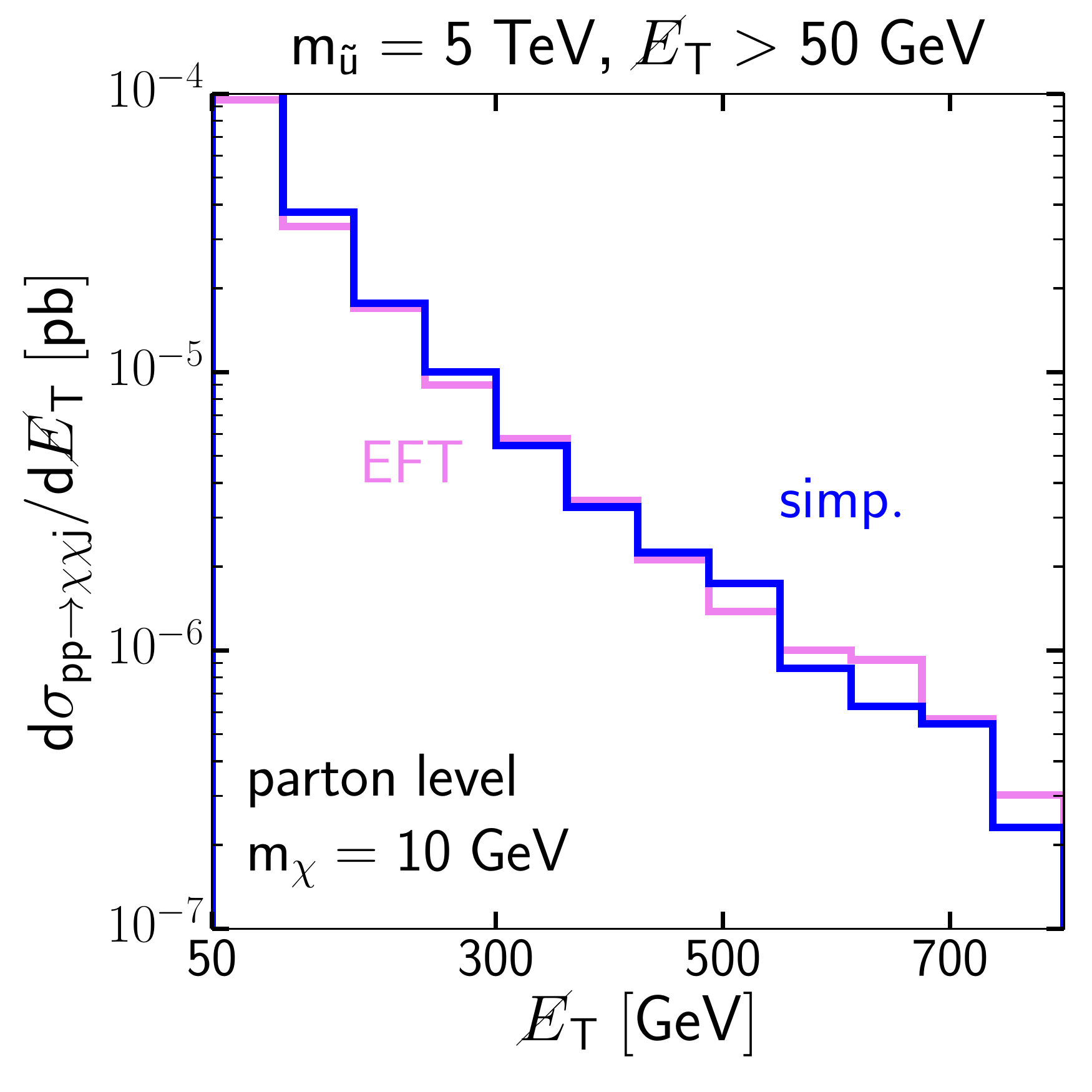} \\
  \includegraphics[width=0.33\textwidth]{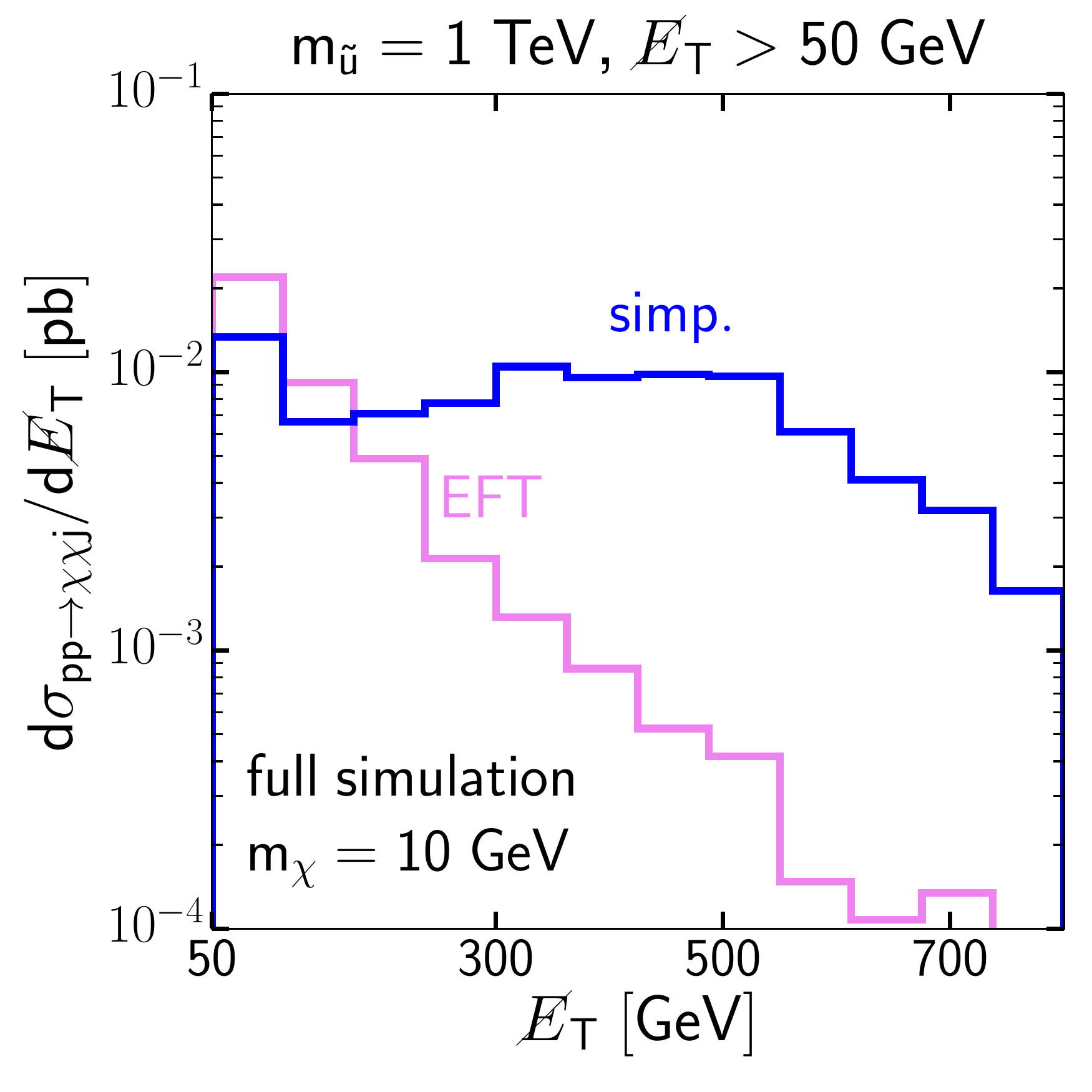}\hspace*{-0.2cm}
  \includegraphics[width=0.33\textwidth]{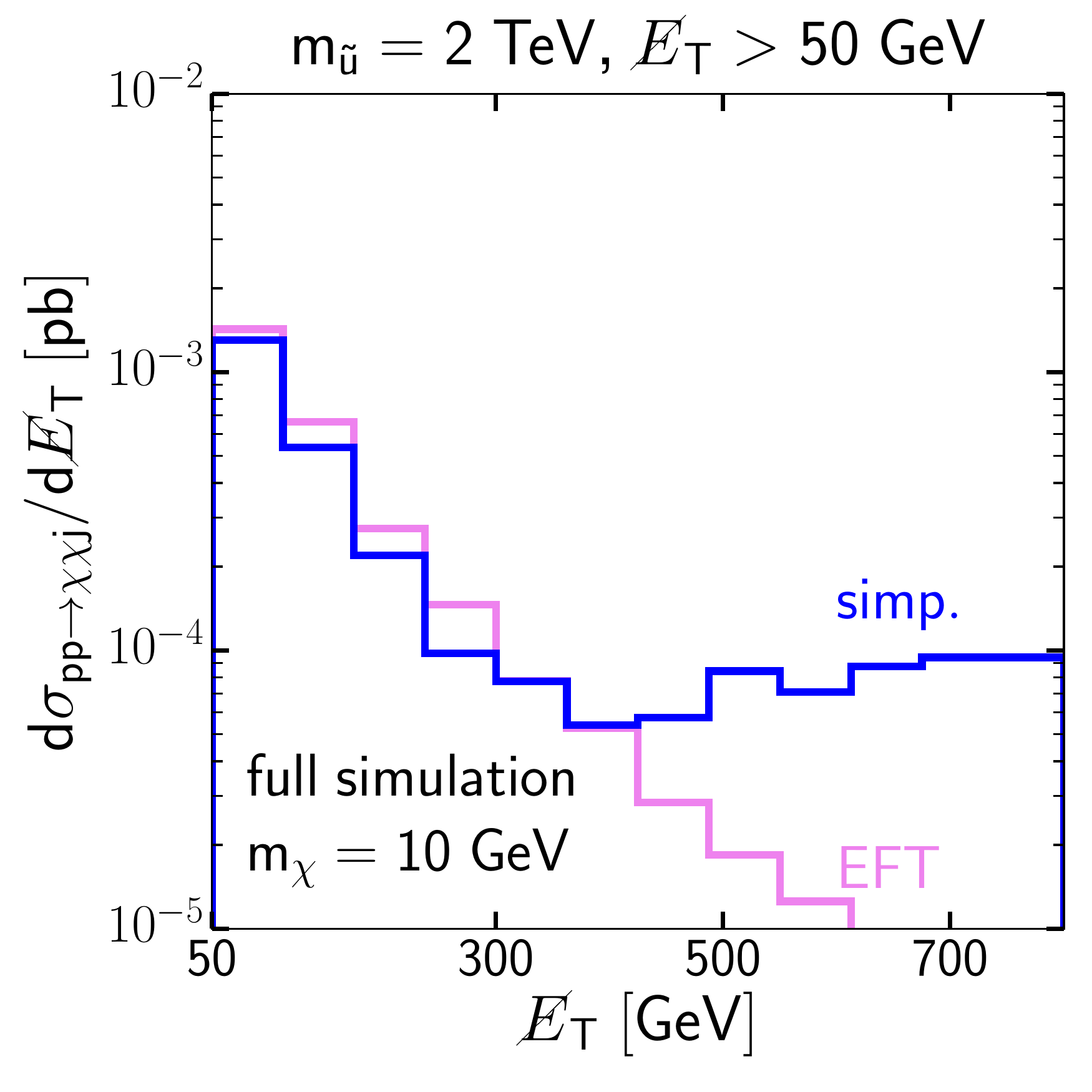}\hspace*{-0.12cm}
  \includegraphics[width=0.33\textwidth]{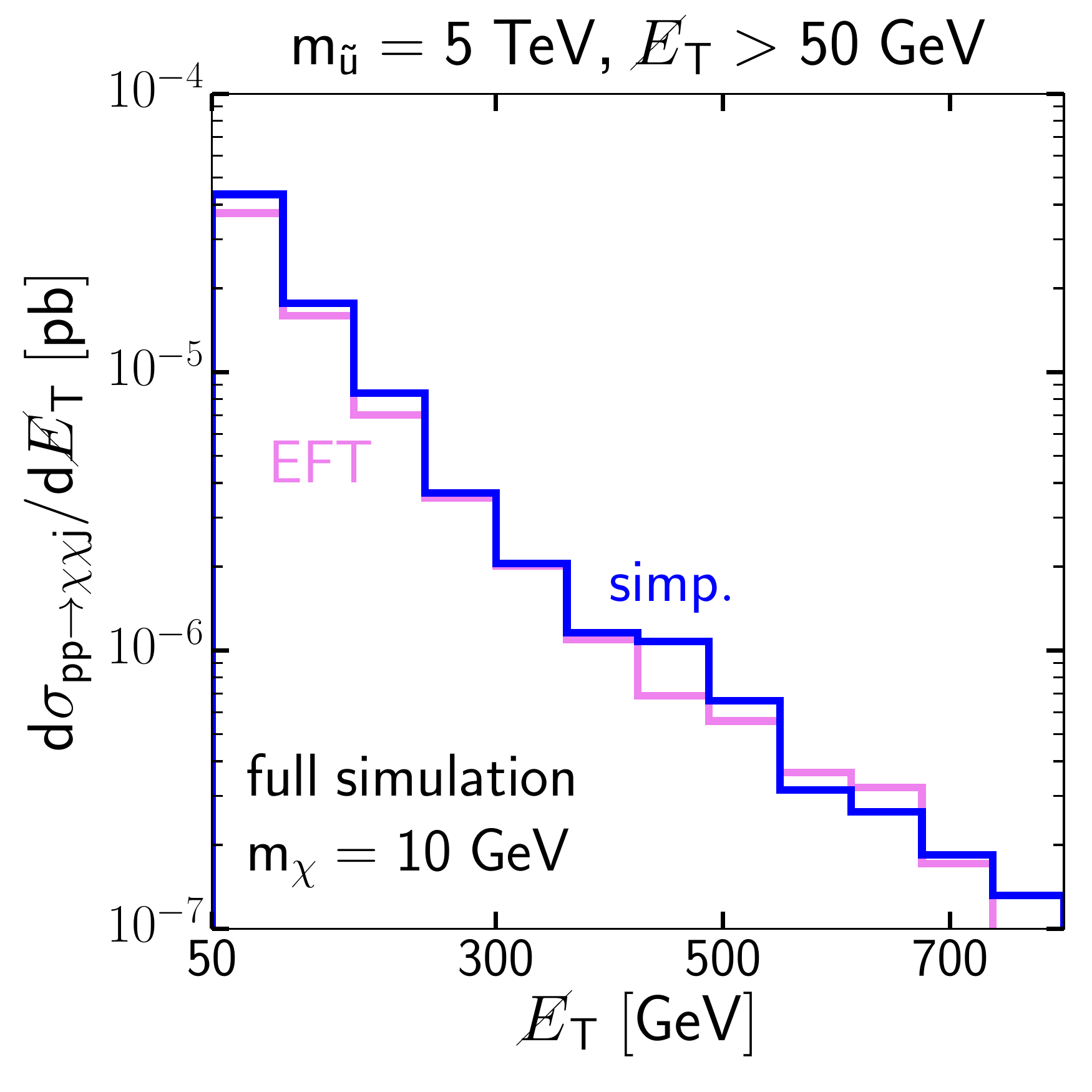}
\caption{$\met$ distributions in the $t$-channel
    model. In the upper panels we show the results obtained from
  the parton level simulations. The lower plots depict the distributions after the full
  simulation procedure.}
\label{fig:t_ptj_full}
\end{figure}

Finally, even though Fig.~\ref{fig:t_cross_full} already points to
unaltered shapes of the $\met$ distributions, we show a selection of
such distributions in Fig.~\ref{fig:t_ptj_full}. We fix the missing
transverse momentum cut to $\met > 50$~GeV, and we show the
distributions for both the simplified model and the effective
Lagrangian approximation for $m_\chi=10$~GeV, and mediator masses
$\msu=1,\ 2$ and 5~TeV.

We see exactly the same patterns as in Sec.~\ref{sec:t_channel}.  For
the lightest mediator masses considered, $\msu=1$~TeV, the
single-resonant diagrams dominate the simplified model cross sections,
The $\met$ distribution in the simplified model is very different from
the EFT distribution. The latter has a similar shape as the pure
$t$-channel contributions.  For larger mediator masses the impact of
the single-resonant channel is reduced.  The deviations of the
effective Lagrangian from the simplified model are only visible for
the larger $\met$ values.  For the largest mediator mass, $\msu=5$
TeV, the simplified model and the effective approximations are in full
agreement, for the total cross section and for the $\met$
distribution.  Our parton-level simulation in the main body of the
paper is entirely justified.


\end{document}